\documentclass[aps,prb,reprint,amsmath,amssymb,floatfix]{revtex4-2}
\usepackage[T1]{fontenc}
\usepackage[utf8]{inputenc}
\usepackage{bbm}
\usepackage{physics}
\usepackage{graphicx,color}
\usepackage{bm,mathrsfs,mathdots}
\usepackage[caption=false]{subfig}
\usepackage[export]{adjustbox}
\usepackage{comment}
\usepackage{amscd}
\usepackage{hyperref}
\hypersetup{colorlinks=true,linkcolor=blue,urlcolor=blue,citecolor=blue,
  pdftitle={Topological phases protected by half-chiral symmetry}}
\newcommand{\bea}{\begin{equation}}
\newcommand{\eea}{\end{equation}}


\begin{document}

\title{Topological phases protected by half-chiral symmetry}
\author{Chen-Shen Lee}
\email{clee3039@gatech.edu}
\author{Itamar Kimchi}
\email{ikimchi3@gatech.edu}
\affiliation{School of Physics, Georgia Institute of Technology, Atlanta, GA}
\begin{abstract}
Chiral (or sublattice) symmetry describes bipartite systems such as the Su-Schrieffer-Heeger (SSH) chain where same-sublattice A-A and B-B hoppings are forbidden, enabling the protection of topological phases.  
Here we identify new topological phases protected by half-chiral symmetry wherein chiral symmetry is preserved on only one sublattice, as may occur with sublattice imbalance (e.g.\ graphene-hBN) or localized orbitals (e.g.\ zincblende). Unlike conventional symmetry operators described by group theory, half-chiral symmetry is an algebraic constraint on the sublattice-projected Hamiltonian such as $PHP=0$, requiring a new framework for topological protection. We develop such a framework, identifying real-space bulk topological invariants and corresponding topological boundary modes that stick to zero energy. In two dimensions we construct a half-chiral Kekule honeycomb model and show that its unusual higher-order topology protects a single zero-energy mode at a single corner.
\end{abstract}

\maketitle
\tableofcontents

\section{Introduction}
\begin{figure*}[t]
\centering
\begin{minipage}[c]{0.48\textwidth}
\centering
\subfloat[]{
\includegraphics[width=\linewidth]{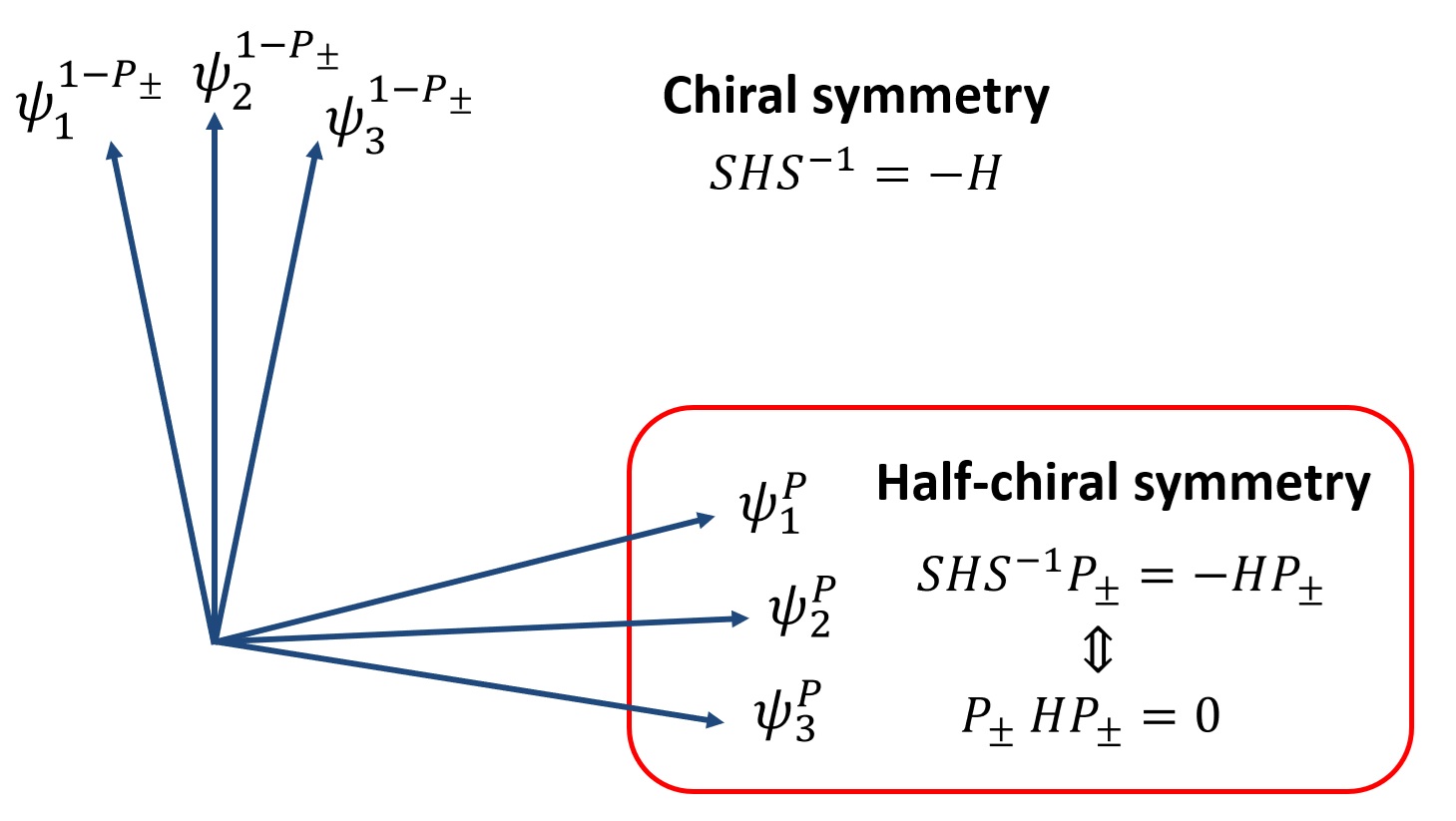}
\label{schematic_half_chiral_subspaces}
}
\end{minipage}
\hfill
\begin{minipage}[c]{0.50\textwidth}
\centering
\subfloat[]{
\includegraphics[width=\linewidth]{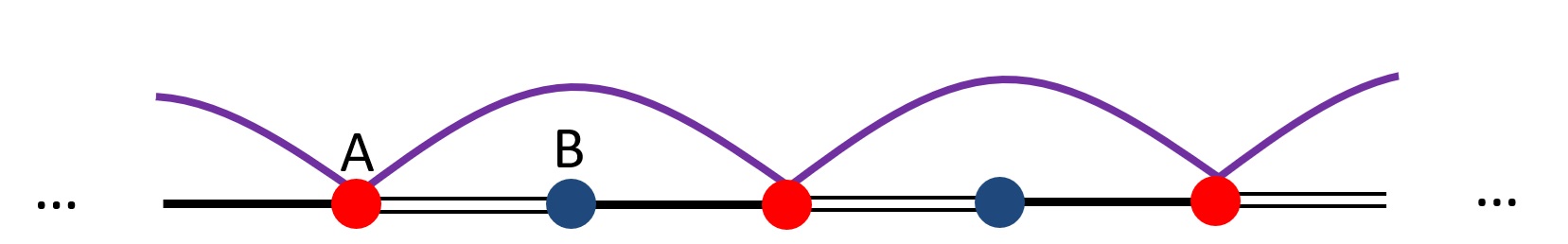}
\label{schematic_half_chiral_SSH_with_on-site}
}

\subfloat[]{
\includegraphics[width=\linewidth]{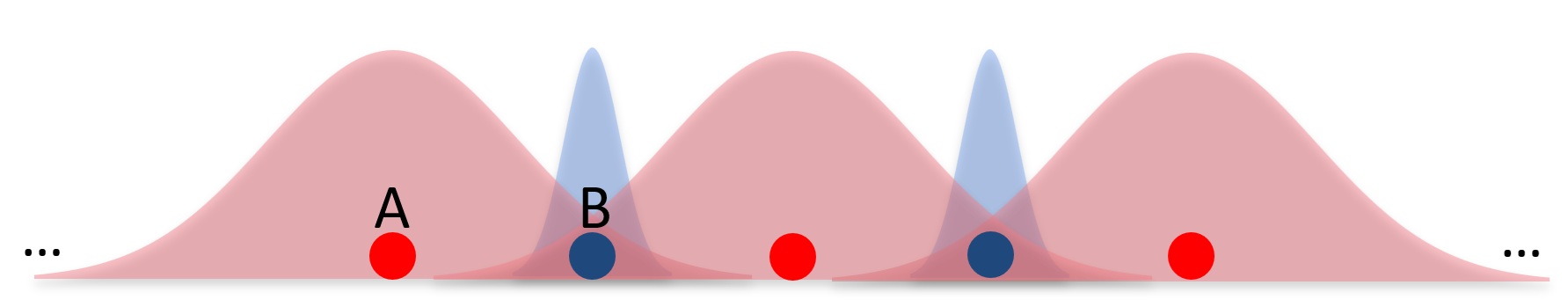}
\label{schematic_half_chiral_orbitals}
}
\end{minipage}
\caption{Chiral and half-chiral symmetries. (a) Illustration in the Hilbert space. The vectors $\psi_i^{P_\pm}$ and $\psi_i^{1-P_\pm}$ span the selected chiral subspace (red box) and its complement, respectively, with $P_\pm=(1\pm S)/2$. Chiral symmetry requires $SHS^{-1}=-H$ on the full Hilbert space, whereas half-chiral symmetry preserves this relation on the selected subspace: $SHS^{-1}P_\pm=-HP_\pm$, or equivalently $P_\pm HP_\pm=0$. (b) Half-chiral SSH chain, where the two chiral subspaces correspond to the A and B sublattices. The A and B sublattices can exhibit a chemical potential imbalance and long-range couplings within one of the sublattices, while preserving half-chiral symmetry and its associated topological phases with boundary modes. (c) Such models could arise from sublattice imbalance or from structures such as zincblende where one sublattice hosts a more localized atomic orbital. The shaded profiles schematically represent the spatial extent of the atomic orbitals, with the blue orbitals more localized than the red ones.}
\label{fig:schematic_half_chiral}
\end{figure*}

Symmetry has long played a central role in the classification of phases of matter. Recently, this importance has been further highlighted by the discovery of topological phases, which fall outside the conventional Landau paradigm of spontaneous symmetry breaking. Symmetry-protected topological (SPT) phases~\cite{gu_tensorentanglementfiltering_2009,chen_local_2010,chiu_classification_2016,fu_topological_2007a,hasan_colloquium_2010,sato_topological_2017} represent a significant category within this framework, where the bulk topology is inextricably linked to robust boundary states through the bulk–boundary correspondence. In free-fermion systems, SPT phases are systematically classified by the ten-fold way based on internal symmetries such as time-reversal, particle–hole, and chiral (sublattice) symmetries~\cite{schnyder_classification_2008,kitaev_periodic_2009,ryu_topological_2010,chiu_classification_2016}, and this framework has also been extended to include crystalline symmetries, leading to crystalline SPT phases~\cite{fu_topological_2007,fu_topological_2011,ando_topological_2015,morimoto_topological_2013,chiu_classification_2013,chiu_classification_2014,shiozaki_topology_2014}.
Bulk topological invariants are often discussed through momentum space constructions but can also be constructed in real space, such as the local Chern marker
\cite{bianco_mapping_2011}, Kitaev's three-region construction
\cite{kitaev_anyons_2006}, and recently Loring's spectral localizer which combines the Hamiltonian and position operators with a Clifford algebra \cite{
loring_theory_2015,loring_finite_2017,loring_guide_2019,
cerjan_local_2022}.

The scope of topological phases has further broadened with the emergence of higher-order topological (HOT) phases~\cite{benalcazar_quantized_2017,langbehn_reflectionsymmetric_2017,schindler_higherorder_2018,khalaf_higherorder_2018}. Unlike conventional SPT phases that host $(d-1)$-dimensional boundary states, a $d$-dimensional $n$-th order topological phase possesses protected states on $(d-n)$-dimensional boundaries. For instance, a 2D second-order topological phase manifests as 0D corner states instead of 1D edge modes. The underlying mechanism of many HOT phases can be understood in terms of boundary-localized mass domains~\cite{langbehn_reflectionsymmetric_2017,geier_secondorder_2018,trifunovic_higherorder_2019}. Because this description is based on the topology of the gapped boundaries, it does not directly yield a corresponding bulk topological invariant. Nevertheless, bulk topological invariants can still be constructed to characterize certain HOT phases. One route uses additional crystalline symmetries to define crystalline bulk invariants~\cite{khalaf_symmetry_2018,shiozaki_variants_2019,benalcazar_quantization_2019}. A few exceptions, such as the multipole chiral number (MCN) for chiral-symmetric HOT phases~\cite{benalcazar_chiralsymmetric_2022}, provide a bulk characterization without requiring crystalline symmetry. Another approach to characterizing HOT phases uses the spectral localizer, whose applications to HOT phases include systems protected by chiral or crystalline symmetries~\cite{cerjan_local_2022,cerjan_local_2024,cerjan_classifying_2024,dureth_probing_2025}.

Recently, there has been growing interest in topological phenomena that fall beyond the protection of conventional symmetry notions, with the two main developments involving the concepts of ``subsymmetry''~\cite{wang_subsymmetryprotected_2023} and of a subspace-property~\cite{shimomura_subspaceprotected_2026}. 
The concept of ``subsymmetry''~\cite{wang_subsymmetryprotected_2023} involves a symmetry that is broken globally remains preserved when restricted to a specific subspace of the Hilbert space. In particular, subsymmetry can protect specific zero-energy ($E=0$) boundary states even when the conventional bulk topological invariant loses its quantization. Subsequent studies have explored this protection in 1D topological insulators and superconductors~\cite{kang_subsymmetry_2024} and established a bulk-boundary correspondence using a non-Bloch winding number~\cite{verma_nonbloch_2024}. For 2D systems, this framework has been extended to higher-order topology, where a state-resolved quadrupole moment characterizes the corresponding topological corner states~\cite{kang_higherorder_2026}. A related study considered a Hamiltonian matrix structure that can be described in terms of chiral subsymmetry and rigorously established a bulk-boundary correspondence, relating a winding number defined from the off-diagonal block to the number of $E=0$ edge states in a semi-infinite 1D chain~\cite{lee_linear_2024}. 

The recently proposed subspace-property framework~\cite{shimomura_subspaceprotected_2026} is formulated in momentum space through constraints on how the Bloch Hamiltonian maps between Hilbert subspaces. Systems with the subspace property can support topological boundary states confined to specific subspaces in the Hermitian case and subspace skin-effect states in the non-Hermitian case. The skin effect is a phenomenon in which eigenstates accumulate at the system boundaries due to the non-Hermitian nature of the Hamiltonian~\cite{hatano_localization_1996,yao_edge_2018,kawabata_symmetry_2019,zhang_correspondence_2020,okuma_topological_2020,okuma_nonhermitian_2023}. The associated topology is characterized by momentum-space invariants constructed from restricted Hamiltonians. These unconventional topological phenomena are remarkable because they reveal new mechanisms for realizing robust boundary states, which could lead to novel physical effects and potential applications in engineering.

In this work we introduce half-chiral symmetry and construct a framework for its protection of topological phases. Half-chiral symmetry is related to the concepts above, but by specializing to its setting we are able to derive a more general set of results including the relation to conventional chiral symmetry, real space invariants, and higher order topology. Given a chiral operator $S$, half-chiral symmetry requires the chiral symmetry relation to hold on one of its two eigenspaces (sublattices), selected by $P_\pm=(1\pm S)/2$, while allowing it to be broken on the other, as illustrated in Fig.~\ref{schematic_half_chiral_subspaces}. This requirement takes the form $SHS^{-1}P_\pm=-HP_\pm$ and is equivalent to $PHP=0$, where $P$ denotes the selected projector to one of the sublattices (See Sec.~\ref{Definition of half-chiral symmetry}). It is thus specified by a projected algebraic constraint on the Hamiltonian rather than by a unitary or antiunitary symmetry operation on the full Hilbert space. We show that this constraint supports new classes of topological and HOT phases including a 2D HOT phase with a very unusual single topological $E=0$ corner state. We construct real-space half-chiral invariants to characterize these phases. We also identify 2D HOT phases with full chiral or half-chiral symmetry whose topological $E=0$ corner states are characterized by spectral-localizer invariants despite vanishing MCNs.
Note that throughout this paper, when we refer to $E=0$, this can correspond to any other specified energy $E_0$ by an energy shift, with corresponding shifts in all expressions such as $PHP = E_0 P$.

Approximate half-chiral symmetry may naturally arise in materials with a bipartite nearest-neighbor lattice when sublattice imbalance is introduced and when the atomic orbital on one of the sublattices is more spatially localized such that its same-sublattice longer-ranged hoppings can be neglected, as illustrated schematically in Figs.~\ref{schematic_half_chiral_SSH_with_on-site} and~\ref{schematic_half_chiral_orbitals}.
This enables chiral symmetry to persist for one of the sublattices, preventing long-range same-sublattice hopping and keeping its chemical potential fixed (at a reference energy $E_0$ which we use as the zero-energy measure in the rest of this manuscript). 
(For details of the definition of half-chiral symmetry see Sec.~\ref{Definition of half-chiral symmetry} below.)
This situation could arise in multiorbital systems where one type of atomic orbital is more spatially localized than the other, when the two types of orbitals reside on two interpenetrating sublattices. This occurs in many crystals, including zincblende structure materials such as the zinc chalcogenides ZnX where Zn and X atoms form two interpenetrating sublattices and the Zn 3d orbitals are more spatially localized than the chalcogen p orbitals. Another route proposed for the subspace property in Ref.~\cite{shimomura_subspaceprotected_2026} also applies to half-chiral symmetry: in heavy-fermion systems, strong electron-electron correlations can renormalize the electronic structure, yielding an effective single-particle description in which more localized electrons form a nearly flat band that hybridizes with a dispersive band of more itinerant electrons. 
More broadly, in systems such as photonic lattices~\cite{ozawa_topological_2019}, topoelectric circuits~\cite{lee_topolectrical_2018,dong_topolectric_2021}, and microwave resonator lattices~\cite{peterson_quantized_2018}, lattice positions and hopping amplitudes can be controlled independently and generate models with exact half-chiral symmetry. 
The 1D half-chiral SSH model (Fig.~\ref{fig:schematic_half_chiral}) has already been realized in photonic systems~\cite{wang_subsymmetryprotected_2023}. 
\newpage

\section{Overview of main results}\label{sec:overview_main_results}

\begin{table*}[t]

\centering
\small
\setlength{\tabcolsep}{4pt}
\renewcommand{\arraystretch}{1}

\resizebox{\textwidth}{!}{%
\begin{tabular}{lccccc}
\hline\hline
& \begin{tabular}[c]{@{}c@{}}1D chiral\\SPT phases\\(Sec.~\ref{1D chiral symmetric topological phases})\end{tabular}
 & \begin{tabular}[c]{@{}c@{}}1D half-chiral\\topological phases\\(Sec.~\ref{1D half-chiral topological phases})\end{tabular}
 & \begin{tabular}[c]{@{}c@{}}2D chiral HOT phases\\(MCN-characterized: Sec.~\ref{2D Chiral-symmetric Higher-order topological phases};\\SL with MCN $=0$: Sec.~\ref{2D Chiral-symmetric Higher-order topological phases without MCNs})\end{tabular}
 & \begin{tabular}[c]{@{}c@{}}2D HOT phases\\protected by crystalline\\and internal symmetries\\(e.g., Ref.~\cite{schindler_higherorder_2018})\end{tabular}
 & \begin{tabular}[c]{@{}c@{}}2D half-chiral HOT phases$^\dagger$\\(MCN-characterized: Sec.~\ref{2D half-chiral HOT phases with MCNs};\\SL with MCN $=0$: Sec.~\ref{2D half-chiral HOT phases without MCNs})\end{tabular} \\[6pt]
\hline
\noalign{\vskip 6pt}
\shortstack[l]{Topological boundary \\ states in an \\ elementary phase}
 & \shortstack{Paired $E=0$\\edge states}
 & \shortstack{A single unpaired\\$E=0$ edge state}
 & \shortstack{Paired $E=0$\\corner states}
 & \shortstack{$E=0$ states at\\$n\geq2$ corners}
 & \shortstack{A single unpaired\\$E=0$ corner state} \\[12pt]
\shortstack[l]{Spectral behavior\\at topological\\phase transitions}
 & \shortstack{Bulk-gap\\closing}
 & \shortstack{Bulk spectrum\\touches $E=0$}
 & \shortstack{Boundary- or\\bulk-gap closing}
 & \shortstack{Boundary- or\\bulk-gap closing}
 & \shortstack{Boundary or bulk\\spectrum touches $E=0$} \\[4pt]
\hline\hline
\end{tabular}%
}

\medskip
\begin{minipage}{\textwidth}
\footnotesize
\raggedright
$^\dagger$ Among the models studied here, the MCN-characterized 2D half-chiral HOT models host at least two topological $E=0$ corner states, whereas those characterized by spectral-localizer invariants despite vanishing MCNs can host a single topological $E=0$ corner state.
\end{minipage}
\caption{Comparison of the elementary topological phases considered in this work. ``SL with MCN $=0$'' denotes phases characterized by spectral-localizer invariants despite vanishing MCNs. The first row summarizes the number and pairing of protected $E=0$ boundary states in an elementary nontrivial phase of each class. The second row describes the spectral behavior of the physical Hamiltonian at transitions between topological phases. Relevant sections are indicated in the column headings.}
\label{tab:comparison_topological_phases}
\end{table*}

We give an overview of our main results and outline the structure of this paper.

We begin by reviewing conventional chiral symmetry and the topological phases it protects in Sec.~\ref{Review of Chiral Symmetry and Topological Phases}. In 1D, we discuss chiral-symmetric SPT phases and show how momentum-space winding numbers and real-space invariants characterize their topological $E=0$ edge states, with particular emphasis on the spectral localizerh~\cite{loring_theory_2015,loring_finite_2017,loring_spectral_2020,cerjan_classifying_2024} in real space. In 2D, we turn to chiral-symmetric higher-order topological (HOT) phases and show how multipole chiral numbers (MCNs) and spectral-localizer invariants can characterize their topological $E=0$ corner states. These conventional phases and their invariants provide the starting point for our half-chiral construction.

In Sec.~\ref{Hermitian Subspace-Chiral Symmetry and Topological Phases}, we define half-chiral symmetry and establish its bulk-boundary correspondence. A Hamiltonian $H$ satisfies half-chiral symmetry if the chiral relation $SHS^{-1}P_\pm=-HP_\pm$ holds for one of the two projectors $P_\pm=(1\pm S)/2$. For any such $H$, we can define an associated chiralized Hamiltonian $H_{\text{chiralized}}$ that obeys conventional chiral symmetry. The key observation is that when the half-chiral condition is imposed with $P_\pm$, any $E=0$ state of $H_{\text{chiralized}}$ with $S=\pm1$ is also an $E=0$ state of $H$. Through this mechanism, $H$ inherits topologically protected $E=0$ boundary states from $H_{\text{chiralized}}$, and we define half-chiral topological phases by the presence of these robust states. To characterize these phases, we define each half-chiral invariant as the corresponding conventional chiral invariant evaluated for $H_{\text{chiralized}}$. The resulting bulk-boundary correspondence relates these invariants to the protected $E=0$ boundary states of the physical Hamiltonian. The framework applies in arbitrary spatial dimensions, accommodates both momentum-space and real-space formulations, and provides a half-chiral counterpart for every conventional chiral topological phase.

We then construct 1D half-chiral topological phases from conventional chiral-symmetric SPT phases in Sec.~\ref{1D half-chiral topological phases}. 
We show how to characterize these phases using a real-space half-chiral spectral localizer invariant inspired by Loring's spectral localizer~\cite{loring_theory_2015}, as the signature of 
$\Tilde{L}_{\text{half-chiral}}(x,0,H) =\frac{i}{2}(HS-SH)-\kappa(X-x\mathbbm{1})S$ with $S$ the chiral symmetry (see Eq.~\ref{half-chiral reduece spectral localizer} in Sec.~\ref{Hermitian Subspace-Chiral Symmetry and Topological Phases}).
This invariant, which for clean systems becomes equal to a known momentum space winding number invariant, gives the number of topological $E=0$ edge states. In their momentum space formulations these types of phases have already been studied in the context of subsymmetry~\cite{wang_subsymmetryprotected_2023,lee_linear_2024,kang_subsymmetry_2024,verma_nonbloch_2024,cheng_subsymmetryprotected_2026} and the subspace property~\cite{shimomura_subspaceprotected_2026}. Under appropriate conditions, the constraints in these two formulations are equivalent to half-chiral symmetry, as explained in Sec.~\ref{Definition of half-chiral symmetry}. However, the topological invariants derived in earlier work for the 1D phases considered there are formulated in momentum space. For the 1D half-chiral phases constructed here, we formulate topological invariants in both momentum and real space. The real-space formulation yields a bulk-boundary correspondence without relying on translational symmetry, making it applicable to disordered lattices, quasicrystals, and amorphous structures.

In Sec.~\ref{2D half-chiral HOT phases with MCNs}, we turn to 2D and construct half-chiral HOT phases from conventional chiral-symmetric HOT phases characterized by MCNs~\cite{benalcazar_chiralsymmetric_2022}. We characterize the resulting phases using the half-chiral MCN, whose nonzero values indicate the presence of topological $E=0$ corner states. These phases can also be obtained by extending the real-space construction of 1D half-chiral topological phases to 2D.

Moreover, in this section we go beyond half-chiral symmetry to identify a class of 2D chiral-symmetric HOT phases whose topological $E=0$ corner states are characterized by spectral-localizer invariants despite vanishing MCNs. We demonstrate this in Sec.~\ref{2D Chiral-symmetric Higher-order topological phases without MCNs} using a modified Kekulé-honeycomb model whose corner states persist when crystalline symmetry is broken while chiral symmetry is preserved. The spectral-localizer invariant for chiral-symmetric HOT phases was conjectured to be a local version of the corresponding MCN~\cite{cerjan_local_2022}. Our example shows that the spectral localizer can resolve topological corner states even when the MCN vanishes, revealing higher-order topological information beyond that encoded by the MCN.

In Sec.~\ref{2D half-chiral HOT phases without MCNs}, we construct half-chiral counterparts of the 2D chiral-symmetric HOT phases characterized by spectral-localizer invariants despite vanishing MCNs. The resulting phases can be characterized by a half-chiral spectral-localizer invariant. We illustrate these phases by adding terms to the modified Kekulé-honeycomb model that break full chiral symmetry while preserving half-chiral symmetry. Interestingly, the resulting half-chiral Kekulé model hosts a single topological $E=0$ corner state. Its protection does not require crystalline symmetry. This simple model may also be experimentally realizable.

The framework of topological protection for Hamiltonian constraints such as half-chiral symmetry requires us to clarify the definition of ``topological phase'' in the half-chiral setting: it refers to phases characterized by robust boundary states whose energies remain pinned at $E=0$. We do not restrict this terminology to phases whose transitions require a bulk-gap closing, as in conventional SPT phases. Nevertheless, topological transitions in half-chiral systems have a distinct signature in the spectrum of the physical Hamiltonian. For (one-dimensional) half-chiral topological phases, the bulk spectrum of the physical Hamiltonian touches $E=0$ at a transition, but the bulk band gap can remain open. For half-chiral HOT phases, it is the boundary or bulk spectrum that touches $E=0$, again without necessarily closing the corresponding band gap. In both cases, the corresponding gap closes in the chiralized Hamiltonian. Table~\ref{tab:comparison_topological_phases} compares conventional SPT and HOT phases with their half-chiral counterparts in terms of the protected boundary states supported by elementary nontrivial phases and the spectral behavior at topological transitions.

\section{Review of chiral symmetry and topological phases}\label{Review of Chiral Symmetry and Topological Phases}
In this section, we review chiral symmetry and its associated topological properties. In 1D, chiral symmetry protects SPT phases belonging to the $\mathbb{Z}$ classification, which can be characterized by either momentum-space winding numbers or real-space invariants, as shown in Sec.~\ref{1D chiral symmetric topological phases}. In 2D, chiral symmetry supports higher-order topological phases that can be characterized by real-space multipole chiral numbers, as discussed in Sec.~\ref{2D Chiral-symmetric Higher-order topological phases}. Spectral-localizer invariants provide another characterization of these HOT phases, as discussed in Sec.~\ref{2D Chiral-symmetric Higher-order topological phases and spectral localizer}.

We begin by recalling the definition of chiral symmetry. A non-interacting system has chiral symmetry if there exists a unitary operator \(S\), called the chiral operator, such that the Hamiltonian \(H\) satisfies
\bea
S H S^{-1} = -H,
\eea
with $S^2 = 1$. Here, $H$ denotes the single-particle Hamiltonian. In a basis where the chiral operator is diagonal, the Hamiltonian can be written in an off-diagonal form,
\bea
H=\left(
\begin{matrix}
0 & M \\
M^{\dag} & 0\\
\end{matrix}
\right),\quad \text{with}\;
S=\left(
\begin{matrix}
\mathbbm{1} & 0 \\
0 & -\mathbbm{1}\\
\end{matrix}
\right),
\eea
which naturally decomposes the system into two chiral subspaces corresponding to the eigenvalues $S=\pm1$. In a bipartite lattice, these subspaces correspond to the two sublattices.

\subsection{Chiral-symmetric topological phases in 1D}\label{1D chiral symmetric topological phases}
\begin{figure*}[t]
\centering
\begin{adjustbox}{max totalsize={\textwidth}{0.72\textheight},center}

\begin{minipage}[c]{\textwidth}
\centering
\begin{minipage}[c]{0.35\textwidth}
\centering
\subfloat[]{
\includegraphics[width=0.95\linewidth]{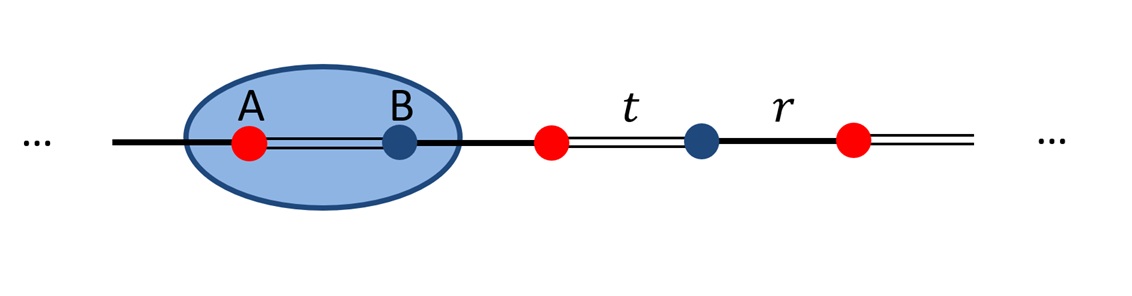}
\label{SSH_figure}}
\end{minipage}\hfill
\begin{minipage}[c]{0.65\textwidth}
\centering
\subfloat[]{
\includegraphics[width=0.48\linewidth,valign=c]{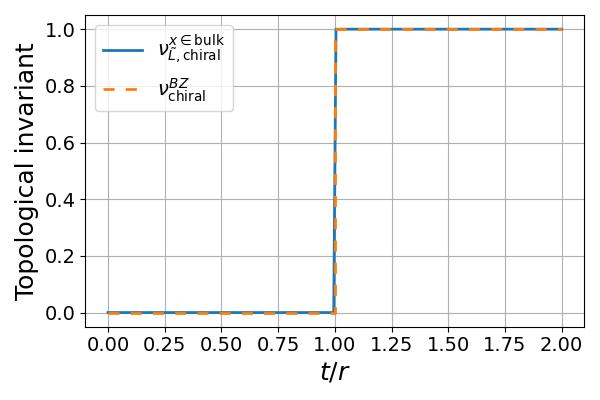}
\label{SSH_phase}}\hfill
\subfloat[]{
\includegraphics[width=0.48\linewidth,valign=c]{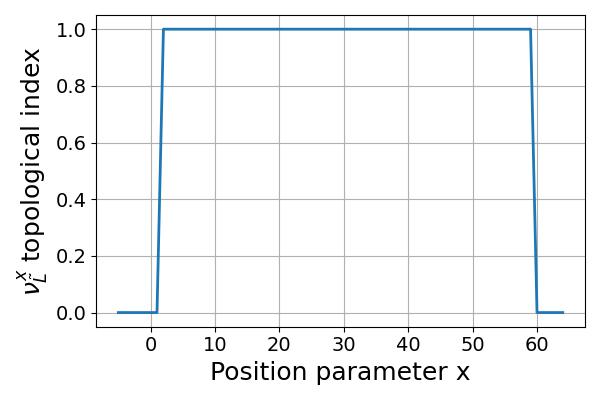}
\label{SSH_localizer_spatial}}\\[6pt]
\subfloat[]{
\includegraphics[width=0.48\linewidth,valign=c]{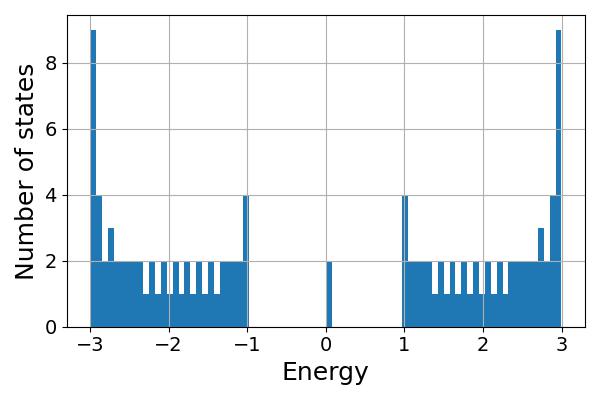}
\label{SSH_En_distribution}}\hfill
\subfloat[]{
\includegraphics[width=0.48\linewidth,valign=c]{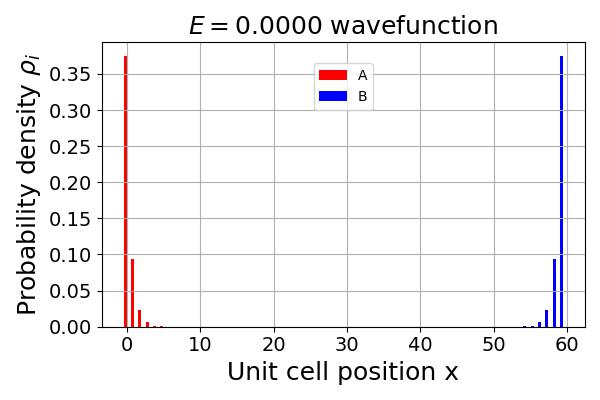}
\label{SSH_wavefunction}}
\end{minipage}
\end{minipage}

\end{adjustbox}
\par
\caption{(a) The conventional Su-Schrieffer-Heeger (SSH) model of Eq.~\eqref{SSH} which obeys conventional (full) chiral symmetry. Lines with different styles denote different hopping terms, and the shaded area represents the unit cell. (b) Phase diagram showing the winding number $\nu^{BZ}_{\text{chiral}}$ and the spectral-localizer invariant $\nu^{x}_{\Tilde{L},\text{chiral}}$ for the SSH model. To evaluate $\nu^{x\in\text{bulk}}_{\Tilde{L},\text{chiral}}$, we choose $\kappa=0.05$ and take $x$ at the center of the chain, i.e., $x=N_{\text{cell}}/2$. (c) Spatial dependence of $\nu^{x}_{\Tilde{L},\text{chiral}}$ for the SSH model under OBCs with $t=1$, $r=2$, $N_{\text{cell}}=60$, and $\kappa=0.05$. The unit increase near the left edge and unit decrease near the right edge predict one $S=+1$ and one $S=-1$ topological $E=0$ edge state, respectively. (d) Energy distribution of states for the same SSH system as in (c), showing two $E=0$ edge states. (e) Total probability density of the $E=0$ states, $\rho_i=\sum_n|\langle i|\psi_{0,n}\rangle|^2$, where $\ket{\psi_{0,n}}$ denotes the $n$-th zero-energy eigenstate and $\ket{i}$ denotes the real-space basis state localized at site $i$. Red and blue indicate the probability density on the $A$ and $B$ sublattices, respectively, corresponding to the chiral subspaces $S=+1$ and $S=-1$. The left and right edge states occupy the $A$ and $B$ sublattices, respectively, as predicted in (c).}
\end{figure*}

1D chiral-symmetric topological phases belong to class AIII in the Altland–Zirnbauer (AZ) classification. In this class, gapped Hamiltonians are topologically distinguished by an integer $\mathbb{Z}$ invariant, which characterizes distinct topological phases that cannot be smoothly deformed into each other without closing the bulk gap while preserving chiral symmetry. This invariant can be formulated either in momentum space or, equivalently, in real space.

In momentum space, the topological invariant can be expressed as a winding number associated with the off-diagonal block of the chiral-symmetric Hamiltonian (see Ref.~\cite{chiu_classification_2016} and references therein). In the chiral basis where the Bloch Hamiltonian takes the form
\bea\label{bulk and chiral}
\mathscr{H}(k)=\left(
\begin{matrix}
0 & M(k) \\
M^{\dag}(k) & 0\\
\end{matrix}
\right)
\eea
the winding number is given by
\bea\label{winding number}
\begin{aligned}
\nu^{BZ}_{\text{chiral}}(\mathscr{H}) &= \frac{1}{2\pi i} \int_{BZ}dk\;\text{Tr}[(M^{\dag})^{-1}\partial_kM^{\dag}],\\
&=\frac{1}{2\pi i} \int_{BZ}dk\; \partial_k \text{log}(\text{det}[M^{\dag}]),
\end{aligned}
\eea
where $BZ$ denotes the first Brillouin zone, the notation $\text{Tr}$ is trace, and $\text{det}$ represents the determinant.

In real space, several equivalent formulations of the topological invariant exist, all of which do not require translational symmetry and can therefore be applied to disordered lattices, quasicrystals, and amorphous structures. In this work, we adopt the spectral-localizer approach~\cite{loring_theory_2015,loring_finite_2017,loring_spectral_2020,cerjan_classifying_2024}, as it is more general and also less susceptible to finite-size effects than other formulations, which we discuss further in the Appendix.~\ref{Real-space invariants for 1D chiral symmetric topological phases}. We now briefly recall the spectral localizer formalism and then focus on its realization in 1D chiral symmetric systems. For a $d$-dimensional system, the spectral localizer is a Hermitian operator defined as
\bea
\begin{aligned}
L(\mathbf{x},E,H)
&=\sum_{j=1}^{d}\kappa(X_j-x_j\mathbbm{1})\otimes\Gamma_j\\
&\quad +(H-E\mathbbm{1})\otimes\Gamma_{d+1}.
\end{aligned}
\eea
where $\kappa > 0$ is a dimensionful coefficient, $X$ is the position operator, $H$ is the system Hamiltonian with open boundary conditions (OBCs), $\Gamma_j$ are the generators of an irreducible Clifford algebra, and the energy and position spectra of the system are shifted to be centered at $(\mathbf{x}, E)$. The corresponding index, such as the signature, is invariant under continuous gap-preserving deformations, thereby providing a real-space topological invariant. 

However, in odd spatial dimensions, an irreducible Clifford representation necessarily involves an additional anticommuting generator, which leads to an off-block-diagonal form of the spectral localizer in an appropriate basis. This structure implies that the two blocks encode opposite spectral information, resulting in a trivial contribution to the associated index. To obtain a nontrivial invariant, one therefore constructs a reduced spectral localizer using a single block, with a symmetry constraint ensuring that the resulting operator admits a well-defined Hermitian structure. In 1D with chiral symmetry, the reduced spectral localizer used in Ref.~\cite{cerjan_classifying_2024} takes the form
\bea
\Tilde{L}^{\text{Ref.~\cite{cerjan_classifying_2024}}}_{\text{chiral}}(x,0,H)=\Big[\kappa(X-x\mathbbm{1})-iH\Big]S,
\eea
where $S$ is the chiral operator. Here, we adopt the opposite overall sign and define
\bea\label{spectral localizer}
\begin{aligned}
\Tilde{L}_{\text{chiral}}(x,0,H)
&=-\Tilde{L}^{\text{Ref.~\cite{cerjan_classifying_2024}}}_{\text{chiral}}(x,0,H)\\
&=\Big[iH-\kappa(X-x\mathbbm{1})\Big]S.
\end{aligned}
\eea
The construction is restricted to $E=0$, where the reduced localizer is Hermitian. This Hermitian structure enables the definition of the topological invariant through its signature:
\bea\label{local winding number}
\nu^{x}_{\Tilde{L},\text{chiral}}(H)=\frac{1}{2}\text{sig}\Big[\Tilde{L}_{\text{chiral}}(x,0,H)\Big].
\eea
Here, $\text{sig}[A]$ denotes the signature of a matrix $A$ whose eigenvalues are real, defined as the difference between the number of its positive and negative eigenvalues. The prefactor $1/2$ accounts for the fact that when an eigenvalue of the matrix changes sign, its signature changes by two. Our overall sign convention for the reduced spectral localizer is chosen so that this invariant agrees in sign with the winding number defined in Eq.~\eqref{winding number}.

The bulk-boundary correspondence relates bulk topology to boundary states. In 1D chiral-symmetric systems under OBCs, it relates the topological invariant to the number of topological $E=0$ edge states. For the two invariants introduced above, the winding number $\nu^{BZ}_{\text{chiral}}$ and the spectral-localizer invariant $\nu^{x}_{\Tilde{L},\text{chiral}}$ with $x$ in the bulk, the absolute value of each equals half the number of topological $E=0$ edge states. 

For the spectral-localizer invariant, the spatial dependence provides further information about the positions and chiralities of the topological $E=0$ edge states. To understand how this information is encoded, consider a topological edge state $\ket{\psi_0}$ localized near a boundary with localization center $x_0$. Since chiral symmetry leaves the zero-energy subspace invariant under $S$, we can choose this state to have a definite chirality $s_0=\pm1$, satisfying $H\ket{\psi_0}=0$ and $S\ket{\psi_0}=s_0\ket{\psi_0}$. Together with the localization approximation $X\ket{\psi_0}\simeq x_0\ket{\psi_0}$, these relations allow Eq.~\eqref{spectral localizer} to be written as
\bea\label{1D localizer spectral-localizer counting relation}
\Tilde{L}_{\text{chiral}}(x,0,H)\ket{\psi_0}
\simeq s_0\kappa(x-x_0)\ket{\psi_0}.
\eea
Thus, the localized edge state produces a localizer eigenvalue that crosses zero as the position parameter $x$ passes through $x_0$. The direction of this crossing is determined by the state's chirality: the eigenvalue increases through zero for $s_0=+1$ and decreases through zero for $s_0=-1$. These crossings change the localizer signature by $+2$ and $-2$, respectively, so a single localized edge state contributes $\Delta\nu_{\Tilde{L},\text{chiral}}=s_0$. If several $E=0$ states are localized near the same position $x_0$, their contributions add, giving the general relation
\bea
\begin{aligned}
\Delta\nu_{\Tilde{L},\text{chiral}}
&=
\nu^{x_+}_{\Tilde{L},\text{chiral}}(H)
-\nu^{x_-}_{\Tilde{L},\text{chiral}}(H)\\
&=\sum_{i=1}^{n}s_i=n_+-n_-,
\end{aligned}
\eea
where $s_i=\pm1$ is the chirality of the $i$-th localized state, $n_\pm$ count the states with chirality $S=\pm1$, and $x_-<x_0<x_+$ are chosen on the invariant plateaus on either side of the localized states. In 1D chiral-symmetric SPT phases, the topological protection of boundary zero modes is characterized by their net chirality. Zero modes of opposite chirality at the same boundary can couple and shift away from zero energy without breaking chiral symmetry, forming topologically trivial pairs. Therefore, the number of topological $E=0$ edge states is given by the absolute value of the net chirality of all $E=0$ states at that boundary. The sign of this net chirality determines the common chirality of the topological states. Connecting this physical picture with the invariant-jump relation derived above, we find that the position of the change identifies where the topological states are localized, while its magnitude and direction give their number and chirality, respectively. Specifically, when $n$ topological $E=0$ states with common chirality $s_0$ are localized near $x_0$, the invariant change across this position satisfies
\bea\label{1D localizer invariant jump}
\operatorname{sgn}\!\left(\Delta\nu_{\Tilde{L},\text{chiral}}\right)=s_0,
\qquad
|\Delta\nu_{\Tilde{L},\text{chiral}}|=n,
\eea
where $\operatorname{sgn}$ denotes the sign function.

To illustrate the bulk-boundary correspondence and the relation between spatial changes in the spectral-localizer invariant and boundary-state chirality, we consider the paradigmatic Su–Schrieffer–Heeger (SSH) model~\cite{su_solitons_1979}, whose Hamiltonian is given by
\bea\label{SSH}
H_{\text{SSH}} = \sum_{n=1}^{N_{\text{cell}}} \Big(
t \; c_{n,A}^\dagger c_{n,B} 
+ r \; c_{n,B}^\dagger c_{n+1,A}
+ \text{h.c.} \Big).
\eea
Here, $c_{n,A}^\dagger$ and $c_{n,B}^\dagger$ are the fermion creation operators corresponding to the sublattices $A$ and $B$ (red and blue sites in Fig.~\ref{SSH_figure}), respectively, and $N_{\text{cell}}$ denotes the number of unit cells. The systems respects the chiral symmetry with the operator
\bea
S = \sum_{n} \left(
c_{\mathbf n,A}^\dagger c_{\mathbf n,A}
- c_{\mathbf n,B}^\dagger c_{\mathbf n,B}
\right).
\eea
The $A$ and $B$ sublattices therefore correspond to the chiral sectors $S=+1$ and $S=-1$, respectively.

\begin{figure*}[t]
\centering
\begin{minipage}[t]{0.27\textwidth}
\vspace{0pt}
\centering
\subfloat[]{\includegraphics[width=0.9\linewidth]{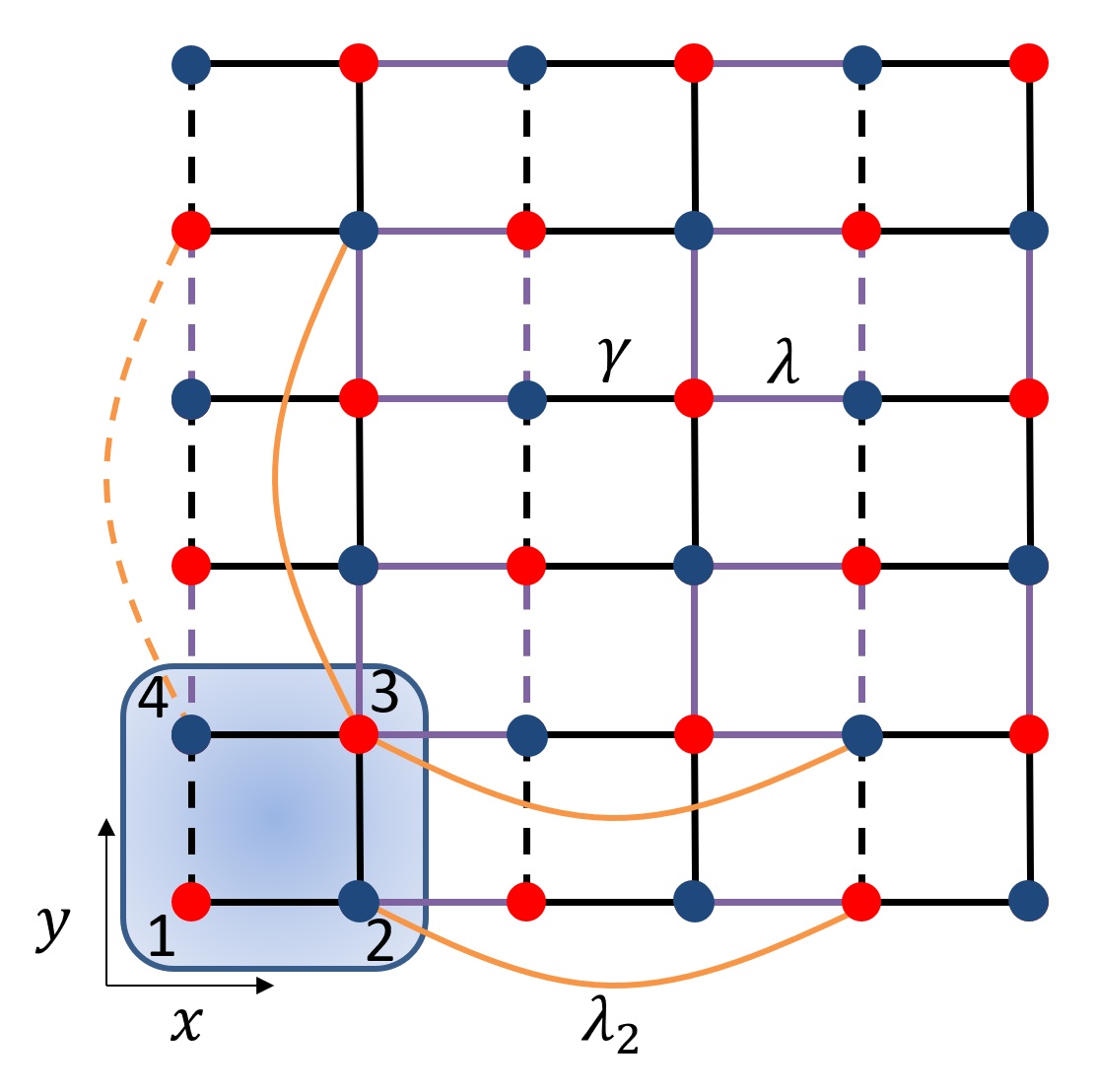}\label{LRQTI_figure}}\\[6pt]
\subfloat[]{\includegraphics[width=0.9\linewidth]{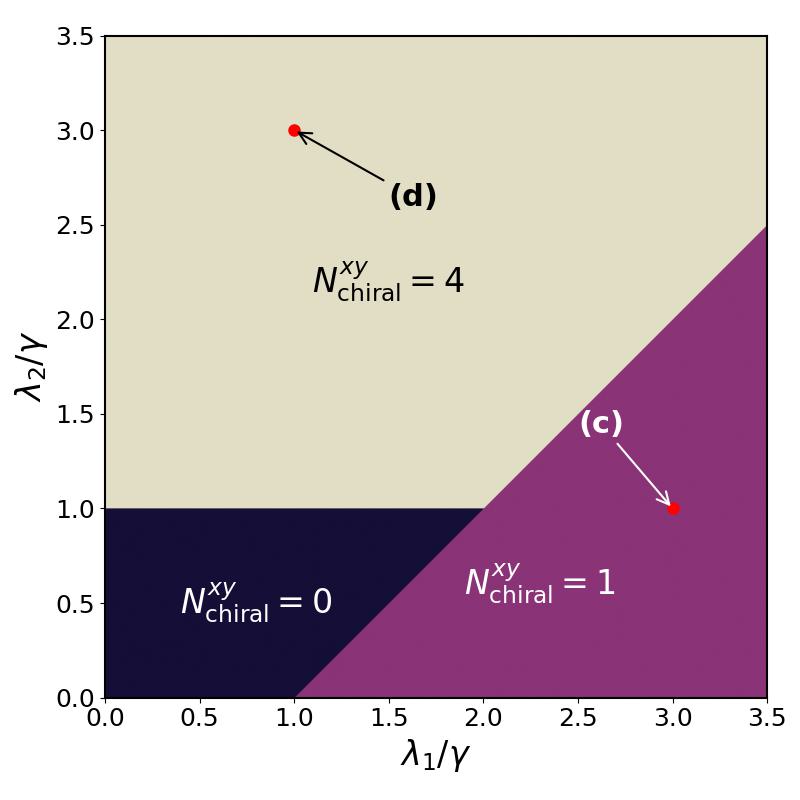}\label{LRQTI_phase}}
\end{minipage}\hskip 0.cm
\begin{minipage}[t]{0.68\textwidth}
\vspace{0pt}
\centering
\subfloat[]{
\includegraphics[width=0.42\linewidth,valign=c]{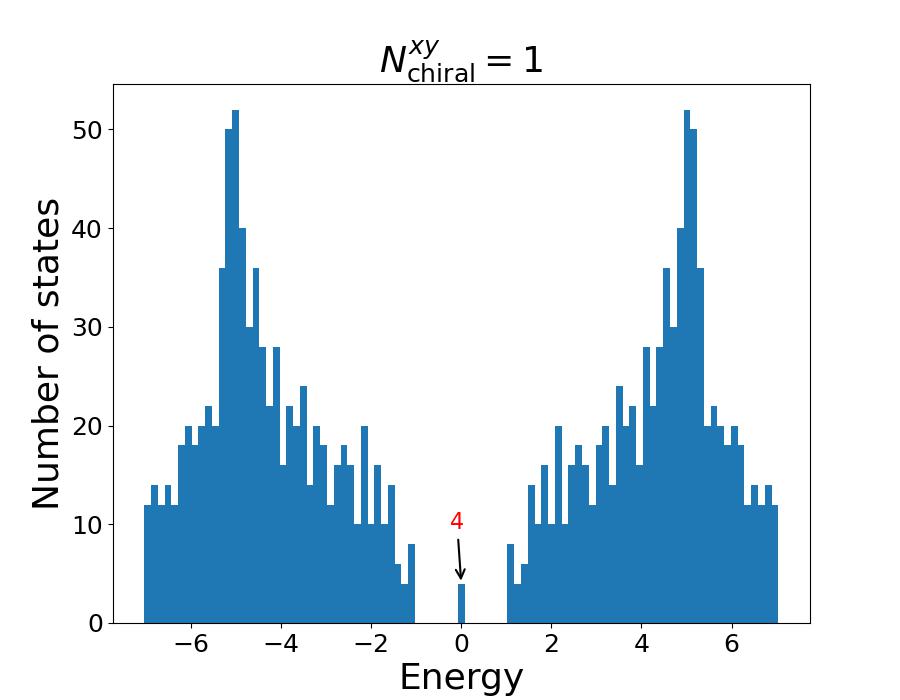}\hfill
\includegraphics[width=0.42\linewidth,valign=c]{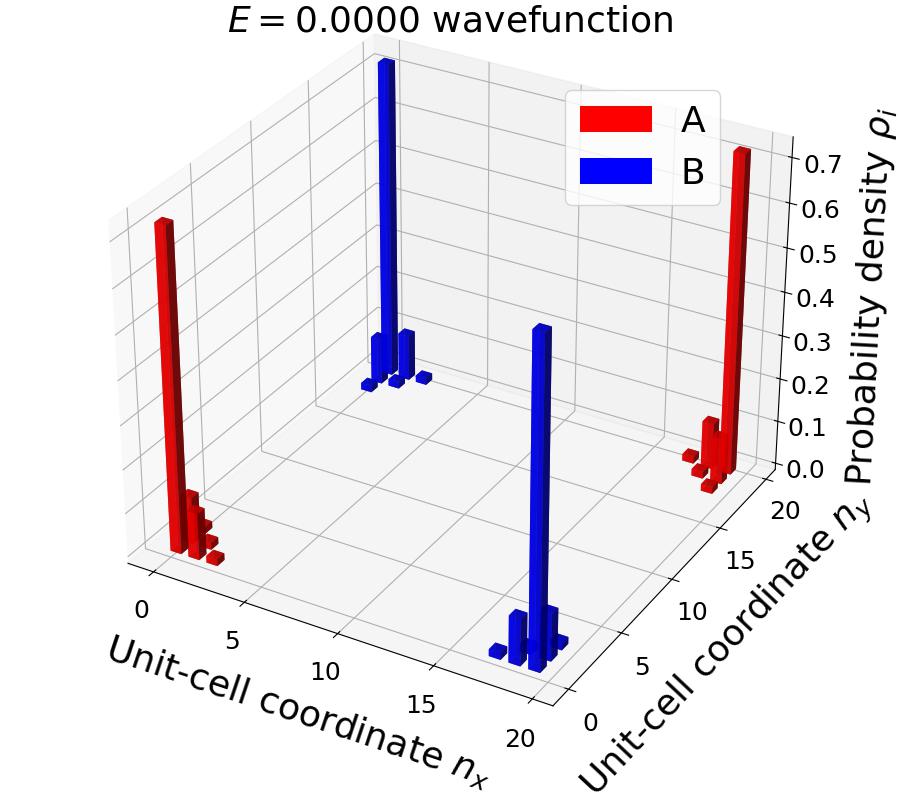}
\label{LRQTI_N=1}}\\[6pt]
\subfloat[]{
\includegraphics[width=0.42\linewidth,valign=c]{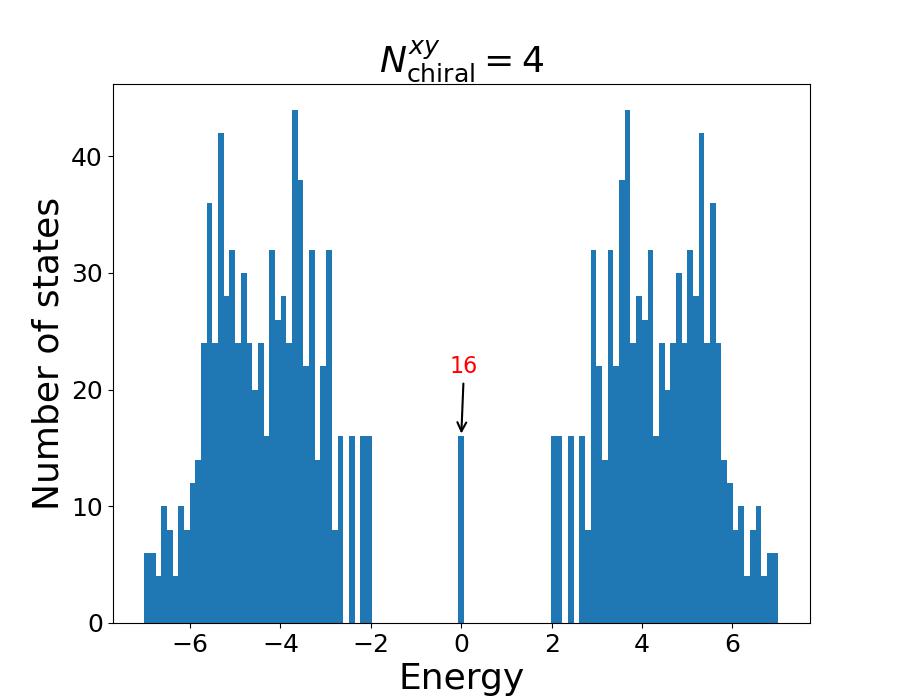}\hfill
\includegraphics[width=0.42\linewidth,valign=c]{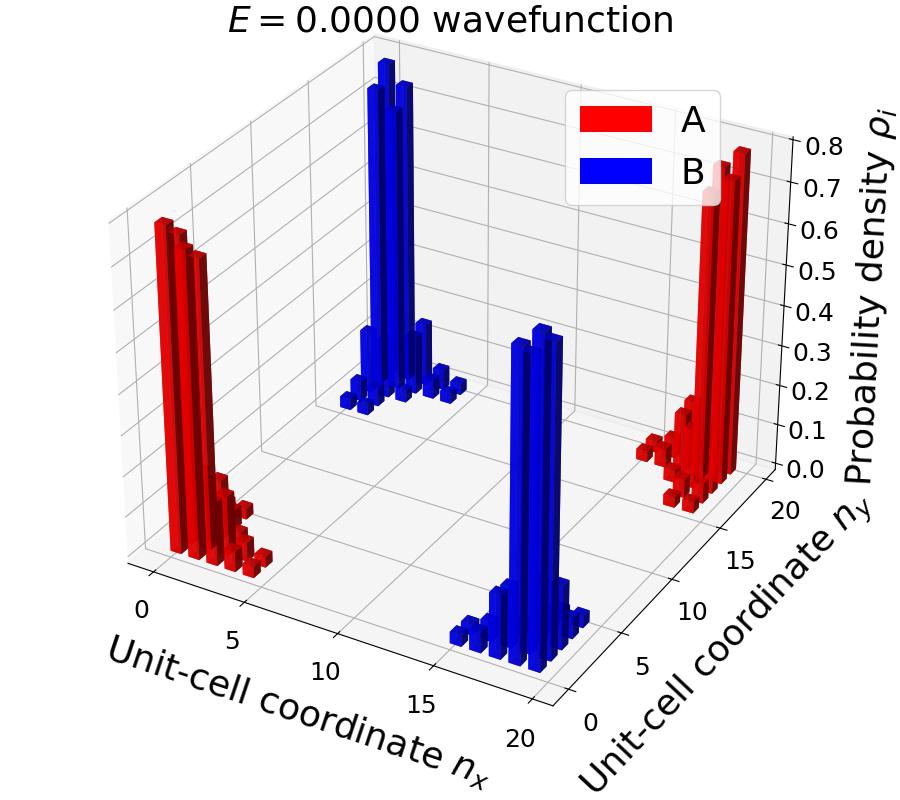}
\label{LRQTI_N=4}}
\end{minipage}
\par
\caption{(a) Schematic illustration of the longer-range
quadrupole topological insulator (LRQTI) in Eq.~\eqref{LRQTI}. Lines with different colors denote different hoppings, and dashed lines indicate hopping terms with negative signs, such that each plaquette carries a uniform $\pi$ flux. For clarity, only a few longer-range hoppings ($\lambda_2$) are shown. (b) Phase diagram showing the multipole chiral number $N^{xy}_{\text{chiral}}$ for the LRQTI. (c) Energy distribution of states (left) and total probability density of the $E=0$ states (right) for a $20\times20$-unit-cell LRQTI in the $N^{xy}_{\text{chiral}}=1$ phase. The parameters are $\gamma=1$, $\lambda_1=3$, and $\lambda_2=1$. Four $E=0$ corner states are present and all are localized at the corners. (d) Same as (c) but in the $N^{xy}_{\text{chiral}}=4$ phase with $\gamma=1$, $\lambda_1=1$, and $\lambda_2=3$. Sixteen $E=0$ corner states are present and localized at the corners.}
\end{figure*}

As shown in Fig.~\ref{SSH_phase}, the system lies in the nontrivial phase with $\nu^{BZ}_{\text{chiral}}=\nu^{x\in\text{bulk}}_{\Tilde{L},\text{chiral}}=1$ when $r>t$, predicting two topological $E=0$ edge states through the bulk-boundary correspondence. For $t=1$, $r=2$, $N_{\text{cell}}=60$, and $\kappa=0.05$, the spatial profile of $\nu^{x}_{\Tilde{L},\text{chiral}}$ in Fig.~\ref{SSH_localizer_spatial} increases from $0$ to $1$ near the left edge and decreases from $1$ to $0$ near the right edge. According to Eq.~\eqref{1D localizer invariant jump}, these unit changes predict one $S=+1$ topological $E=0$ state at the left edge and one $S=-1$ topological $E=0$ state at the right edge. The energy distribution in Fig.~\ref{SSH_En_distribution} confirms the presence of two $E=0$ edge states. Their probability density in Fig.~\ref{SSH_wavefunction} further shows that the left and right edge states occupy the $A$ and $B$ sublattices, respectively, as predicted by the spectral-localizer invariant.

\subsection{Chiral-symmetric 2D higher-order topological phases and multipole chiral number}\label{2D Chiral-symmetric Higher-order topological phases}
Beyond the 1D phases discussed above, chiral symmetry also supports higher-order topological phases in 2D. These phases host toppological $E=0$ corner states whose origin can be understood through the topology of the gapped 1D boundaries. Specifically, a junction between topologically distinct boundary segments forms a domain wall that supports $E=0$ localized states~\cite{langbehn_reflectionsymmetric_2017,geier_secondorder_2018,trifunovic_higherorder_2019}. From this boundary perspective, these HOT phases admit a $\mathbb{Z}$ classification, corresponding to that of 1D chiral-symmetric SPT phases. However, this boundary description does not by itself yield a corresponding bulk topological invariant.

Nevertheless, bulk invariants associated with these phases do exist. There are two commonly used invariants. The first is the quantized quadrupole moment~\cite{benalcazar_quantized_2017}, which provides a $\mathbb{Z}_2$ characterization and is therefore insufficient to capture the full $\mathbb{Z}$ classification of these HOT phases. The second is the multipole chiral number (MCN)~\cite{benalcazar_chiralsymmetric_2022}, which is $\mathbb{Z}$-valued and can distinguish topological phases that the quadrupole moment cannot. We will discuss the MCN below.

The construction of the MCN originates from a real-space reformulation of the topological winding number. In the standard AZ classification, 2D chiral-symmetric systems are topologically trivial, and the conventional momentum-space winding number is ill-defined. However, the 1D winding number can be reformulated in real space using sublattice-resolved operators. This real-space formulation can be extended to higher dimensions, leading to the $\mathbb{Z}$-valued MCN. In 2D, the MCN is defined as
\bea\label{MCN}
N^{xy}_{\text{chiral}}(H) = \frac{1}{2\pi i} \text{Tr} \log \left(\bar{Q}^{+}_{xy} \bar{Q}^{-\dagger}_{xy}\right),
\eea
where $\bar{Q}^{\pm}_{xy} = U_{\pm}^\dagger Q^{\pm}_{xy} U_{\pm}$. Here, the superscripts $+$ and $-$ label the chiral subspaces corresponding to $S = +1$ and $S = -1$, respectively. The operator $Q^{\pm}_{xy}$ is the sublattice-resolved quadrupole moment operator restricted to each chiral sector,
\bea\label{Q_xy def}
Q^{\pm}_{xy} = \sum_{\mathbf{R}, \alpha \in \pm} \ket{\mathbf{R}, \alpha} \exp \left(-i \frac{2\pi xy}{L_x L_y}\right) \bra{\mathbf{R}, \alpha},
\eea
where $\mathbf{R}=(x,y)$ labels the unit cells and $\alpha$ denotes internal degrees of freedom on each sublattice. The unitary matrices $U_{\pm}$ are constructed from the singular value decomposition of the off-diagonal block of the chiral-symmetric Hamiltonian. Specifically, in the chiral basis, the Hamiltonian takes the form
\bea
H = \begin{pmatrix} 0 & M \\ M^\dagger & 0 \end{pmatrix},
\eea
and we perform the decomposition
\bea
M = U_{+} \Sigma U_{-}^\dagger,
\eea
where
\bea
\begin{gathered}
U_{+} = \left( \ket{\phi^{+}_1}, \ket{\phi^{+}_2}, \dots \right), \\ U_{-} = \left( \ket{\phi^{-}_1}, \ket{\phi^{-}_2}, \dots \right).
\end{gathered}
\eea
The states $\ket{\phi^{+}_n}$ and $\ket{\phi^{-}_n}$ are obtained from the eigenvectors of $MM^\dagger$ and $M^\dagger M$, respectively. In fact, the eigenstate of $H$ with energy $\epsilon_n$ can be written as $\ket{\psi_{n}} = \frac{1}{\sqrt{2}}(\ket{\phi^{+}_n}, \ket{\phi^{-}_n})^T$. The corresponding state with energy $-\epsilon_n$ then takes the form $S\ket{\psi_{n}} = \frac{1}{\sqrt{2}}(\ket{\phi^{+}_n}, -\ket{\phi^{-}_n})^T$.

To illustrate the bulk–boundary correspondence characterized by the MCNs, we consider the longer-range quadrupole topological insulator (LRQTI) model introduced in Ref.~\cite{benalcazar_chiralsymmetric_2022}. In this system, the correspondence manifests as the presence of $4|N^{xy}_{\text{chiral}}|$ zero-energy corner states. The Hamiltonian is given by
\bea\label{LRQTI}
H_{\text{LRQTI}}=H_{\text{QTI}}+H_{\text{LR}}.
\eea

Here, $H_{\text{QTI}}$ is the standard quadrupole topological insulator (QTI), also known as the Benalcazar–Bernevig–Hughes (BBH) model~\cite{benalcazar_quantized_2017},
\begin{widetext}
\bea
\begin{aligned}
H_{\text{QTI}}
&=\sum_{\mathbf R}\Bigl\{
\gamma\bigl(c_{\mathbf R,1}^\dagger c_{\mathbf R,2}+c_{\mathbf R,3}^\dagger c_{\mathbf R,4}-c_{\mathbf R,1}^\dagger c_{\mathbf R,4}+c_{\mathbf R,2}^\dagger c_{\mathbf R,3}\bigr)\\
&\hphantom{=\sum_{\mathbf R}\Bigl\{}{}
+\lambda_1\bigl(c_{\mathbf R,2}^\dagger c_{\mathbf R+\hat{x},1}+c_{\mathbf R,3}^\dagger c_{\mathbf R+\hat{x},4}-c_{\mathbf R,4}^\dagger c_{\mathbf R+\hat{y},1}+c_{\mathbf R,3}^\dagger c_{\mathbf R+\hat{y},2}\bigr)+\text{h.c.}\Bigr\},
\end{aligned}
\eea
\end{widetext}
and $H_{\text{LR}}$ describes the longer-range hoppings,
\bea
\begin{aligned}
H_{LR}
&=\lambda_2\sum_{\mathbf R}\Big(
c_{\mathbf R,2}^\dagger c_{\mathbf R+2\hat{x},1}
+c_{\mathbf R,3}^\dagger c_{\mathbf R+2\hat{x},4}\\
&\qquad
-c_{\mathbf R,4}^\dagger c_{\mathbf R+2\hat{y},1}
+c_{\mathbf R,3}^\dagger c_{\mathbf R+2\hat{y},2}
\Big)\\
&\quad+\text{h.c.}
\end{aligned}
\eea
The system respects chiral symmetry with the operator
\bea
\begin{aligned}
S&=\sum_{\mathbf R}\Bigl(
c_{\mathbf R,1}^\dagger c_{\mathbf R,1}-c_{\mathbf R,2}^\dagger c_{\mathbf R,2}\\
&\hphantom{=\sum_{\mathbf R}\Bigl(}{}
+c_{\mathbf R,3}^\dagger c_{\mathbf R,3}-c_{\mathbf R,4}^\dagger c_{\mathbf R,4}\Bigr),
\end{aligned}
\eea
where $c_{\mathbf R,1}$ and $c_{\mathbf R,3}$ belong to sublattice $A$ with $S=+1$, while $c_{\mathbf R,2}$ and $c_{\mathbf R,4}$ belong to sublattice $B$ with $S=-1$ (red and blue dots in Fig.~\ref{LRQTI_figure}, respectively). As shown in Ref.~\cite{benalcazar_chiralsymmetric_2022}, the system exhibits three distinct phases (Fig.~\ref{LRQTI_phase}). In the nontrivial phase with $N^{xy}_{\text{chiral}}=1$, it hosts four $E=0$ corner states (Fig.~\ref{LRQTI_N=1}). In contrast, in the nontrivial phase with $N^{xy}_{\text{chiral}}=4$, it hosts sixteen $E=0$ corner states (Fig.~\ref{LRQTI_N=4}). Remarkably, these $E=0$ corner states are protected solely by chiral symmetry, without relying on the $C_4$ symmetry of the clean model. As demonstrated in Ref.~\cite{benalcazar_chiralsymmetric_2022}, they persist under disorder that breaks $C_4$ symmetry while preserving chiral symmetry.

\subsection{Chiral-symmetric 2D higher-order topological phases and spectral localizer}
\label{2D Chiral-symmetric Higher-order topological phases and spectral localizer}

\begin{figure*}[t]
\centering
\begin{adjustbox}{max totalsize={\textwidth}{0.8\textheight},center}
\begin{minipage}{\textwidth}
\centering

\subfloat[]{
\includegraphics[height=0.25\textwidth]{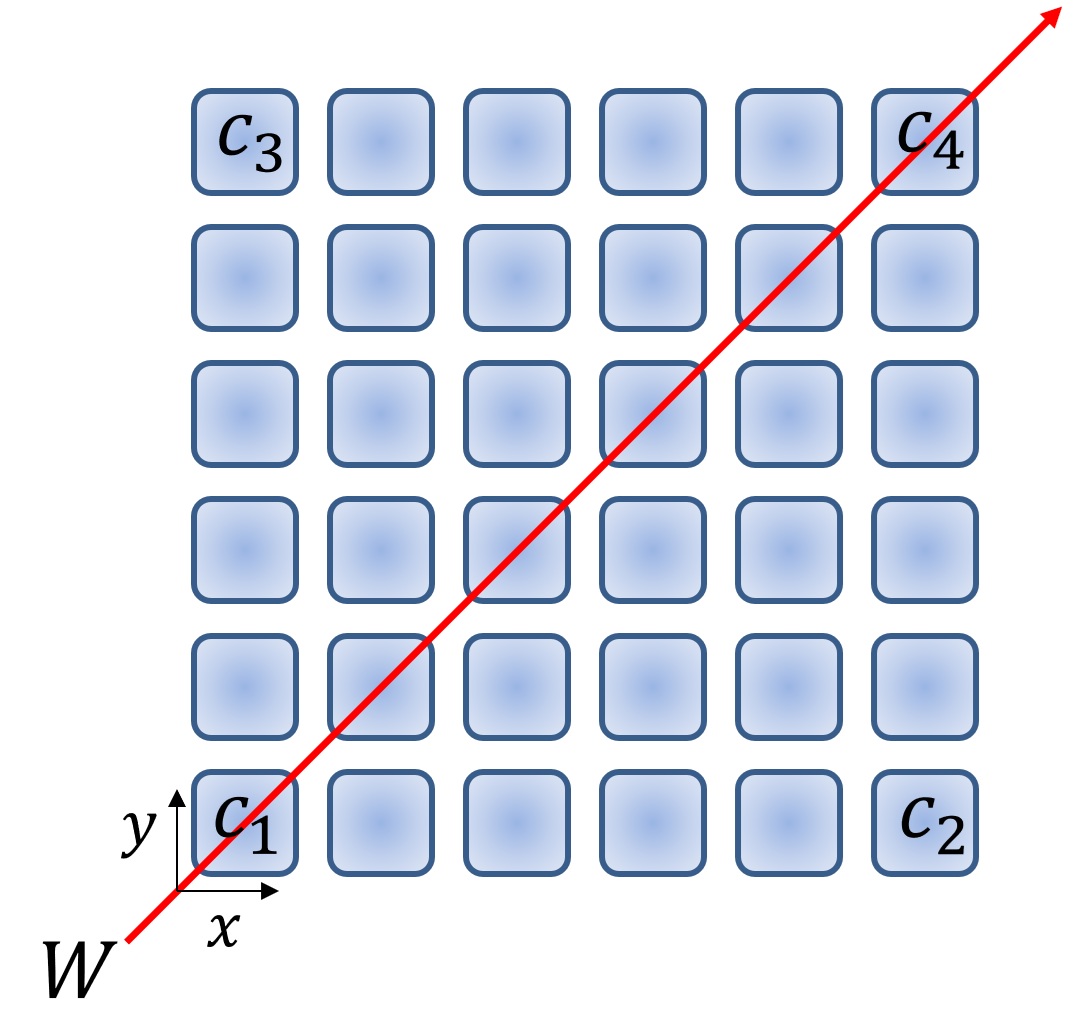}\hspace{0.5cm}
\includegraphics[height=0.25\textwidth]{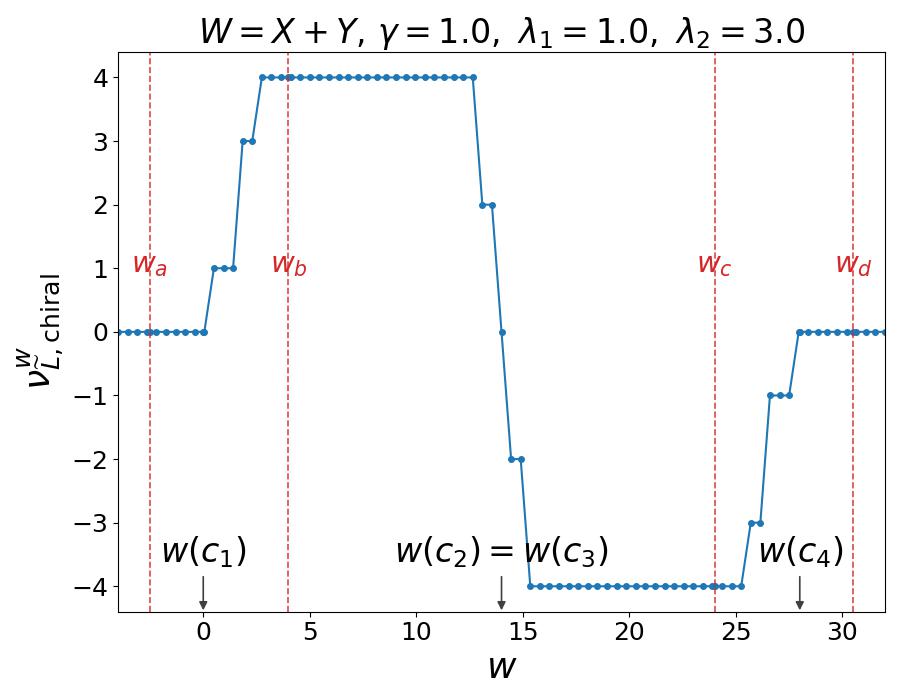}
\label{LRQTI_SL_plus}
}\\

\subfloat[]{
\includegraphics[height=0.25\textwidth]{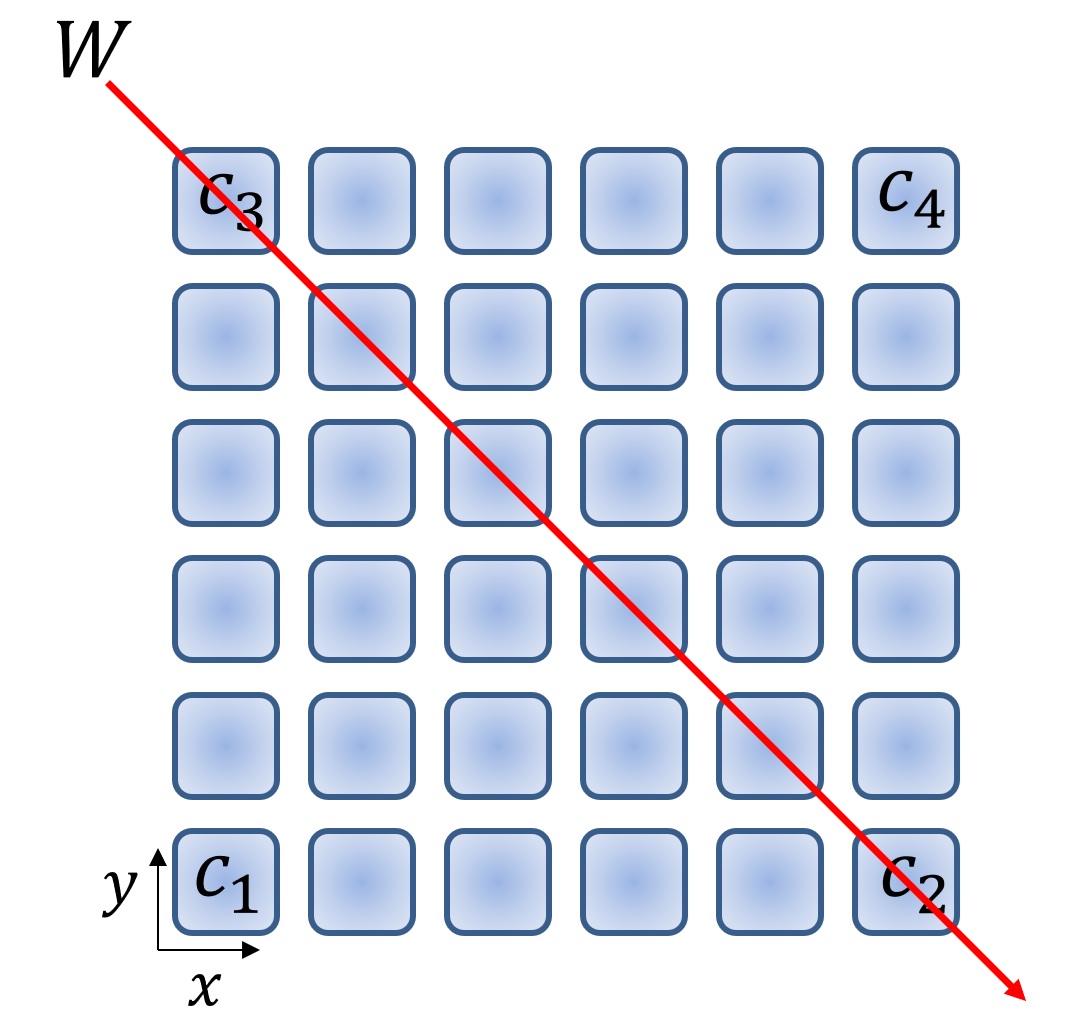}\hspace{0.5cm}
\includegraphics[height=0.25\textwidth]{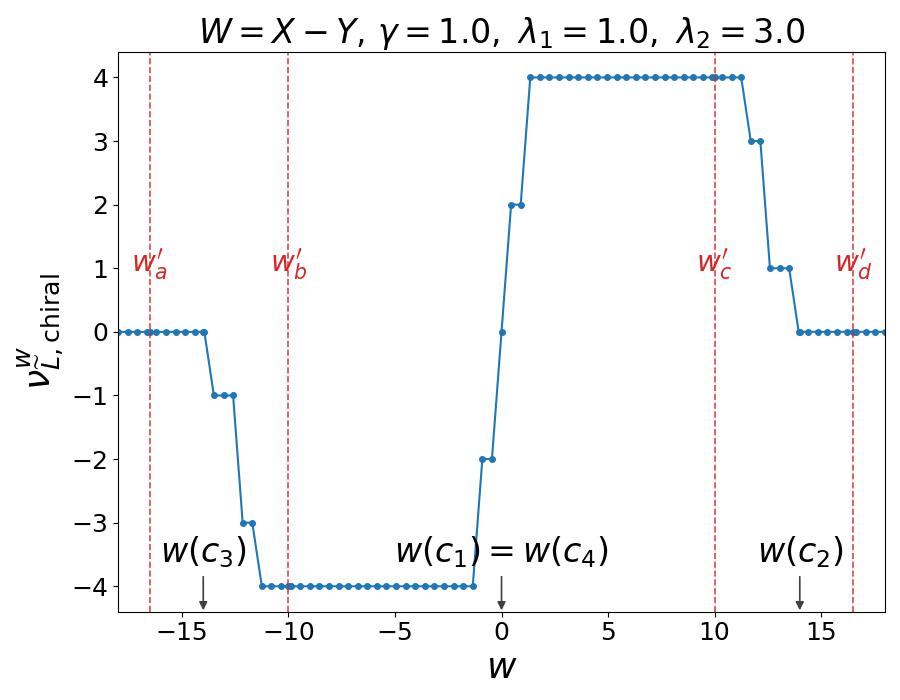}
\label{LRQTI_SL_minus}
}

\end{minipage}
\end{adjustbox}
\par
\caption{ Spectral-localizer characterization of a $15\times15$-unit-cell LRQTI described by the Hamiltonian in Eq.~\eqref{LRQTI}. The parameters are $\gamma=1$, $\lambda_1=1$, and $\lambda_2=3$, the same as those used for the direct numerical results in Fig.~\ref{LRQTI_N=4}, and $\kappa=0.05$. The four corners of the finite sample are labeled as $c_1$, $c_2$, $c_3$, and $c_4$, corresponding respectively to the lower-left, lower-right, upper-left, and upper-right corners. (a) Left: schematic illustration of the projection along the $W=X+Y$ diagonal. Under this projection, the four corners are mapped to the projected coordinates $w(c_1)=0$, $w(c_2)=w(c_3)=14$, and $w(c_4)=28$. Right: the corresponding spectral-localizer invariant $\nu^{w}_{\Tilde{L},\text{chiral}}$ defined in Eq.~\eqref{spectral_localizer invariant for HOT} as a function of the projected coordinate $w$. The red dashed lines indicate $w_a=-2.5$, $w_b=4$, $w_c=24$, and $w_d=30.5$, which are chosen on the two sides of the projected positions of $c_1$ and $c_4$. (b) Left: schematic illustration of the projection along the $W=X-Y$ diagonal. In this case, the corners are mapped to $w(c_3)=-14$, $w(c_1)=w(c_4)=0$, and $w(c_2)=14$. Right: the corresponding spectral-localizer invariant as a function of $w$. The red dashed lines indicate $w_a'=-16.5$, $w_b'=-10$, $w_c'=10$, and $w_d'=16.5$, which are chosen on the two sides of the projected positions of $c_3$ and $c_2$. }

\label{LRQTI_SL}
\end{figure*}

Spectral-localizer invariants have also been used to characterize higher-order topological phases~\cite{cerjan_local_2022,cerjan_local_2024,cerjan_classifying_2024,dureth_probing_2025}. Specifically, one can project the spatial coordinates of a higher-dimensional system onto a lower-dimensional space and apply the spectral-localizer invariant appropriate to that dimension, while retaining the full Hamiltonian. For 2D chiral-symmetric HOT phases, the 1D spectral-localizer invariant defined in Eq.~\eqref{local winding number} can characterize the topological $E=0$ corner states with an appropriate projection onto a 1D direction.

We now describe the construction of this invariant explicitly for 2D chiral-symmetric HOT phases.
The invariant relies on an arbitrary choice of a 1D axis within the 2D system, but any generic 1D axis is sufficient to produce a meaningful invariant. 
Starting from the position operators $X$ and $Y$, we define the projected position operator
\bea
W=aX+bY,
\eea
where $a$ and $b$ are the arbitrarily chosen parameters that determine the projection direction. Correspondingly, a spatial point $(x,y)$ is assigned the projected coordinate
\bea
w=ax+by.
\eea
Then, by replacing $(X,x)$ with $(W,w)$ in the 1D chiral-symmetry-reduced spectral localizer defined in Eq.~\eqref{spectral localizer} but retaining the full 2D Hamiltonian $H$, we obtain
\bea
\Tilde{L}_{\text{chiral}}(w,0,H)
=\Big[iH-\kappa(W-w\mathbbm{1})\Big]S,
\eea
with the corresponding invariant
\bea\label{spectral_localizer invariant for HOT}
\nu^{w}_{\Tilde{L},\text{chiral}}(H)
=\frac{1}{2}\text{sig}\Big[\Tilde{L}_{\text{chiral}}(w,0,H)\Big].
\eea
When the projected position operator $W$ maps the corners of interest onto distinct values of $w$, the number of topological $E=0$ states localized at the corner of interest can be determined from the change in $\nu^{w}_{\Tilde{L},\text{chiral}}$ across that corner's projected position. This occurs for generic axes (any axis that is not exactly perpendicular to the line that connects the two corners). Specifically, for a corner at the projected position $w_c$, we define this integer change as
\bea
\Delta \nu_{\Tilde{L},\text{chiral}}
=
\nu^{w_{+}}_{\Tilde{L},\text{chiral}}
-
\nu^{w_{-}}_{\Tilde{L},\text{chiral}},
\eea
where $w_{-}<w_c<w_{+}$ are chosen sufficiently far from the corner that the invariant has reached the plateaus on either side. The magnitude $|\Delta \nu_{\Tilde{L},\text{chiral}}|$ gives the number of topological $E=0$ states localized at that corner. The sign of this change identifies the chirality of the corner states. As $w$ increases, an increase in the invariant corresponds to states with chirality $S=+1$, whereas a decrease corresponds to states with chirality $S=-1$, following the relation established in Eq.~\eqref{1D localizer invariant jump}, with $(X,x)$ replaced by $(W,w)$.

To illustrate this characterization, we consider the LRQTI described by the Hamiltonian in Eq.~\eqref{LRQTI}. We choose two complementary projected position operator, $W=X+Y$ and $W=X-Y$, corresponding to the two diagonal directions shown in the left panels of Figs.~\ref{LRQTI_SL_plus} and~\ref{LRQTI_SL_minus}, respectively. For $W=X+Y$, the corners $c_1$ and $c_4$ are mapped onto distinct values of $w$ and can therefore be characterized individually, whereas $c_2$ and $c_3$ are projected onto the same value of $w$ and cannot be resolved separately. Conversely, for $W=X-Y$, the corners $c_2$ and $c_3$ are mapped onto distinct values of $w$, while $c_1$ and $c_4$ share the same projected position. Thus, the two projections provide complementary information and, together, allow the corner states at all four corners to be characterized.

The corresponding spectral-localizer invariants are shown in the right panels of Figs.~\ref{LRQTI_SL_plus} and~\ref{LRQTI_SL_minus} for $\gamma=1$, $\lambda_1=1$, and $\lambda_2=3$. For the $W=X+Y$ projection, we choose $w_a$ and $w_b$ on the two sides of the projected position of $c_1$, and $w_c$ and $w_d$ on the two sides of that of $c_4$. The changes of the invariant are
\bea
\begin{gathered}
\nu^{w_b}_{\Tilde{L},\text{chiral}}
-\nu^{w_a}_{\Tilde{L},\text{chiral}}
=4 \quad\text{for }c_1,
\\
\nu^{w_d}_{\Tilde{L},\text{chiral}}
-\nu^{w_c}_{\Tilde{L},\text{chiral}}
=4 \quad\text{for }c_4.
\end{gathered}
\eea
These increases identify four topological $E=0$ states with chirality $S=+1$ at each of the corners $c_1$ and $c_4$. For the $W=X-Y$ projection, $w_a'$ and $w_b'$ are chosen on the two sides of the projected position of $c_3$, while $w_c'$ and $w_d'$ are chosen on the two sides of that of $c_2$. We obtain
\bea
\begin{gathered}
\nu^{w_b'}_{\Tilde{L},\text{chiral}}
-\nu^{w_a'}_{\Tilde{L},\text{chiral}}
=-4 \quad\text{for }c_3,
\\
\nu^{w_d'}_{\Tilde{L},\text{chiral}}
-\nu^{w_c'}_{\Tilde{L},\text{chiral}}
=-4 \quad\text{for }c_2.
\end{gathered}
\eea
These decreases identify four topological $E=0$ states with chirality $S=-1$ at each of the corners $c_3$ and $c_2$. Thus, the two projections determine both the number and chirality of the corner states: four states at each corner, giving sixteen in total, consistent with Fig.~\ref{LRQTI_N=4}.

Here, we clarify why the counting relation above requires the corners of interest to be resolved separately by the projection. When a projected position corresponds to a single corner, the invariant change can be assigned to the states localized at that corner. If several corners share the same projected position, their contributions enter the change simultaneously and cannot be resolved separately.

At an isolated corner, topological $E=0$ states protected by chiral symmetry alone must share the same chirality, because opposite-chirality zero modes at the same corner can couple and shift away from zero energy without breaking the symmetry. For a corner hosting $n_c$ such states with common chirality $s_c$, the spectral-localizer counting relation discussed above gives
\bea
\Delta\nu_{\Tilde{L},\text{chiral}}=s_c n_c,
\qquad
|\Delta\nu_{\Tilde{L},\text{chiral}}|=n_c.
\eea
Thus, the magnitude and sign of the invariant change determine the number and chirality of the topological states at that corner.

States at different corners, however, need not share the same chirality. When several corners are projected onto the same position $w_c$, the combined change is
\bea
\Delta\nu_{\Tilde{L},\text{chiral}}
=
\sum_j s_j n_j,
\eea
where the sum runs over those corners, and $n_j$ and $s_j$ denote the number and common chirality of the topological $E=0$ states at the $j$-th corner. Contributions from corners with the same chirality add, whereas those from corners with opposite chiralities partially or completely cancel. Consequently, the magnitude of the combined change does not generally give the total number of topological $E=0$ states at those corners.

In the LRQTI example above, the corner states at $c_2$ and $c_3$ share chirality $S=-1$, while those at $c_1$ and $c_4$ share chirality $S=+1$. Their contributions therefore add when projected onto the same position. This is why the magnitudes of the changes between $w_b$ and $w_c$ for $W=X+Y$ and between $w_b'$ and $w_c'$ for $W=X-Y$ reproduce the total numbers of topological $E=0$ states at the respective pairs of corners.

\section{Half-chiral symmetry and its topological protection of boundary states}\label{Hermitian Subspace-Chiral Symmetry and Topological Phases}
In this section, we define half-chiral symmetry and establish its bulk-boundary correspondence. This correspondence provides the basis for constructing half-chiral topological and HOT phases in the following sections. In Sec.~\ref{Definition of half-chiral symmetry}, we first define half-chiral symmetry and discuss its relation to subsymmetry~\cite{wang_subsymmetryprotected_2023} and the subspace property~\cite{shimomura_subspaceprotected_2026}. Then, in Sec.~\ref{bulk-boundary correspondecn for Hermitian half-chiral}, we establish the bulk-boundary correspondence for systems with half-chiral symmetry. This correspondence links the bulk chiral topology of an associated ``chiralized Hamiltonian'' to the topological $E=0$ boundary states of the physical Hamiltonian. Our analysis also reveals the distinctive spectral behavior of half-chiral systems at topological transitions. Finally, in Sec.~\ref{Example: real space topological invariants for half-chiral topological phases using the spectral localizer}, we explain how to construct the half-chiral topological invariants used in the following sections, using 1D systems as concrete examples.

\subsection{Definition of half-chiral symmetry}\label{Definition of half-chiral symmetry}

We now introduce half-chiral symmetry, which can be viewed as a partial relaxation of conventional chiral symmetry. Let $S$ be a unitary operator satisfying $S=S^\dagger$ and $S^2=1$. Its eigenspaces with eigenvalues $S=\pm1$ define the two chiral sectors, which correspond to the two sublattices in a bipartite lattice. We assume these sectors to have equal dimension, corresponding to an equal number of sites on the two sublattices. The corresponding projection operators are
\bea\label{projection op}
P_\pm=\frac{1}{2}(1\pm S),
\eea
and satisfy
\bea
P_\pm^2=P_\pm,\qquad P_+P_-=0,\qquad P_++P_-=1.
\eea

Starting from a Hermitian Hamiltonian $H_0$ with conventional chiral symmetry, $SH_0S^{-1}=-H_0$, we consider modifications that preserve the chiral relation on one sector while allowing it to be broken on the other. The resulting Hermitian Hamiltonian $H$ satisfies the half-chiral symmetry
\bea\label{type1SCS}
SHS^{-1}P_\pm=-HP_\pm,
\eea
where the choice of $P_+$ or $P_-$ specifies the sector on which the chiral relation is preserved. The two choices are related by $S\to -S$, which interchanges $P_+$ and $P_-$ and thus relabels the preserved chiral sector. If both conditions hold for the same operator $S$, the Hamiltonian respects full chiral symmetry. Using $S=2P_+-1=1-2P_-$, the half-chiral symmetry can equivalently be written in terms of the selected projector alone as
\bea
P_\pm HP_\pm=0.
\eea

In a chiral basis ordered by the $S=+1$ and $S=-1$ sectors, a Hamiltonian satisfying half-chiral symmetry takes one of the following two block forms:
\bea\label{Block_form_Hermitian_half-chiral topological}
\begin{aligned}
&H=\left(
\begin{matrix}
0 & M \\
M^{\dag} & H_{-}\\
\end{matrix}
\right)\quad \text{for}\;
SHS^{-1}P_{+}=-HP_{+},\\
&H=\left(
\begin{matrix}
H_{+} & M \\
M^{\dag} & 0\\
\end{matrix}
\right)\quad \text{for}\;
SHS^{-1}P_{-}=-HP_{-}.
\end{aligned}
\eea
Thus, half-chiral symmetry requires the Hamiltonian block within the selected chiral subspace to vanish, while allowing a nonzero block within the complementary subspace. When the two subspaces correspond to physical sublattices, this condition means that on-site potentials and hoppings within the selected sublattice are absent, whereas hoppings between the two sublattices and terms within the complementary sublattice are allowed. This formulation makes the symmetry condition directly identifiable from the coupling structure of a lattice model.

Half-chiral symmetry can also be formulated directly in terms of a projection operator, without first specifying a chiral-symmetric parent Hamiltonian. Suppose that a Hermitian Hamiltonian $H$ admits an orthogonal projector $P=P^\dagger=P^2$ onto half of the Hilbert space such that
\bea\label{PHP=0}
PHP=0.
\eea
We can then construct a chiral operator in either of two ways,
\bea
S=2P-1\qquad\text{or}\qquad S=1-2P.
\eea
For $S=2P-1$, we have $P=P_+$ and $SHS^{-1}P_+=-HP_+$. Alternatively, choosing $S=1-2P$ gives $P=P_-$ and $SHS^{-1}P_-=-HP_-$. Both constructions therefore establish half-chiral symmetry as defined in Eq.~\eqref{type1SCS}. We further impose $[P,X_j]=0$ for each spatial direction $j$, where $X_j$ is the position operator defined by $X_j\ket{\mathbf r,\alpha}=r_j\ket{\mathbf r,\alpha}$, with $\mathbf r$ labeling spatial locations and $\alpha$ labeling internal degrees of freedom at each location. Because either choice of $S$ is constructed directly from $P$, it also satisfies $[S,X_j]=0$. Thus, neither operator mixes states at different spatial locations, giving the resulting half-chiral symmetry the same on-site character as conventional internal chiral symmetry. 

The reference energy in the definition of half-chiral symmetry need not be zero. More generally, the symmetry relation in Eq.~\eqref{type1SCS} can be written at a real, spatially uniform reference energy $E_0$ as
\bea
S(H-E_0\mathbbm{1})S^{-1}P_\pm
=-(H-E_0\mathbbm{1})P_\pm,
\eea
which is equivalent to
\bea
P_\pm HP_\pm=E_0P_\pm,
\eea
generalizing Eq.~\eqref{PHP=0}. For example, choosing $P_-$ gives the following Hamiltonian in the chiral basis:
\bea
H=
\begin{pmatrix}
H_+&M\\
M^\dagger&E_0\mathbbm{1}
\end{pmatrix}.
\eea
Introducing the shifted Hamiltonian
\bea
H'=H-E_0\mathbbm{1}
=
\begin{pmatrix}
H_+-E_0\mathbbm{1}&M\\
M^\dagger&0
\end{pmatrix}
\eea
therefore recovers the half-chiral block form in Eq.~\eqref{Block_form_Hermitian_half-chiral topological}, satisfying $SH'S^{-1}P_-=-H'P_-$ and $P_-H'P_-=0$. This uniform energy shift leaves the eigenstates unchanged and maps states at $E=E_0$ in $H$ to states at $E=0$ in $H'$. Our zero-energy formulation can therefore be applied at any chosen reference energy $E_0$ by using $H'$.

The half-chiral symmetry defined in Eq.~\eqref{type1SCS} is related to two previously introduced concepts, with equivalence holding under specific conditions. The first is the notion of subsymmetry~\cite{wang_subsymmetryprotected_2023}, under which a symmetry is preserved only within a subspace of the Hilbert space. When applied to chiral symmetry, the subsymmetry condition takes the form $SHS^{-1}P_{Hs} = -HP_{Hs}$. Here, $P_{Hs}$ projects onto a selected Hilbert subspace and need not coincide with either chiral-sector projector $P_\pm$. When $P_{Hs}=P_+$ or $P_{Hs}=P_-$, the subsymmetry condition becomes equivalent to half-chiral symmetry. Existing studies on subsymmetry investigate both the equivalent cases~\cite{wang_subsymmetryprotected_2023,kang_subsymmetry_2024,verma_nonbloch_2024,cheng_subsymmetryprotected_2026} and non-equivalent cases \cite{wang_subsymmetryprotected_2023,kang_higherorder_2026}. The second concept is the subspace property~\cite{shimomura_subspaceprotected_2026}, formulated in momentum space as $H(k)\mathcal{M}\subseteq\mathcal{M}'$. It becomes equivalent to half-chiral symmetry when $\mathcal{M}$ occupies half of the Hilbert space and $\mathcal{M}'=\mathcal{M}^{\perp}$.

In settings where these concepts are equivalent to half-chiral symmetry, previous studies have either focused on boundary-state protection without constructing a bulk topological invariant or characterized the boundary states using momentum-space invariants. Here, we develop a general framework based on the algebraic relation between a half-chiral Hamiltonian and an associated chiralized Hamiltonian. This framework establishes a bulk-boundary correspondence in arbitrary spatial dimensions and accommodates both momentum-space and real-space formulations.

\subsection{Bulk-boundary correspondence for half-chiral symmetry}\label{bulk-boundary correspondecn for Hermitian half-chiral}

To demonstrate the bulk–boundary correspondence in systems with the half-chiral symmetry, we introduce the chiralized Hamiltonian
\bea\label{chiralized Hamiltonian}
H_{\text{chiralized}}[H]=\frac{1}{2}(H - SHS^{-1}),
\eea
where $S$ is the chiral operator used to define half-chiral symmetry in Eq.~\eqref{type1SCS} and $H$ is the system Hamiltonian. We refer to it as the \textit{chiralized Hamiltonian} because $\frac{1}{2}(H - SHS^{-1})$ effectively projects out the chiral-symmetry-breaking part of a Hamiltonian. Regardless of whether $H$ possesses any symmetry, the chiralized Hamiltonian $H_{\text{chiralized}}$ always respects chiral symmetry. This can be verified using $S^2=1$,
\bea
\begin{aligned}
&SH_{\text{chiralized}}S^{-1}=\frac{1}{2}(SHS^{-1} - H)\\
&=-H_{\text{chiralized}}.
\end{aligned}
\eea
We will show that a Hamiltonian $H$ satisfying the half-chiral symmetry~\eqref{type1SCS} inherits at least half of the $E=0$ states of its corresponding chiralized Hamiltonian $H_{\text{chiralized}}$. The inherited states remain unchanged and are simultaneous $E=0$ eigenstates of both Hamiltonians.

To establish this result, we first review the structure of the $E=0$ subspace of a Hamiltonian satisfying conventional chiral symmetry. In the chiral basis, the Hamiltonian can be brought into an off-diagonal form,
\bea
H=\left(
\begin{matrix}
0 & M \\
M^{\dag} & 0\\
\end{matrix}
\right),
\eea
The $E=0$ subspace is then explicitly given by
\bea\label{chiral_zero-energy_subspace}
\mathcal{H}_0 = \ker(M^\dagger) \oplus \ker(M)
\eea
which means that one can choose an appropriate basis in which each $E=0$ state is fully located within a single chiral subspace,
\bea\label{chiral paired E=0 states}
\begin{aligned}
&\psi^{+}_{0,i} =
\left(
\begin{matrix}
\phi^+_{i}\\
0\\
\end{matrix}
\right) \quad \text{with} \quad \phi^+_{i} \in \ker(M^\dagger),\\
&\psi^{-}_{0,i} =
\left(
\begin{matrix}
0\\
\phi^-_{i}\\
\end{matrix}
\right)\quad \text{with} \quad \phi^-_{i} \in \ker(M),
\end{aligned}
\eea
where $\psi^{\pm}_{0,i}$ denotes the $E=0$ states with amplitude only in the chiral subspace $S = \pm 1$, and $i$ labels the different $E=0$ states. For a finite system with equal-dimensional chiral sectors, $M$ is a square matrix, so the numbers of $E=0$ states in the two sectors are equal,
\bea
\dim \ker(M^\dagger) = \dim \ker(M) = n,
\eea
so that the total number of $E=0$ states is $2n$, with $n$ states in each chiral sector, a structure that will be important for the subsequent generalization to half-chiral systems.

We now use this structure of the chiral $E=0$ subspace to analyze a Hamiltonian satisfying the half-chiral symmetry
\bea
SHS^{-1}P_-=-HP_-.
\eea
In the chiral basis, the Hamiltonian can be written as
\bea
H=\left(
\begin{matrix}
H_{+} & M \\
M^{\dag} & 0\\
\end{matrix}
\right),
\eea
with
\bea
S=\left(
\begin{matrix}
\mathbbm{1} & 0 \\
0 & -\mathbbm{1}\\
\end{matrix}
\right)\;\text{and}\;
P_-=\left(
\begin{matrix}
0 & 0 \\
0 & \mathbbm{1}\\
\end{matrix}
\right).
\eea
The corresponding chiralized Hamiltonian~\eqref{chiralized Hamiltonian} is given by
\begin{widetext}
\bea\label{chiralized_block}
\begin{aligned}
H_{\text{chiralized}}=\frac{1}{2}(H - SHS^{-1})
&=\frac{1}{2}\left(
\begin{matrix}
H_{+} & M \\
M^{\dag} & 0\\
\end{matrix}
\right)-\frac{1}{2}
\left(
\begin{matrix}
\mathbbm{1} & 0 \\
0 & -\mathbbm{1}\\
\end{matrix}
\right)
\left(
\begin{matrix}
H_{+} & M \\
M^{\dag} & 0\\
\end{matrix}
\right)
\left(
\begin{matrix}
\mathbbm{1} & 0 \\
0 & -\mathbbm{1}\\
\end{matrix}
\right)\\
&=\frac{1}{2}\left(
\begin{matrix}
H_{+} & M \\
M^{\dag} &0\\
\end{matrix}
\right)-
\frac{1}{2}\left(
\begin{matrix}
H_{+} & -M \\
-M^{\dag} & 0\\
\end{matrix}
\right)\\
&=\left(
\begin{matrix}
0 & M \\
M^{\dag} & 0\\
\end{matrix}
\right).
\end{aligned}
\eea
\end{widetext}
We can therefore separate $H$ into its chiralized part and a chiral-breaking term acting only within the $S=+1$ sector,
\bea
H=H_{\text{chiralized}}+\begin{pmatrix}H_+&0\\0&0\end{pmatrix}.
\eea
This decomposition allows us to determine which $E=0$ states survive the chiral-breaking term. From the chiral-subspace structure established above, $H_{\text{chiralized}}$ has $2n$ zero-energy states. These states can be chosen as $n$ states $\psi^+_{0,i}$ in the $S=+1$ sector and $n$ states $\psi^-_{0,i}$ in the $S=-1$ sector. They satisfy
\bea
\begin{aligned}
H_{\text{chiralized}}\psi^+_{0,i}
&=\begin{pmatrix}0&M\\M^\dagger&0\end{pmatrix}
\begin{pmatrix}\phi^+_i\\0\end{pmatrix}\\
&=\begin{pmatrix}0\\M^\dagger\phi^+_i\end{pmatrix}=0,\\[4pt]
H_{\text{chiralized}}\psi^-_{0,i}
&=\begin{pmatrix}0&M\\M^\dagger&0\end{pmatrix}
\begin{pmatrix}0\\\phi^-_i\end{pmatrix}\\
&=\begin{pmatrix}M\phi^-_i\\0\end{pmatrix}=0.
\end{aligned}
\eea
which implies that
\bea\label{zero edge states and M}
\begin{aligned}
&M^\dagger\phi^+_i=0,\\
&M\phi^-_i=0.
\end{aligned}
\eea
The effect of the chiral-breaking term on these states is then obtained directly by applying $H$:
\bea\label{E=0 states for half-chiral}
\begin{aligned}
H\psi^+_{0,i}
&=H_{\text{chiralized}}\psi^+_{0,i}
+\begin{pmatrix}H_+&0\\0&0\end{pmatrix}
\begin{pmatrix}\phi^+_i\\0\end{pmatrix}\\
&=\begin{pmatrix}H_+\phi^+_i\\0\end{pmatrix},\\[4pt]
H\psi^-_{0,i}
&=H_{\text{chiralized}}\psi^-_{0,i}
+\begin{pmatrix}H_+&0\\0&0\end{pmatrix}
\begin{pmatrix}0\\\phi^-_i\end{pmatrix}\\
&=0.
\end{aligned}
\eea
The chiral-breaking term therefore leaves every $\psi^-_{0,i}$ unchanged as an $E=0$ eigenstate of $H$. In contrast, $\psi^+_{0,i}$ generally does not remain at $E=0$ and may not even remain an eigenstate of $H$, depending on the structure of $H_+$.

The above analysis establishes that a Hamiltonian $H$ satisfying the half-chiral symmetry~\eqref{type1SCS} inherits at least half of the $E=0$ states of $H_{\text{chiralized}}$. This inheritance preserves the states themselves, so that each inherited state $\ket{\psi_0}$ satisfies
\bea\label{state_btw_H_Hchiralized}
H\ket{\psi_0}=0,\qquad H_{\text{chiralized}}\ket{\psi_0}=0.
\eea
The physical Hamiltonian may also possess additional $E=0$ states beyond those guaranteed by this inheritance mechanism. Consequently, denoting the total numbers of $E=0$ states by $N_0(H)$ and $N_0(H_{\text{chiralized}})$, respectively, we obtain
\bea\label{En_btw_H_Hchiralized}
N_0(H)\geq\frac{1}{2}N_0(H_{\text{chiralized}}).
\eea
Although derived above for $SHS^{-1}P_-=-HP_-$, both relations also hold for $SHS^{-1}P_+=-HP_+$ by the same reasoning with the two chiral sectors exchanged.

The inheritance mechanism established above does not depend on the origin of the $E=0$ states and therefore also applies to topological $E=0$ boundary states of $H_{\text{chiralized}}$. Specifically, for the half-chiral symmetry $SHS^{-1}P_\pm=-HP_\pm$, the topological $E=0$ states of $H_{\text{chiralized}}$ in the $S=\pm1$ sector are inherited by $H$ with their wavefunctions and chirality unchanged, satisfying
\bea\label{topo_state_btw_H_Hchiralized}
H\ket{\psi_0^{\text{topo}}}=0,\qquad H_{\text{chiralized}}\ket{\psi_0^{\text{topo}}}=0.
\eea
These inherited states constitute the topological $E=0$ boundary states protected by half-chiral symmetry, while any additional accidental $E=0$ states are excluded from the topological count. Therefore, denoting the numbers of topological $E=0$ boundary states protected by the respective symmetry conditions by $N_0^{\text{topo}}$, we obtain
\bea\label{BBC_half-chiral topological}
N_0^{\text{topo}}(H)=\frac{1}{2}N_0^{\text{topo}}(H_{\text{chiralized}}).
\eea

The relations~\eqref{topo_state_btw_H_Hchiralized} and~\eqref{BBC_half-chiral topological} provide a general mechanism for constructing half-chiral topological phases from conventional chiral topological phases. When $H_{\text{chiralized}}$ hosts topological $E=0$ boundary states, the half-chiral Hamiltonian $H$ inherits half of these states together with their topological protection. We can therefore define half-chiral topological phases based on the presence of these robust $E=0$ boundary states. Moreover, because these states are inherited directly from $H_{\text{chiralized}}$, their existence can be characterized by the chiral topology of $H_{\text{chiralized}}$. This allows us to define the corresponding half-chiral topological invariant as
\bea\label{general half-chiral invariant}
\nu_{\text{half-chiral}}(H)\equiv\nu_{\text{chiral}}(H_{\text{chiralized}}),
\eea
where $\nu_{\text{chiral}}$ is the conventional chiral invariant appropriate to the phase under consideration, such as the winding number in 1D. The resulting half-chiral invariant characterizes the existence of robust $E=0$ boundary states in the half-chiral system, establishing a bulk–boundary correspondence between the bulk topology of $H_{\text{chiralized}}$ and the $E=0$ boundary states of the physical Hamiltonian $H$. These states remain protected under half-chiral-symmetry-preserving perturbations as long as the corresponding topological invariant remains unchanged.

Topological phase transitions are characterized by changes in $\nu_{\text{half-chiral}}(H)$, accompanied by the emergence or disappearance of protected $E=0$ boundary states in the physical Hamiltonian. At these transitions, $H_{\text{chiralized}}$ undergoes a bulk-gap closing at $E=0$, reflecting the change in its conventional chiral topology. For the physical Hamiltonian $H$, the bulk spectrum touches $E=0$ at the transition, but its bulk gap need not close. This touching follows by applying Eq.~\eqref{state_btw_H_Hchiralized} to the $E=0$ bulk states at the gap closing of $H_{\text{chiralized}}$: those within the preserved chiral sector remain $E=0$ bulk states of $H$. An example illustrating this distinction will be presented in Sec.~\ref{1D half-chiral topological phases}.

The framework developed above allows us to construct a half-chiral counterpart of any conventional chiral topological phase supporting $E=0$ boundary states. This generality follows from the algebraic relations between $H$ and $H_{\text{chiralized}}$, which do not depend on a particular momentum-space or real-space formulation or on the spatial dimension.

The framework also extends to half-chiral HOT phases. The MCN and spectral-localizer invariant reviewed in Secs.~\ref{2D Chiral-symmetric Higher-order topological phases} and~\ref{2D Chiral-symmetric Higher-order topological phases and spectral localizer}, respectively, can be extended to their half-chiral counterparts through Eq.~\eqref{general half-chiral invariant}, as detailed in Secs.~\ref{2D half-chiral HOT phases with MCNs} and~\ref{2D half-chiral HOT phases without MCNs}. For half-chiral HOT phases, topological transitions involve the emergence or disappearance of topological $E=0$ higher-order boundary states, such as corner states, in the physical Hamiltonian. These transitions are characterized by changes in the corresponding half-chiral topological invariants. At these transitions, $H_{\text{chiralized}}$ undergoes a boundary- or bulk-gap closing. For the physical Hamiltonian, the boundary or bulk spectrum touches $E=0$ without necessarily closing the corresponding gap, as follows from the same inheritance mechanism discussed above.

\subsection{Application: real space topological invariants for half-chiral topological phases using the spectral localizer}\label{Example: real space topological invariants for half-chiral topological phases using the spectral localizer}
We now use Eq.~\eqref{general half-chiral invariant} to construct the half-chiral topological invariants used in the following sections. For concreteness we specialize to 1D here. In 1D, half-chiral topological phases can be constructed from conventional chiral-symmetric SPT phases in Sec.~\ref{1D chiral symmetric topological phases}, with the corresponding topological invariant given by the winding number of the associated chiralized Hamiltonian,
\bea
\begin{aligned}
&\label{half-chiral winding number}
\nu_{\text{half-chiral}}^{BZ}(\mathscr{H})=
\nu_{\text{chiral}}^{BZ}(\mathscr{H}_{\text{chiralized}})\\
&=
\frac{1}{2\pi i} \int_{BZ}dk\; \partial_k \text{log}(\text{det}[M^{\dag}]),
\end{aligned}
\eea
where $[M^{\dag}]$ denotes the lower-left off-diagonal block of $\mathscr{H}_{\text{chiralized}}$ in the chiral basis (Eq.~\eqref{chiralized_block}). The magnitude $|\nu_{\text{half-chiral}}^{BZ}|$ gives the number of topological $E=0$ edge states of the half-chiral system, which is half the number in its conventional chiral-symmetric counterpart.

In real space, the same construction can be applied within the spectral-localizer framework to obtain the corresponding half-chiral topological invariant. Specifically, we introduce a ``half-chiral symmetry-reduced spectral localizer'' $\Tilde{L}_{\text{half-chiral}}$ by evaluating the usual chiral-symmetry-reduced spectral localizer for the chiralized Hamiltonian. The corresponding real-space half-chiral invariant $\nu^{x}_{\tilde{L},\text{half-chiral}}$ is then given by
\bea\label{half-chiral spectral-localizer invariant}
\begin{aligned}
\nu^{x}_{\tilde{L},\text{half-chiral}}(H)
&=\nu^{x}_{\tilde{L},\text{chiral}}(H_{\text{chiralized}})\\
&=\frac{1}{2}\text{sig}\Big[\Tilde{L}_{\text{half-chiral}}(x,0,H)\Big],
\end{aligned}
\eea
where
\bea\label{half-chiral reduece spectral localizer}
\begin{aligned}
&\Tilde{L}_{\text{half-chiral}}(x,0,H)\\
&=\Tilde{L}_{\text{chiral}}(x,0,H_{\text{chiralized}})\\
&=\Big[iH_{\text{chiralized}}-\kappa(X-x\mathbbm{1})\Big]S\\
&=\frac{i}{2}(HS-SH)-\kappa(X-x\mathbbm{1})S.
\end{aligned}
\eea
The factor $1/2$ can be removed by an overall positive rescaling of the localizer and a redefinition of $\kappa$, without changing its signature. When $x$ is chosen in the bulk, the magnitude $|\nu^{x}_{\tilde{L},\text{half-chiral}}(H)|$ gives the number of topological $E=0$ edge states of the half-chiral system.

The spatial dependence of the half-chiral spectral-localizer invariant also provides information about localized topological $E=0$ boundary states, but its interpretation differs from the conventional chiral case. The spectral-localizer counting relation in Eq.~\eqref{1D localizer invariant jump} still applies, since this invariant equals the chiral invariant evaluated for $H_{\text{chiralized}}$, as expressed in Eq.~\eqref{half-chiral spectral-localizer invariant}. Thus, the invariant retains positive and negative changes associated with the $S=+1$ and $S=-1$ topological states of $H_{\text{chiralized}}$, respectively. To quantify these changes, we define the invariant change across a boundary near $x_0$ as
\bea
\Delta\nu_{\tilde{L},\text{half-chiral}}
=
\nu^{x_+}_{\tilde{L},\text{half-chiral}}(H)
-\nu^{x_-}_{\tilde{L},\text{half-chiral}}(H),
\eea
where $x_-<x_0<x_+$ lie on the invariant plateaus on either side of the boundary. When using this invariant to characterize the physical Hamiltonian $H$, however, only changes associated with the inherited states should be counted. Specifically, for the half-chiral symmetry $SHS^{-1}P_\pm=-HP_\pm$ in Eq.~\eqref{type1SCS}, $H$ inherits the topological $E=0$ states of $H_{\text{chiralized}}$ with chirality $S=\pm1$. We therefore select the corresponding invariant changes based on their sign according to
\bea\label{half-chiral localizer selection rule}
\begin{aligned}
SHS^{-1}P_+=-HP_+ &: \quad
\Delta\nu_{\tilde{L},\text{half-chiral}}>0,\\
SHS^{-1}P_-=-HP_- &: \quad
\Delta\nu_{\tilde{L},\text{half-chiral}}<0.
\end{aligned}
\eea
The position of each selected invariant change identifies where the physical topological $E=0$ states are localized, while its magnitude gives their number.

\begin{figure*}[t]
\centering
\begin{adjustbox}{max totalsize={\textwidth}{0.72\textheight},center}
\begin{minipage}{\textwidth}
\centering
\begin{minipage}[c]{0.35\textwidth}
\centering
\subfloat[]{
\includegraphics[width=0.95\linewidth]{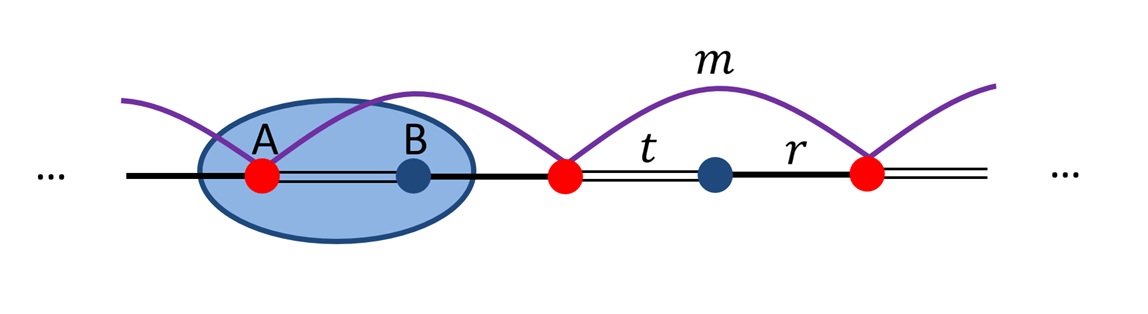}
\label{exSSH_figure}}
\end{minipage}\hfill
\begin{minipage}[c]{0.65\textwidth}
\centering
\subfloat[]{
\includegraphics[width=0.48\linewidth,valign=c]{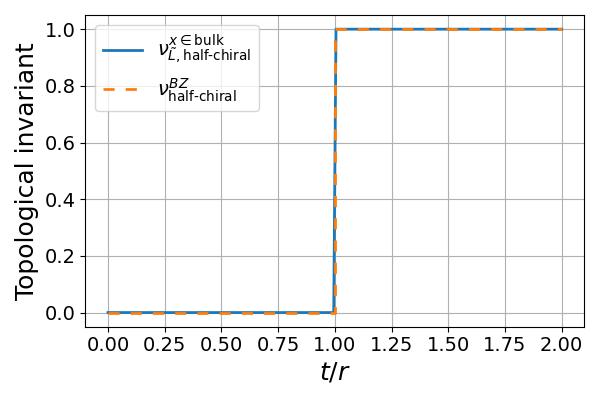}
\label{exSSH_phase}}\hfill
\subfloat[]{
\includegraphics[width=0.48\linewidth,valign=c]{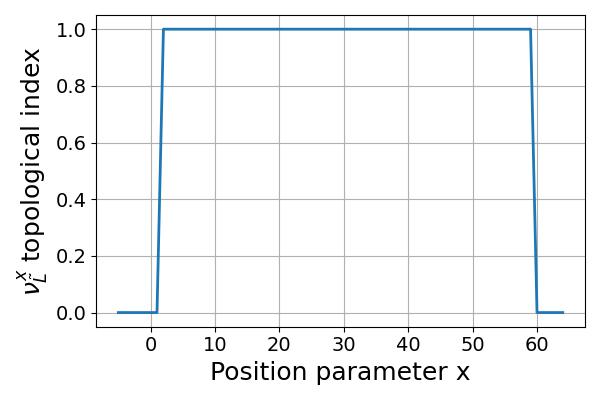}
\label{half_chiral_SSH_localizer_spatial}}\\[6pt]
\subfloat[]{
\begin{minipage}{\linewidth}
\centering
\includegraphics[width=0.48\linewidth,valign=c]{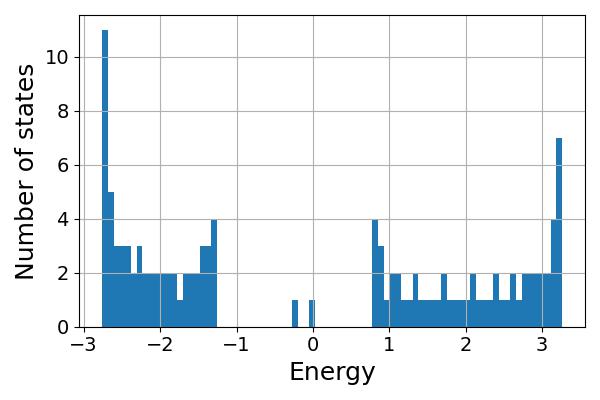}\hfill
\includegraphics[width=0.48\linewidth,valign=c]{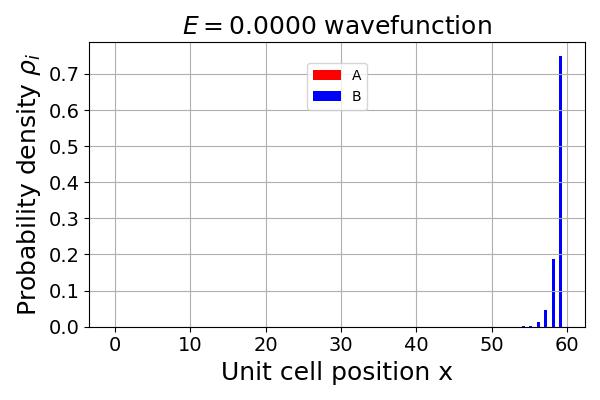}
\end{minipage}
\label{exSSH_En_distribution}
\label{exSSH_wavefunction}}
\end{minipage}
\end{minipage}
\end{adjustbox}
\par
\caption{(a) Schematic illustration of the half-chiral SSH model defined in Eq.~\eqref{half-chiral SSH}. (b) Phase diagram showing the half-chiral winding number $\nu^{BZ}_{\text{half-chiral}}$ and the half-chiral spectral-localizer invariant $\nu^{x\in\text{bulk}}_{\tilde{L},\text{half-chiral}}$ for the half-chiral SSH model. (c) Spatial dependence of $\nu^{x}_{\tilde{L},\text{half-chiral}}$ under OBCs with $t=1$, $r=2$, $m=0.25$, $N_{\text{cell}}=60$, and $\kappa=0.05$. Since the model satisfies $SHS^{-1}P_-=-HP_-$, only the unit decrease near the right edge corresponds to a physical topological $E=0$ state, predicting a single state at the right edge on the $B$ sublattice. (d) Energy distribution of states (left) and total probability density of the $E=0$ state (right) for the same system as in (c), confirming the predicted number and localization of the topological $E=0$ edge state.}
\end{figure*}

\begin{figure*}[t]
\centering
\subfloat[]{\includegraphics[width=0.35\textwidth,valign=c]{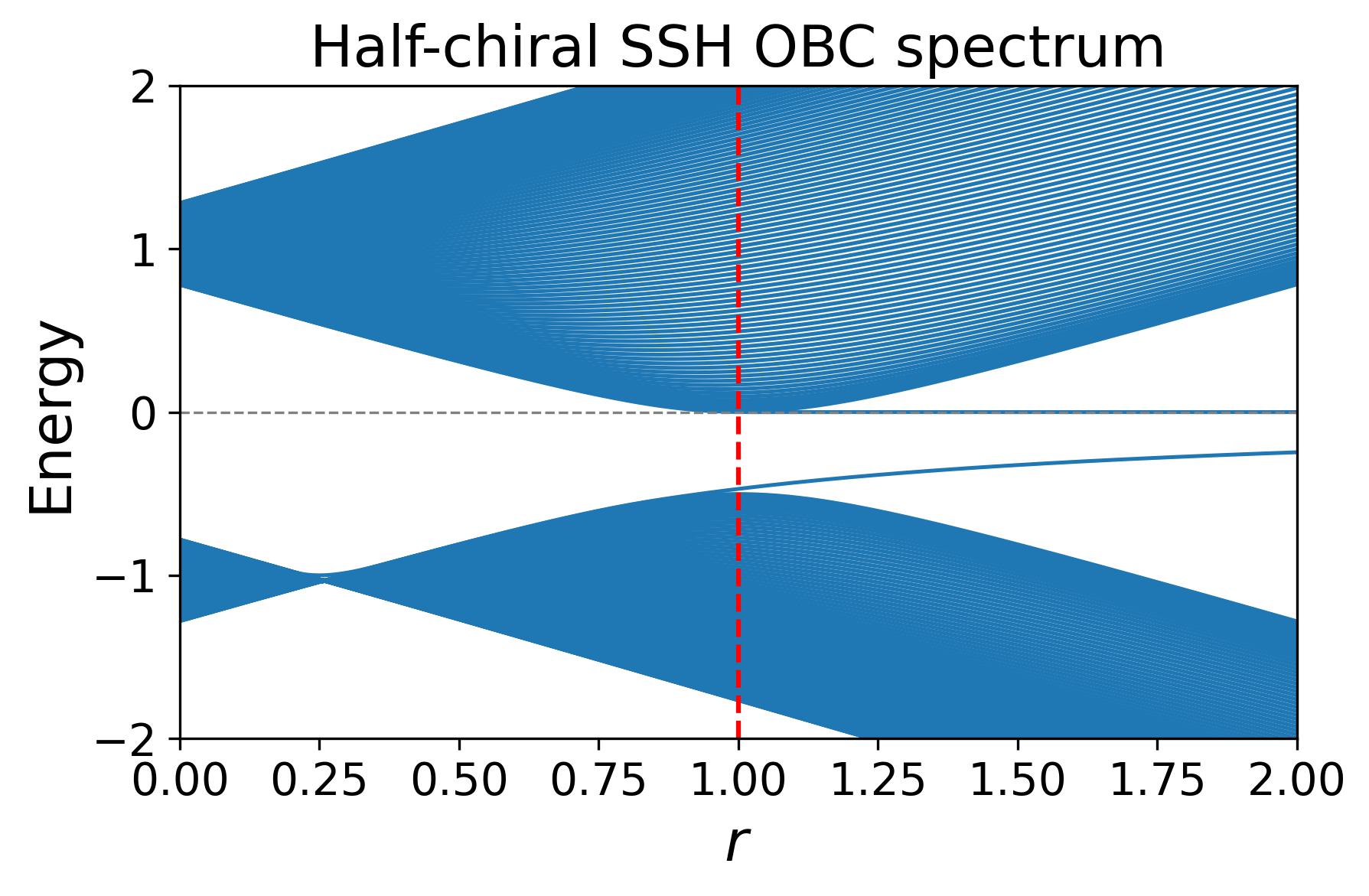}\label{OBCspectrum_HexSSH}}\hskip 0.5cm
\subfloat[]{\includegraphics[width=0.35\textwidth,valign=c]{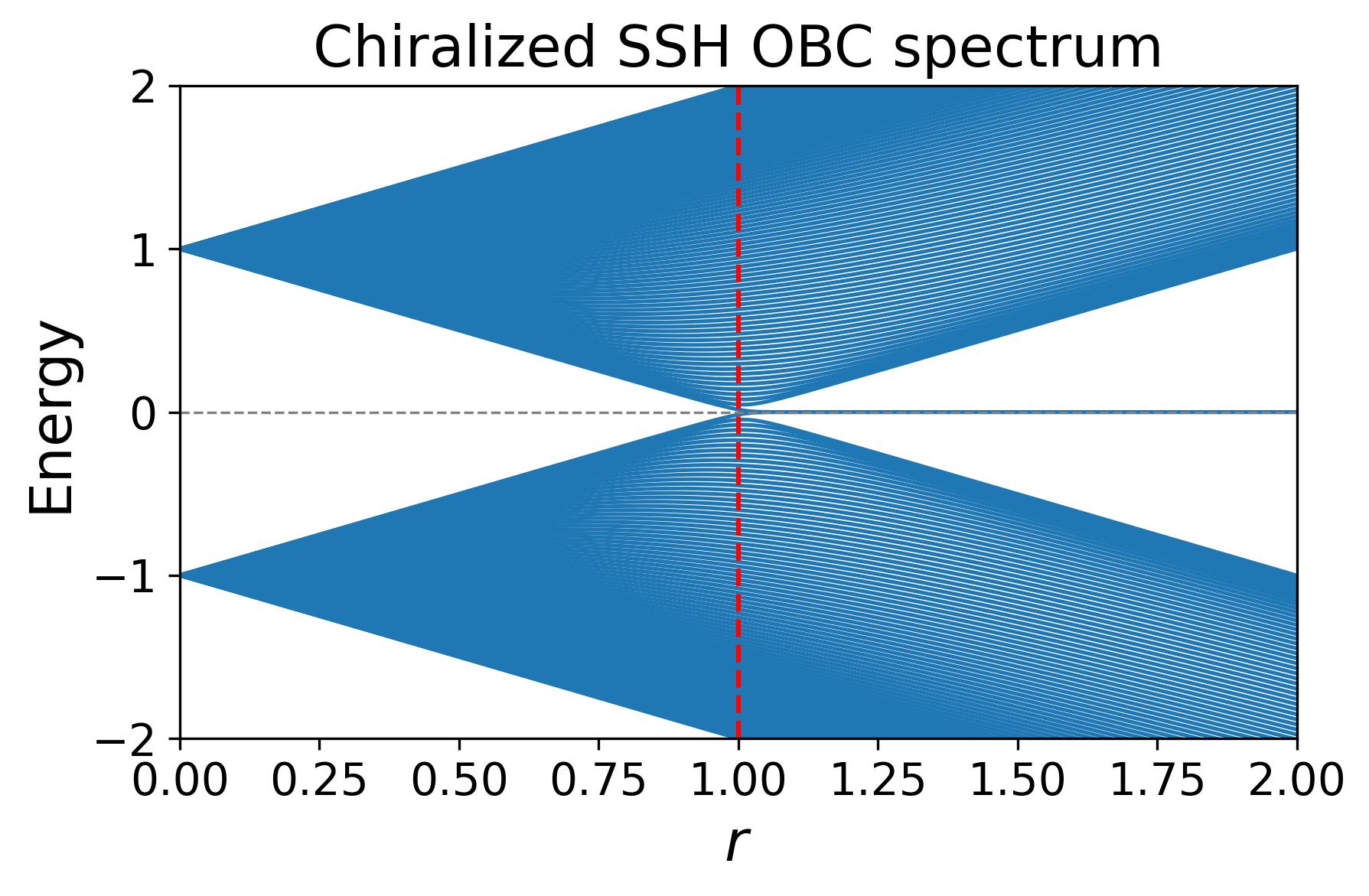}\label{OBCspectrum_Hchiralized}}\\
\subfloat[]{\includegraphics[width=0.35\textwidth,valign=c]{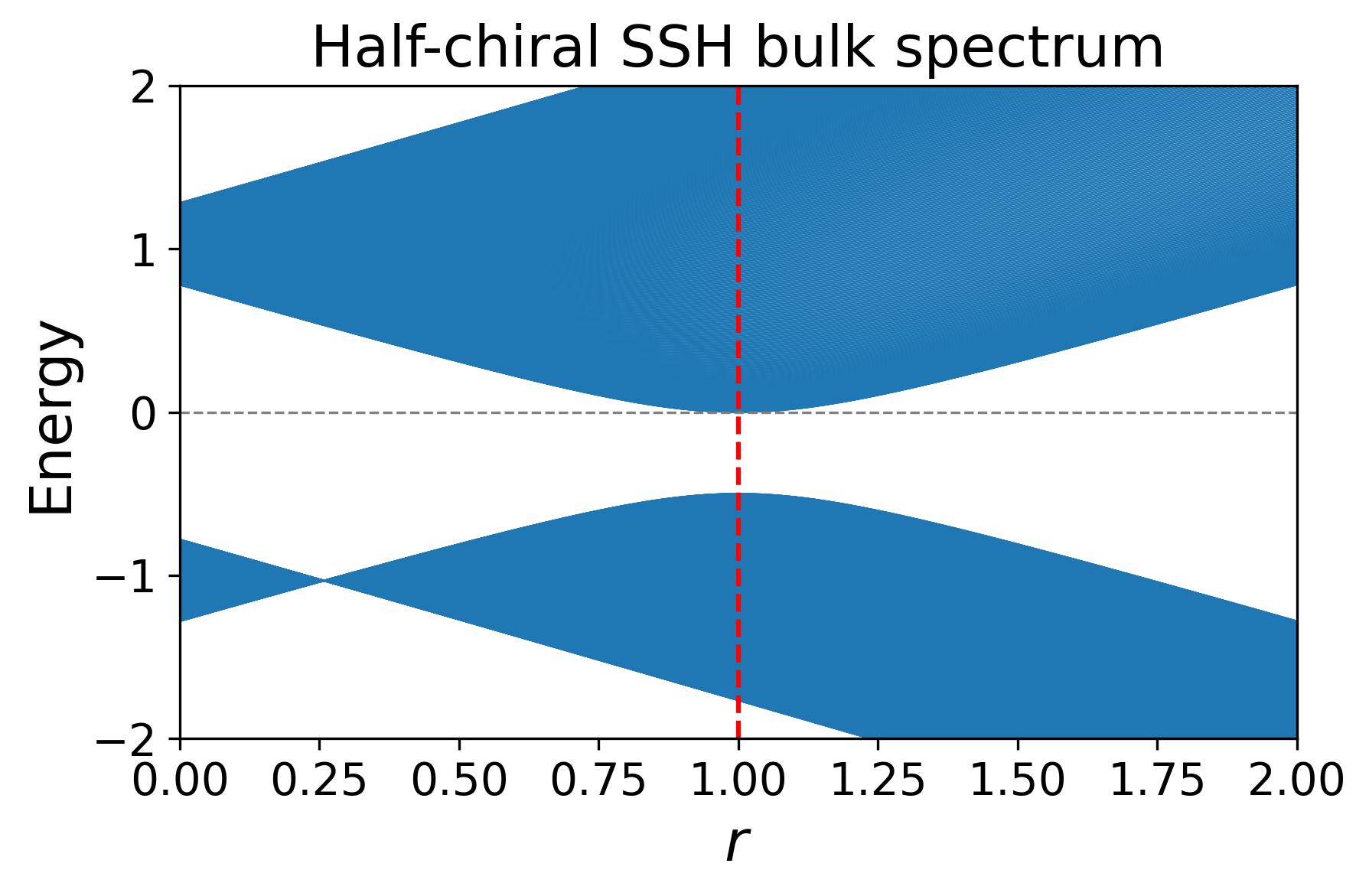}\label{bulk_gap_1d_HexSSH}}\hskip 0.5cm
\subfloat[]{\includegraphics[width=0.35\textwidth,valign=c]{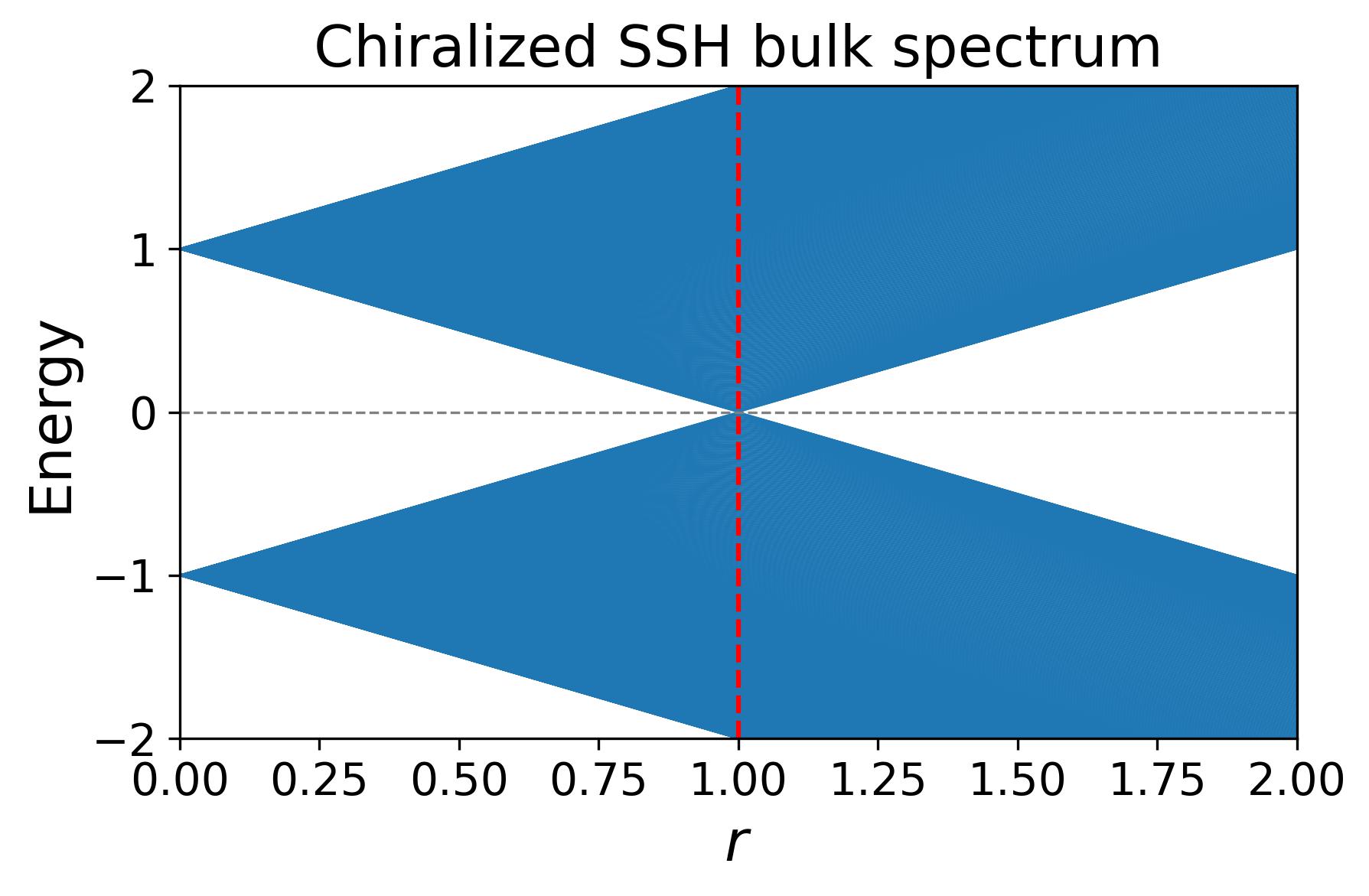}\label{bulk_gap_1d_Hchiralized}}
\caption{OBC and bulk energy spectra of the half-chiral SSH model and its chiralized Hamiltonian $H_{\text{chiralized}}=H_{\text{SSH}}$ as functions of $r/t$, with $t=1$ and $m=0.25$. The red dashed lines indicate the thermodynamic transition point $r/t=1$. (a) OBC spectrum of the half-chiral SSH model for $N_{\text{cell}}=100$, showing the emergence of one topological $E=0$ edge state around $r/t=1$. (b) Corresponding OBC spectrum of $H_{\text{chiralized}}$, showing the emergence of two topological $E=0$ edge states around $r/t=1$. (c) Momentum-space bulk spectrum of the half-chiral SSH model, which touches $E=0$ at the transition while its bulk gap remains open. (d) Corresponding bulk spectrum of $H_{\text{chiralized}}$, whose bulk gap closes at the transition.}
\label{OBC spectrum and bulk gap}
\end{figure*}
\begin{figure*}[t]
\centering
\includegraphics[width=0.35\textwidth,valign=c]{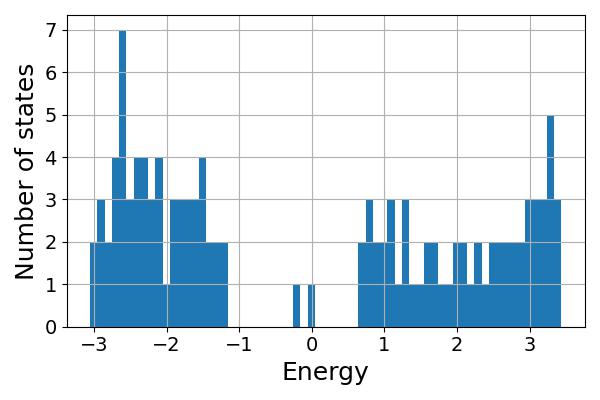}\hskip 0.5cm
\includegraphics[width=0.35\textwidth,valign=c]{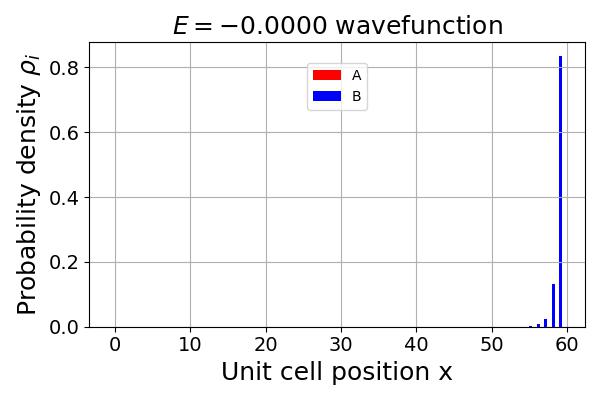}
\caption{Energy distribution of states (left) and total probability density of the $E=0$ state (right) for the disordered half-chiral SSH model in the $\nu^{x \in \text{bulk}}_{\tilde{L},\text{half-chiral}}=1$ half-chiral topological phase. Results are shown under OBC with $N_{\text{cell}}=60$, $t=1$, $r=2$, and $m=0.25$. The disorder terms $\delta t_n$, $\delta r_n$, and $\delta m_n$ are independently sampled from $[-0.2,0.2]$. The system hosts one topological $E=0$ edge state.}
\label{disorderedexSSH_En_distribution}
\label{disorderedexSSH_wavefunction}
\end{figure*}

The correspondence between near-zero localizer modes and physical $E=0$ states for half-chiral systems also differs from that for conventional chiral-symmetric systems. In the conventional case, paired topological $E=0$ states are reflected as paired $\xi\approx0$ modes of the reduced spectral localizer near their localization positions, where $\xi$ denotes the localizer eigenvalues~\cite{cerjan_quadratic_2023}. In the half-chiral case, however, these paired localizer modes reflect the states of $H_{\text{chiralized}}$ in both chiral sectors. Since the physical Hamiltonian $H$ inherits only half of the topological $E=0$ states of $H_{\text{chiralized}}$, only half of the $\xi\approx0$ localizer modes correspond to physical $E=0$ states. A concrete example is presented in Sec.~\ref{Further properties of the spectral localizer in 1D half-chiral topological phases}.

\section{Half-chiral topological phases in 1D}\label{1D half-chiral topological phases}

We now consider the simplest topological phases associated with half-chiral symmetry: 1D half-chiral symmetry-protected topological (half-chiral topological) phases. Using the framework established in Sec.~\ref{bulk-boundary correspondecn for Hermitian half-chiral}, we construct these phases from conventional 1D chiral-symmetric SPT phases. The resulting half-chiral systems inherit half of the topological $E=0$ edge states of their chiralized Hamiltonians, together with their protection. We characterize these states using the half-chiral winding number and spectral-localizer invariant introduced in Sec.~\ref{Example: real space topological invariants for half-chiral topological phases using the spectral localizer}. The bulk-boundary correspondence relates these invariants to the number of topological $E=0$ edge states through $N_0^{\text{topo}}=|\nu^{BZ}_{\text{half-chiral}}|=|\nu^{x\in\text{bulk}}_{\tilde{L},\text{half-chiral}}|$, where $x$ is chosen deep in the bulk. Beyond the total count, the spatial dependence of the half-chiral invariant $\nu^{x}_{\tilde{L},\text{half-chiral}}$ provides information about the localization positions of these edge states. 

In Sec.~\ref{Example: half-chiral SSH model and its boundary mode}, we use the half-chiral SSH model to illustrate the bulk-boundary correspondence and show how the spatial dependence of $\nu^{x}_{\tilde{L},\text{half-chiral}}(H)$ identifies the locations of topological $E=0$ edge states. We also examine the model's spectral behavior at the topological transition and its robustness against disorder. In Sec.~\ref{Further properties of the spectral localizer in 1D half-chiral topological phases}, we investigate the relation between its boundary modes and the spectrum of the half-chiral spectral localizer.

\subsection{
Example: half-chiral SSH model and its boundary mode}\label{Example: half-chiral SSH model and its boundary mode}

To illustrate the bulk-boundary correspondence and the spatial identification of topological $E=0$ edge states in 1D half-chiral topological phases, we consider the half-chiral SSH model
\bea\label{half-chiral SSH}
\begin{aligned}
H_{\text{SSH}}^{\text{half-chiral}}
&=\sum_{n=1}^{N_{\text{cell}}}\Bigl(
t\,c_{n,A}^\dagger c_{n,B}+r\,c_{n,B}^\dagger c_{n+1,A}\\
&\hphantom{=\sum_{n=1}^{N_{\text{cell}}}\Bigl(}{}
+m\,c_{n,A}^\dagger c_{n+1,A}+\text{h.c.}\Bigr).
\end{aligned}
\eea
The schematic illustration of $H_{\text{SSH}}^{\text{half-chiral}}$ is shown in Fig.~\ref{exSSH_figure}. The model satisfies the half-chiral symmetry $SHS^{-1}P_{-} = -HP_{-}$, where
\bea
\begin{aligned}
&S=\sum_n\bigl(c_{\mathbf n,A}^\dagger c_{\mathbf n,A}-c_{\mathbf n,B}^\dagger c_{\mathbf n,B}\bigr),\\
&P_-=\frac{1}{2}(1-S).
\end{aligned}
\eea
With this choice of $S$, the $A$ and $B$ sublattices are assigned to the chiral sectors $S=+1$ and $S=-1$, respectively.

By direct calculation, we find that the corresponding chiralized Hamiltonian is given by
\bea
\begin{aligned}
&H_{\text{chiralized}}[H_{\text{SSH}}^{\text{half-chiral}}]\\
&=\frac{1}{2}(H_{\text{SSH}}^{\text{half-chiral}} - SH_{\text{SSH}}^{\text{half-chiral}}S^{-1})\\
&=H_{\text{SSH}},
\end{aligned}
\eea
The chiralized Hamiltonian is thus the conventional SSH model, which hosts two topological $E=0$ edge states in its nontrivial phase with $r>t$. According to Eq.~\eqref{BBC_half-chiral topological}, the physical Hamiltonian $H_{\text{SSH}}^{\text{half-chiral}}$ inherits one of these states, realizing a half-chiral topological phase. This phase is characterized by the half-chiral winding number and spectral-localizer invariant defined in Eqs.~\eqref{half-chiral winding number} and~\eqref{half-chiral spectral-localizer invariant}, respectively, with $\nu^{BZ}_{\text{half-chiral}}=\nu^{x\in\text{bulk}}_{\tilde{L},\text{half-chiral}}=1$ for $r>t$, as shown in Fig.~\ref{exSSH_phase}. In the topological phase, the spectral-localizer invariant $\nu^{x}_{\tilde{L},\text{half-chiral}}$ exhibits a unit increase near the left edge and a unit decrease near the right edge, as shown in Fig.~\ref{half_chiral_SSH_localizer_spatial}. Since this model satisfies the half-chiral symmetry $SHS^{-1}P_-=-HP_-$, only the decrease corresponds to a physical topological $E=0$ state, following the criterion in Eq.~\eqref{half-chiral localizer selection rule}. This unit decrease therefore predicts a single topological $E=0$ state at the right edge on the $B$ sublattice. The energy distribution and probability density in Fig.~\ref{exSSH_En_distribution} confirm both the predicted number of topological $E=0$ edge states and the localization of the single state at the right edge.

This example also illustrates the spectral behavior at half-chiral topological phase transitions discussed in Sec.~\ref{bulk-boundary correspondecn for Hermitian half-chiral}. Figs.~\ref{OBCspectrum_HexSSH} and~\ref{OBCspectrum_Hchiralized} show the OBC spectra of $H_{\text{SSH}}^{\text{half-chiral}}$ and its chiralized Hamiltonian $H_{\text{chiralized}}=H_{\text{SSH}}$, respectively. In the physical Hamiltonian, a single topological $E=0$ edge state emerges around $r/t=1$ as the bulk spectrum approaches $E=0$, while the bulk bands remain separated. For the chiralized Hamiltonian, a pair of topological $E=0$ edge states emerges around $r/t=1$ as the bulk gap narrows, consistent with the conventional chiral bulk–boundary correspondence. The finite system size rounds the spectral features near the transition. 

To distinguish the bulk spectral behavior from finite-size effects, we also calculate the momentum-space bulk spectra, as shown in Figs.~\ref{bulk_gap_1d_HexSSH} and~\ref{bulk_gap_1d_Hchiralized}. At the transition $r/t=1$, the bulk spectrum of the half-chiral Hamiltonian touches $E=0$ while its bulk gap remains open. In contrast, the bulk gap of $H_{\text{chiralized}}$ closes at this transition.

To examine the robustness of the half-chiral topological phase against disorder, we now introduce disorder into the clean half-chiral SSH model. Since disorder breaks translational symmetry, the momentum-space topological invariant becomes ill-defined in this setting. This further highlights the practical advantage of the real-space formulation. Specifically, we introduce disorder by replacing the clean hopping parameters $\boldsymbol{\eta}$ with $\boldsymbol{\eta}+\delta\boldsymbol{\eta}_n$, where
\bea
\boldsymbol{\eta}=(t,r,m),
\qquad
\delta\boldsymbol{\eta}_n=(\delta t_n,\delta r_n,\delta m_n),
\eea
and $\delta t_n$, $\delta r_n$, and $\delta m_n$ are small, site-dependent random fluctuations in the corresponding hopping amplitudes. The resulting disordered Hamiltonian is
\bea\label{disordered half-chiral SSH}
\Tilde H_{\text{SSH}}^{\text{half-chiral}}
\equiv H_{\text{SSH}}^{\text{half-chiral}}
\left(\boldsymbol{\eta}+\delta\boldsymbol{\eta}_n\right).
\eea
As shown in Fig.~\ref{disorderedexSSH_En_distribution}, even in the presence of disorder, a single $E=0$ edge state remains when the system is in the $\nu^{x \in \text{bulk}}_{\tilde{L},\text{half-chiral}}=1$ phase, demonstrating the robustness of this edge state and the bulk-boundary correspondence without relying on translational symmetry.

\subsection{
Half-chiral spectral localizer: how the boundary modes get halved
}\label{Further properties of the spectral localizer in 1D half-chiral topological phases}
\begin{figure*}[t]
\centering
\begin{adjustbox}{max totalsize={\textwidth}{0.72\textheight},center}
\begin{minipage}{\textwidth}
\centering

\subfloat[]{\includegraphics[width=0.5\textwidth,valign=c]{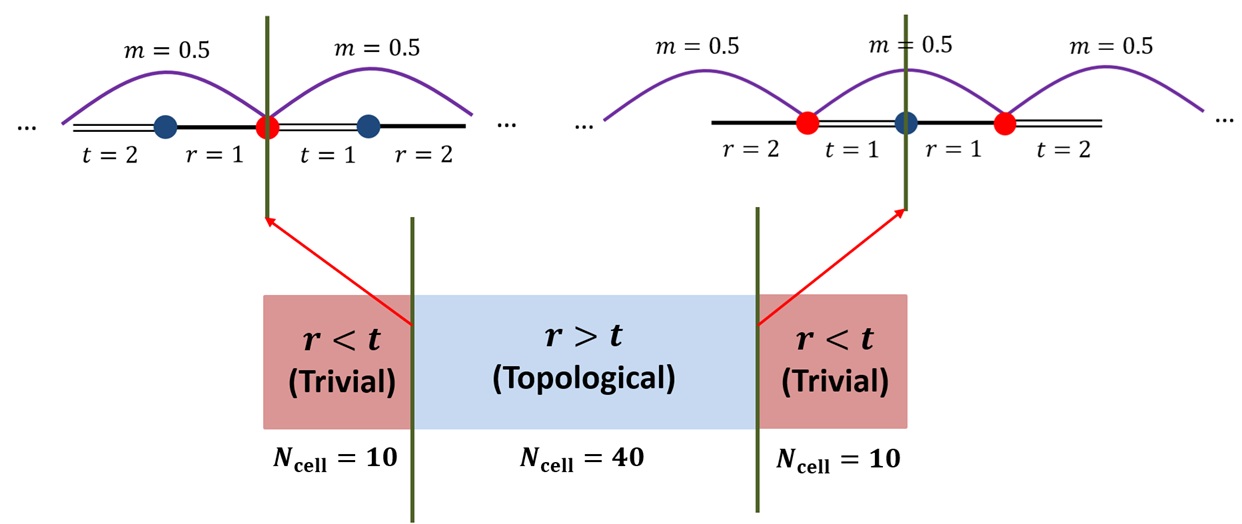}\label{domainwallexSSH_figure}}\\
\subfloat[]{\includegraphics[width=0.32\textwidth,valign=c]{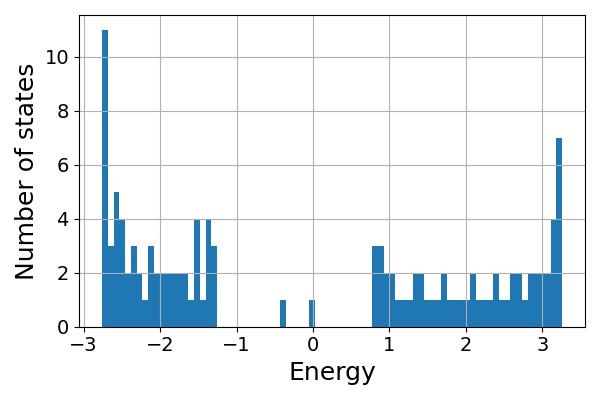}\label{domainwallexSSH_en}}\hskip 0.25cm
\subfloat[]{\includegraphics[width=0.32\textwidth,valign=c]{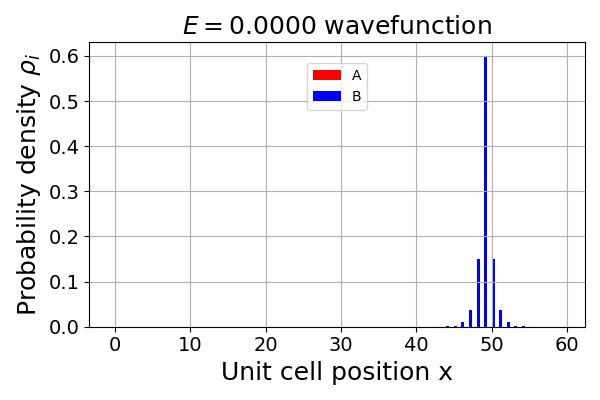}\label{domainwallexSSH_wavefunction}}\hskip 0.25cm
\subfloat[]{\includegraphics[width=0.32\textwidth,valign=c]{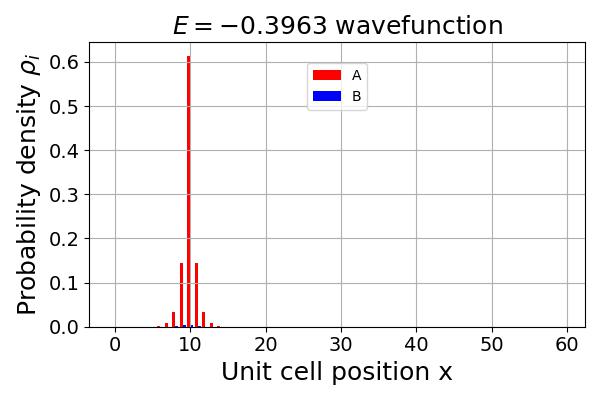}\label{domainwallexSSH_nonzerowavefunction}}\\
\subfloat[]{\includegraphics[width=0.32\textwidth,valign=c]{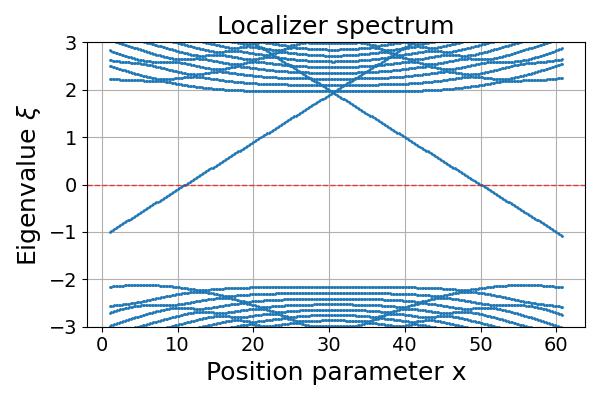}\label{spectrum_L_half_chiral_SSH}}\hskip 0.25cm
\subfloat[]{\includegraphics[width=0.32\textwidth,valign=c]{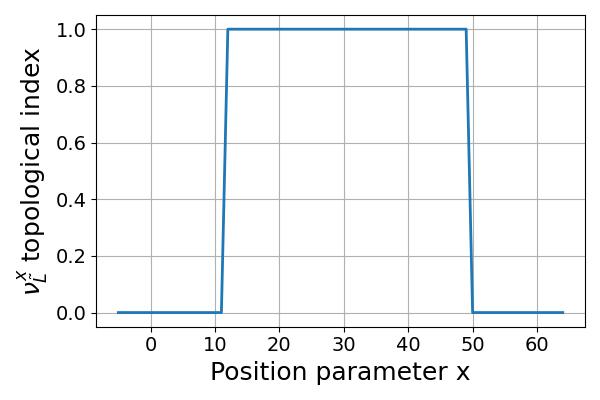}\label{halfchiral_localizer_invariant}}\hskip 0.25cm
\subfloat[]{\includegraphics[width=0.32\textwidth,valign=c]{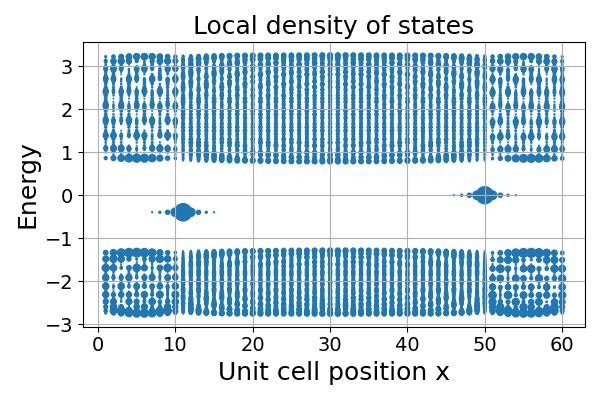}\label{LDS_half_chiral}}\\
\end{minipage}
\end{adjustbox}
\par
\caption{Numerical demonstration of the relation between the $E=0$ state in the domain-wall half-chiral SSH model and the half-chiral spectral localizer $\tilde{L}_{\text{half-chiral}}(x,0,H)$ in Eq.~\eqref{half-chiral reduece spectral localizer}. The results are obtained under OBC. The central topological region has parameters $(t,r,m)=(1,2,0.25)$ and size $N_{\text{cell}}^{C}=40$, while the left and right trivial regions have parameters $(t,r,m)=(2,1,0.25)$ and sizes $N_{\text{cell}}^{L}=N_{\text{cell}}^{R}=10$, respectively. (a) Schematic illustration of the domain-wall half-chiral SSH model. (b) Energy distribution of the eigenstates of $H$, showing a single $E=0$ state. (c) Probability density distribution of the $E=0$ state, showing its localization near the right domain wall. This state is reflected as a $\xi\approx0$ state of $\tilde{L}_{\text{half-chiral}}(x,0,H)$. (d) Probability density distribution of the in-gap state with $E\approx0.77$. This state does not correspond to any $\xi\approx0$ mode of $\tilde{L}_{\text{half-chiral}}(x,0,H)$. (e) Spectrum of the reduced spectral localizer $\tilde{L}_{\text{half-chiral}}(x,0,H)$ as a function of the position parameter $x$, showing paired $\xi\approx0$ modes at two different positions. (f) Spatial dependence of $\nu^{x}_{\tilde{L},\text{half-chiral}}$. The invariant increases near the left domain wall and decreases near the right domain wall. (g) Local density of states of the domain-wall half-chiral SSH model.}
\end{figure*}

We now examine the correspondence between the near-zero modes of the half-chiral reduced spectral localizer and the physical topological $E=0$ boundary states. As discussed in Sec.~\ref{Example: real space topological invariants for half-chiral topological phases using the spectral localizer}, the localizer retains paired $\xi\approx0$ modes, but only half correspond to physical topological $E=0$ states.

To explicitly demonstrate this property, we consider a domain-wall configuration of the half-chiral SSH model, where the hopping parameters are spatially modulated to create interfaces between regions with different half-chiral topological phases, as illustrated in Fig.~\ref{domainwallexSSH_figure}. Specifically, the central region is chosen to be in the nontrivial phase with $r>t$, while the left and right regions are chosen to be in the trivial phase with $r<t$. Compared with an ordinary OBC case, this configuration provides a cleaner setting in which the localized states originate from interfaces between topologically distinct regions rather than the physical boundaries. Moreover, since the domain-wall configuration breaks translational symmetry, the winding number $\nu^{BZ}_{\text{half-chiral}}(H)$ is no longer well defined, and the system can only be characterized by real-space invariants, such as the spectral-localizer invariant $\nu^{x}_{\tilde{L},\text{half-chiral}}(H)$.

As shown in Figs.~\ref{domainwallexSSH_en} and~\ref{domainwallexSSH_wavefunction}, the system hosts a single $E=0$ state localized near the right domain wall at $x\simeq50$. The half-chiral reduced spectral localizer nevertheless exhibits paired $\xi\approx0$ modes near $x\simeq10$ and $x\simeq50$, as shown in Fig.~\ref{spectrum_L_half_chiral_SSH}. Their zero crossings change the localizer signature, producing a unit increase in $\nu^{x}_{\tilde{L},\text{half-chiral}}(H)$ near the left domain wall and a unit decrease near the right domain wall, as shown in Fig.~\ref{halfchiral_localizer_invariant}. Since the model satisfies $SHS^{-1}P_-=-HP_-$, only the decrease corresponds to a physical topological $E=0$ state, following the criterion in Eq.~\eqref{half-chiral localizer selection rule}. Comparing the localizer spectrum in Fig.~\ref{spectrum_L_half_chiral_SSH} with the local density of states in Fig.~\ref{LDS_half_chiral} confirms this correspondence: the near-zero localizer mode associated with the decrease near $x\simeq50$ coincides with the physical $E=0$ state, whereas the mode associated with the increase near $x\simeq10$ does not.

Beyond the numerical comparison, the relation between the localized states of $H_{\text{SSH}}^{\text{half-chiral}}$ and the $\xi\approx0$ modes of $\tilde{L}_{\text{half-chiral}}(x,0,H)$ can be further understood from perturbation theory. In particular, perturbation theory shows that the $E=0$ state of $H_{\text{SSH}}^{\text{half-chiral}}$ is reflected as a $\xi\approx0$ mode of the reduced spectral localizer. In contrast, although the other in-gap state of $H_{\text{SSH}}^{\text{half-chiral}}$ is also localized near the position associated with another $\xi\approx0$ mode of the reduced spectral localizer, it does not correspond to this mode. To understand this difference, we rewrite the domain-wall half-chiral SSH model as the sum of a chiralized part and a chiral-breaking part,
\bea
H_{\text{SSH}}^{\text{half-chiral}}=H_{\text{chiralized}}+H',
\eea
Here, the chiralized part is given by the conventional SSH Hamiltonian,
\bea
H_{\text{chiralized}}=H_{\text{SSH}}
\eea
while the chiral-breaking part is
\bea
H'= \sum_{n=1} 
 m \; c_{n,A}^\dagger c_{n+1,A}
+ \text{h.c.}
\eea
Both Hamiltonians are considered in the same domain-wall configuration.

We first analyze the localized states of the domain-wall half-chiral SSH model. As shown in Figs.~\ref{domainwallexSSH_wavefunction} and~\ref{domainwallexSSH_nonzerowavefunction}, the domain-wall half-chiral SSH model hosts an $E=0$ state with $S=-1$ localized near the right domain wall and an in-gap state localized near the left domain wall, with most of its amplitude distributed on the $S=+1$ sites. More precisely, denoting the left- and right-localized states as $\psi_{+}^{L}$ and $\psi_{-}^{R}$, respectively, we have
\bea
\begin{aligned}
&H_{\text{SSH}}^{\text{half-chiral}}\psi_{-}^{R}=0\quad \quad\quad\;\text{and}\;S\psi_{-}^{R}=-\psi_{-}^{R},\\
&H_{\text{SSH}}^{\text{half-chiral}}\psi_{+}^{L} = E_L\psi_{+}^{L}\quad \text{and}\;S\psi_{+}^{L}\approx\psi_{+}^{L},
\end{aligned}
\eea
where $E_L$ is the energy of the in-gap state $\psi_{+}^{L}$. The approximate relation $S\psi_{+}^{L}\approx\psi_{+}^{L}$ arises because most of the weight of $\psi_{+}^{L}$ is localized on sites belonging to the $S=1$ sector.

We next examine the relation between these two localized states and the chiralized Hamiltonian using perturbation theory. According to the framework developed in Sec.~\ref{bulk-boundary correspondecn for Hermitian half-chiral}, $\psi_-^R$ is also the $E=0$ localized state of the corresponding chiralized Hamiltonian. For the other localized state $\psi_+^L$, its relation to $H_{\text{chiralized}}$ can be obtained using perturbation theory by treating the chiral-breaking term $H'$ as a perturbation. Specifically, the action of $H_{\text{chiralized}}$ on these two localized states is given by
\bea\label{domain wall half-chiral perturbation}
\begin{aligned}
&H_{\text{chiralized}}\psi_{-}^{R}=0;
&&S\psi_{-}^{R}=-\psi_{-}^{R},\\
&H_{\text{chiralized}}\psi_{+}^{L}\approx(E_L-\Delta E)\psi_{+}^{L};
&&S\psi_{+}^{L}\approx\psi_{+}^{L}.
\end{aligned}
\eea
where $\Delta E$ is the energy correction induced by the chiral-breaking term $H'$,
\bea
\Delta E= (\psi_{+}^{L})^{\dag} H' \psi_{+}^{L}.
\eea
In general, $(E_L-\Delta E)$ is non-zero when the chiral-breaking term $H'$ is present. In the special case of $H'=0$, the system recovers the chiral SSH structure, and the in-gap state $\psi_+^L$ becomes an exact $E=0$ state with $E_L=0$. Meanwhile, $\Delta E=0$ because the perturbation term vanishes. We further denote the localization positions of $\psi_{+}^{L}$ and $\psi_{-}^{R}$ as $x_L$ and $x_R$, respectively. Since these states are spatially localized, the action of the position operator can be approximated as
\bea\label{localization position}
\begin{aligned}
&X\psi_{-}^{R}\approx x_R\psi_{-}^{R},\\
&X\psi_{+}^{L}\approx x_L\psi_{+}^{L}.
\end{aligned}
\eea

\begin{figure*}[t]
\centering
\edef\LRQTIrowheight{\the\dimexpr0.30\textwidth\relax}

\begin{minipage}[t]{0.25\textwidth}
\vspace{0pt}
\centering
\begin{minipage}[t][\LRQTIrowheight][t]{\linewidth}
\vspace{0pt}
\centering
\subfloat[]{
\includegraphics[width=0.9\linewidth]{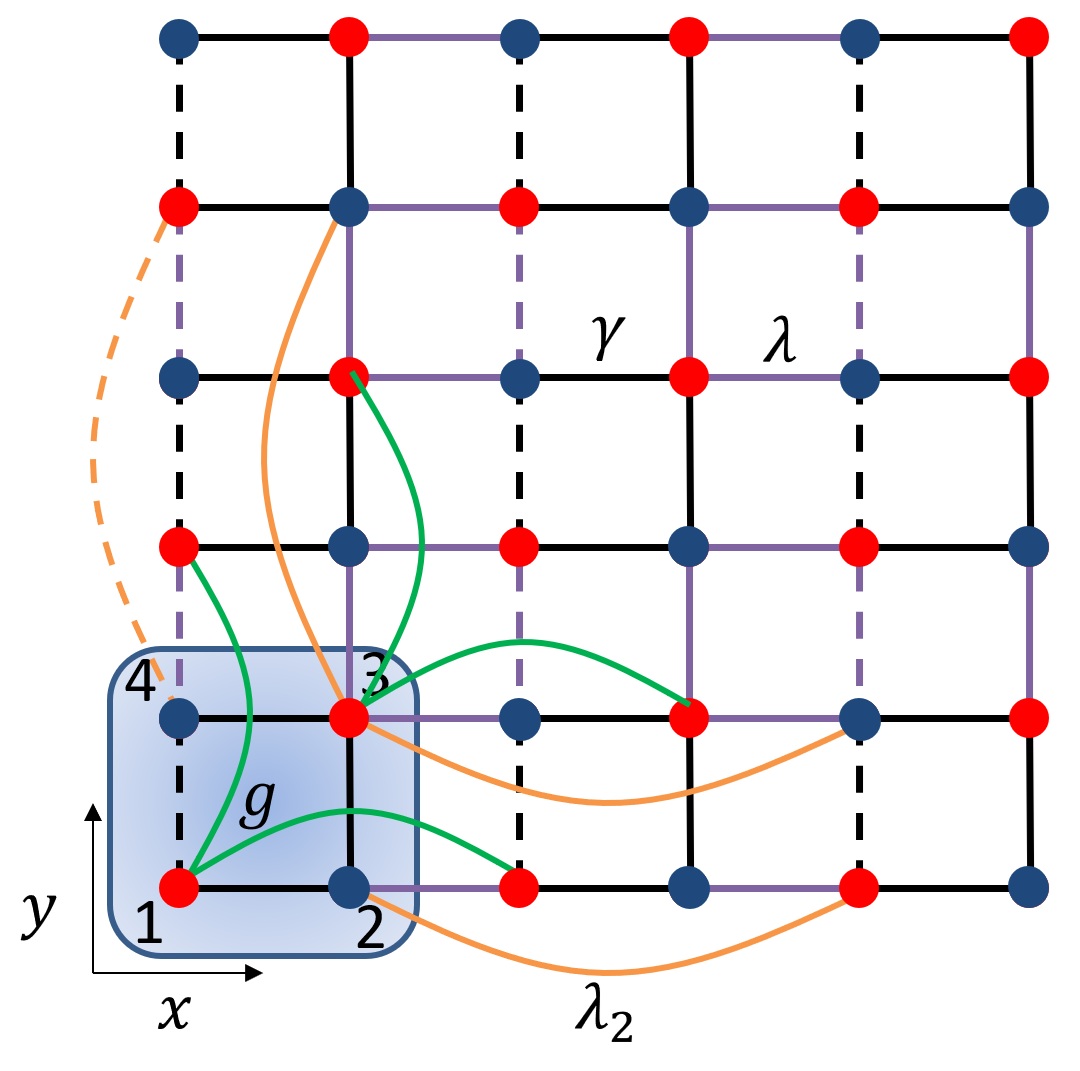}
\label{exLRQTI_figure}
}
\end{minipage}
\par\vspace{6pt}
\subfloat[]{
\includegraphics[width=0.9\linewidth]{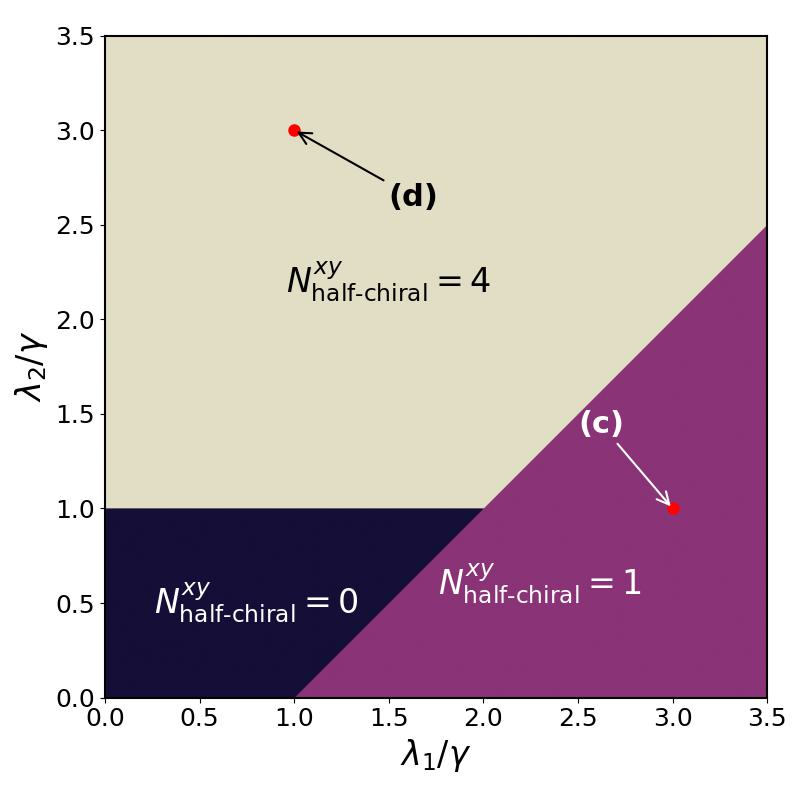}
\label{exLRQTI_phase}
}
\end{minipage}%
\begin{minipage}[t]{0.7\textwidth}
\vspace{0pt}
\centering
\begin{minipage}[t][\LRQTIrowheight][t]{\linewidth}
\vspace{0pt}
\centering
\subfloat[]{
\includegraphics[width=0.42\linewidth,valign=c]{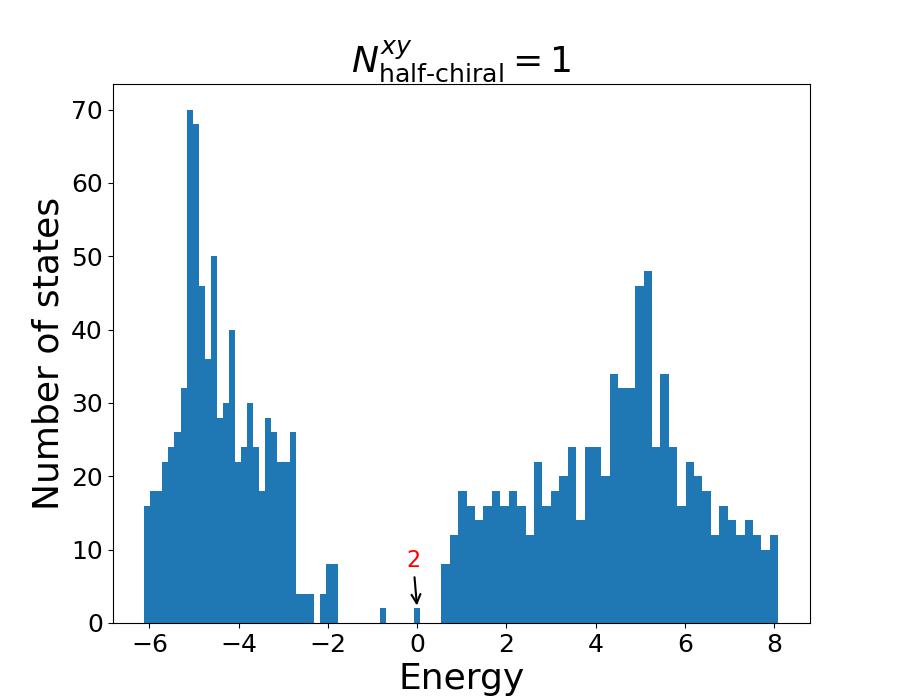}\hfill
\includegraphics[width=0.42\linewidth,valign=c]{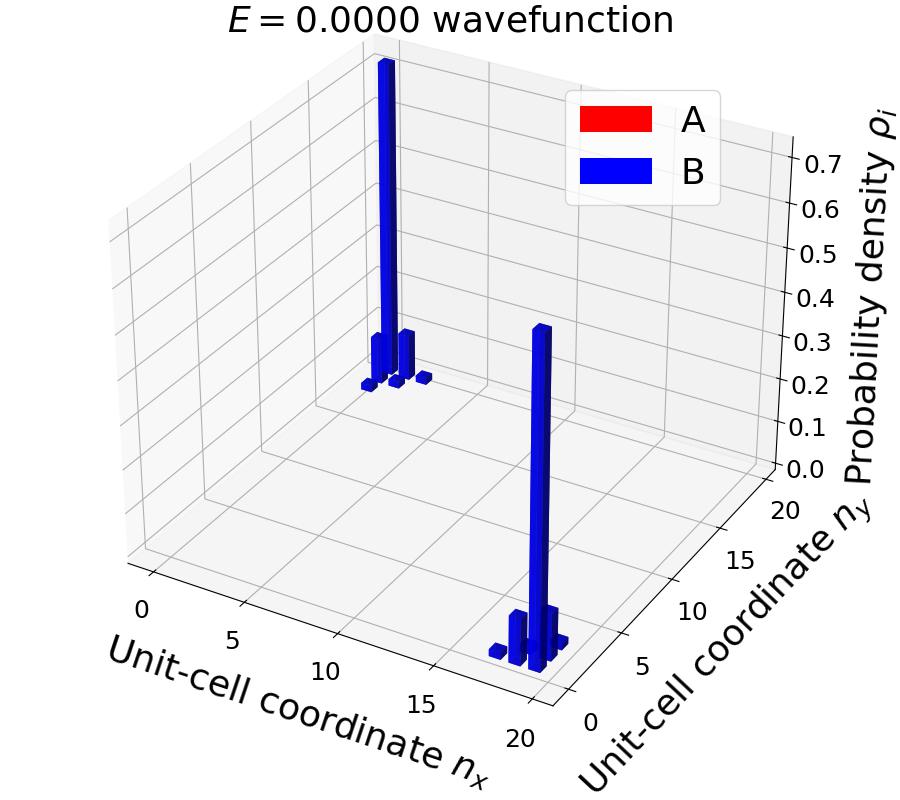}
\label{exLRQTI_N=1}
}
\end{minipage}
\par\vspace{6pt}
\subfloat[]{
\includegraphics[width=0.42\linewidth,valign=c]{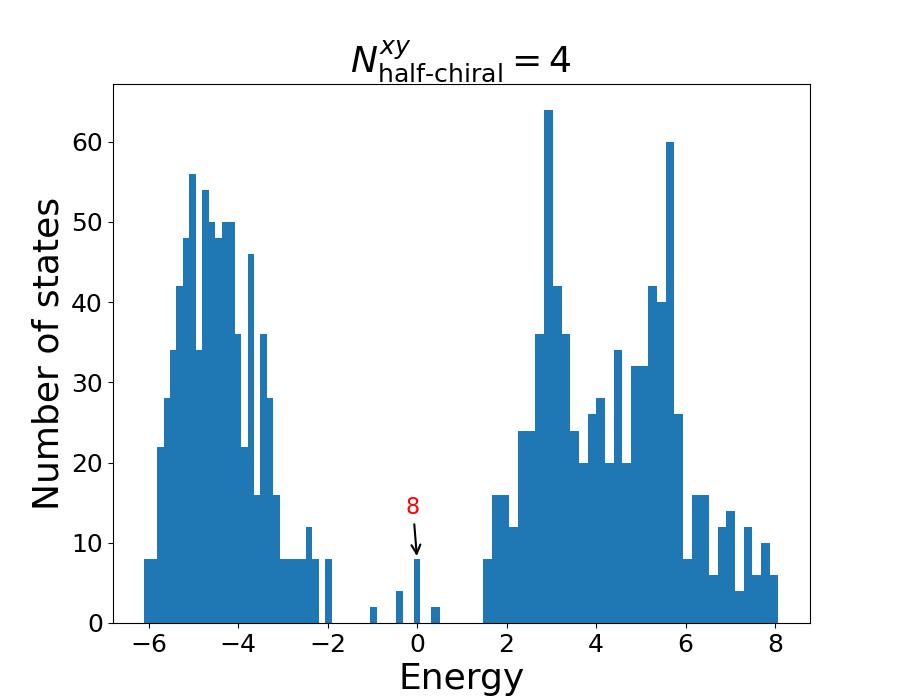}\hfill
\includegraphics[width=0.42\linewidth,valign=c]{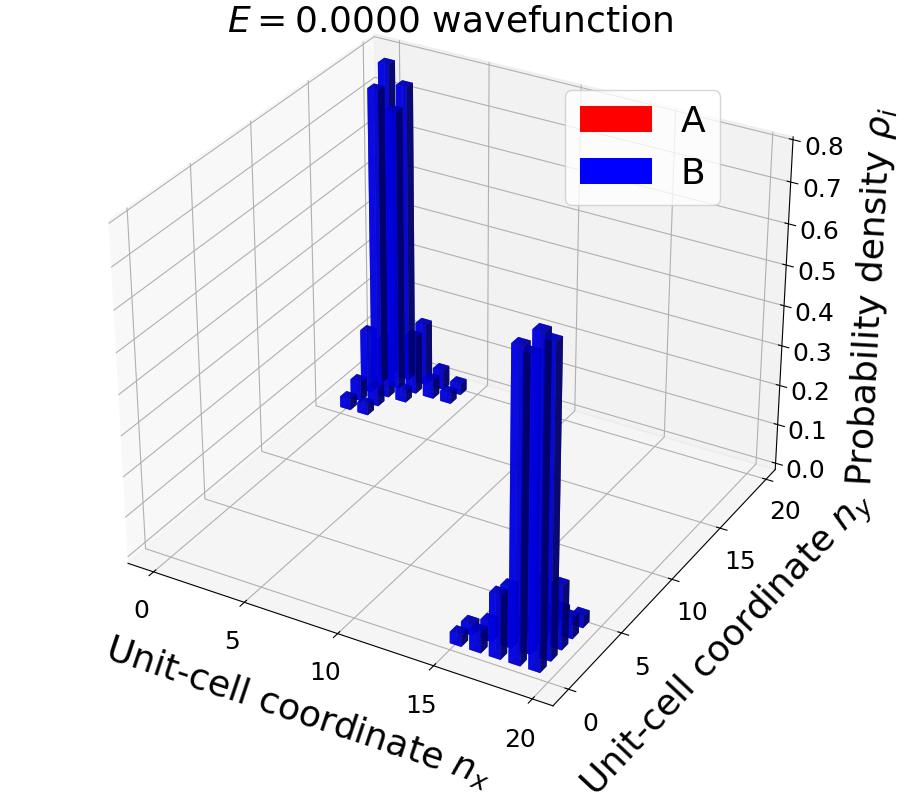}
\label{exLRQTI_N=4}
}
\end{minipage}
\par

\caption{(a) Schematic illustration of the half-chiral LRQTI defined in Eq.~\eqref{exLRQTI}. For clarity, only a few longer-range hoppings ($\lambda_2$) and chiral-breaking hoppings ($g$) are shown. (b) Phase diagram showing the half-chiral invariant $N^{xy}_{\text{half-chiral}}$ for the half-chiral LRQTI. (c) Energy distribution of states (left) and total probability density of the $E=0$ states (right) for a $20\times20$-unit-cell half-chiral LRQTI in the $N^{xy}_{\text{half-chiral}}=1$ phase. The parameters are $\gamma=1$, $\lambda_1=3$, $\lambda_2=1$, and $g=0.5$. Two $E=0$ corner states are present. (d) Same as (c) but in the $N^{xy}_{\text{half-chiral}}=4$ phase with $\gamma=1$, $\lambda_1=1$, $\lambda_2=3$, and $g=0.5$. Eight $E=0$ corner states are present.}
\end{figure*}

\begin{figure*}[t]
\centering
\subfloat[]{\includegraphics[width=0.35\textwidth,valign=c]{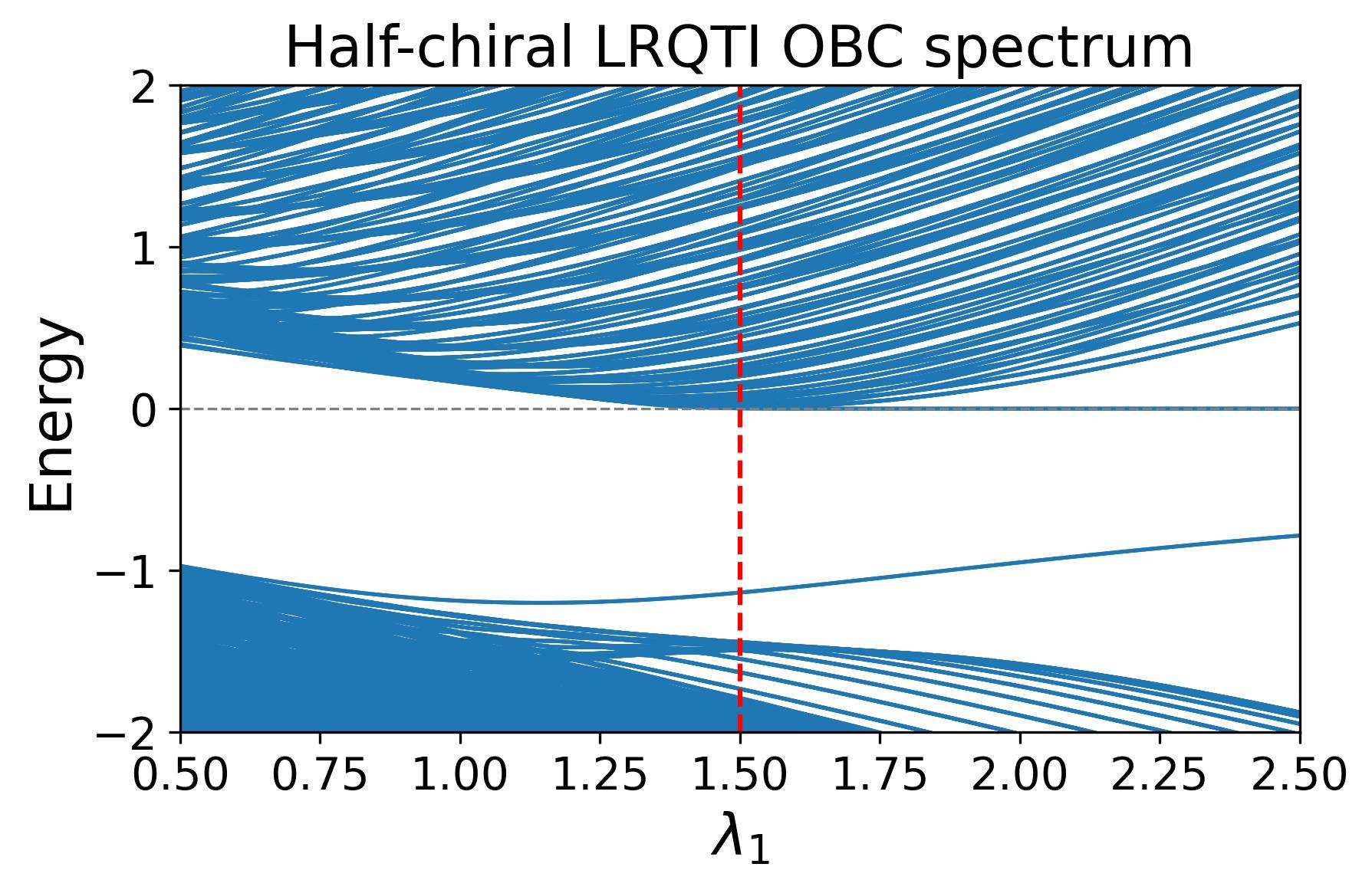}\label{exLRQTI_OBC_spectrum}}\hskip 0.5cm
\subfloat[]{\includegraphics[width=0.35\textwidth,valign=c]{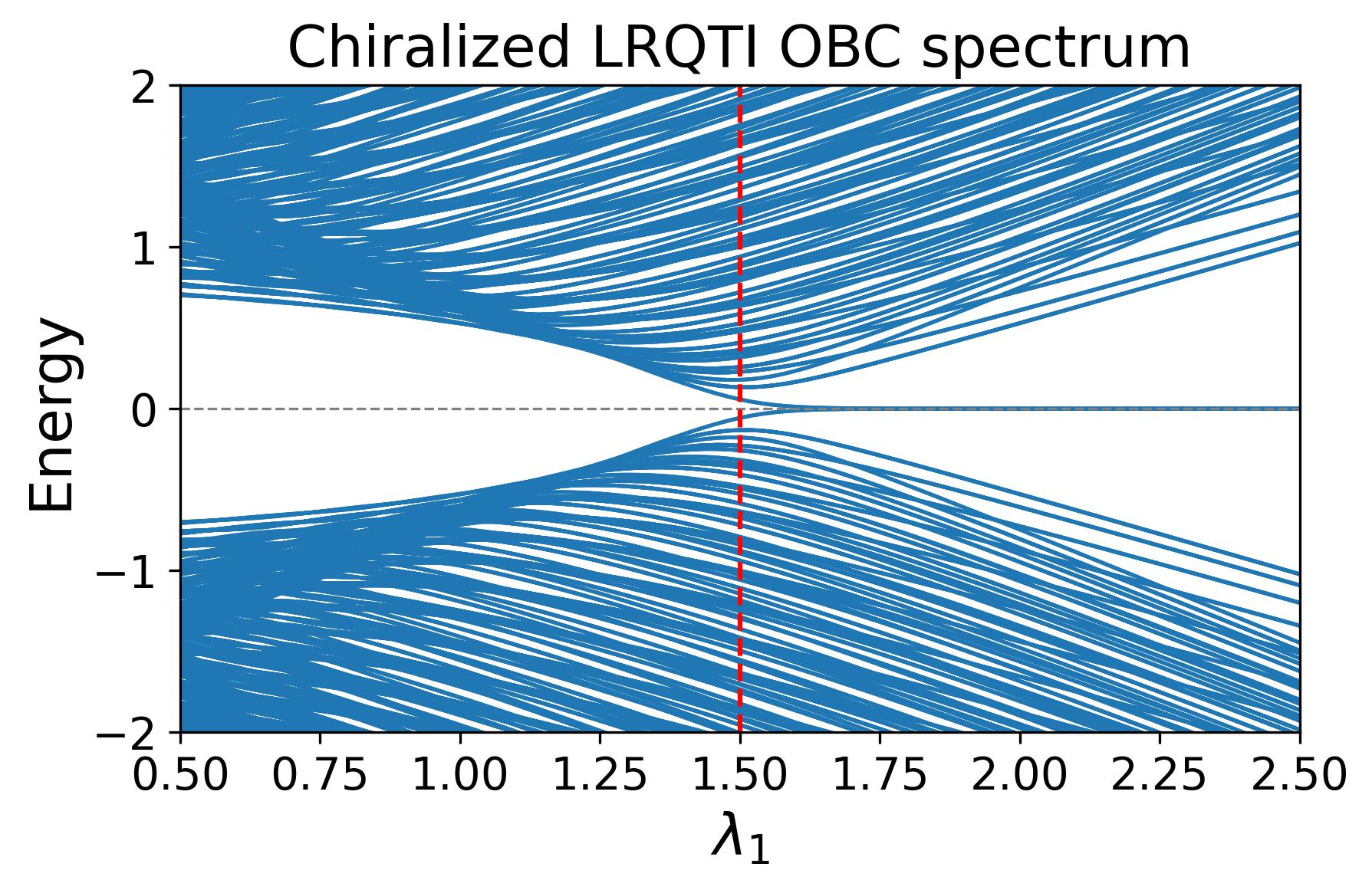}\label{LRQTI_chiralized_OBC_spectrum}}\\
\subfloat[]{\includegraphics[width=0.35\textwidth,valign=c]{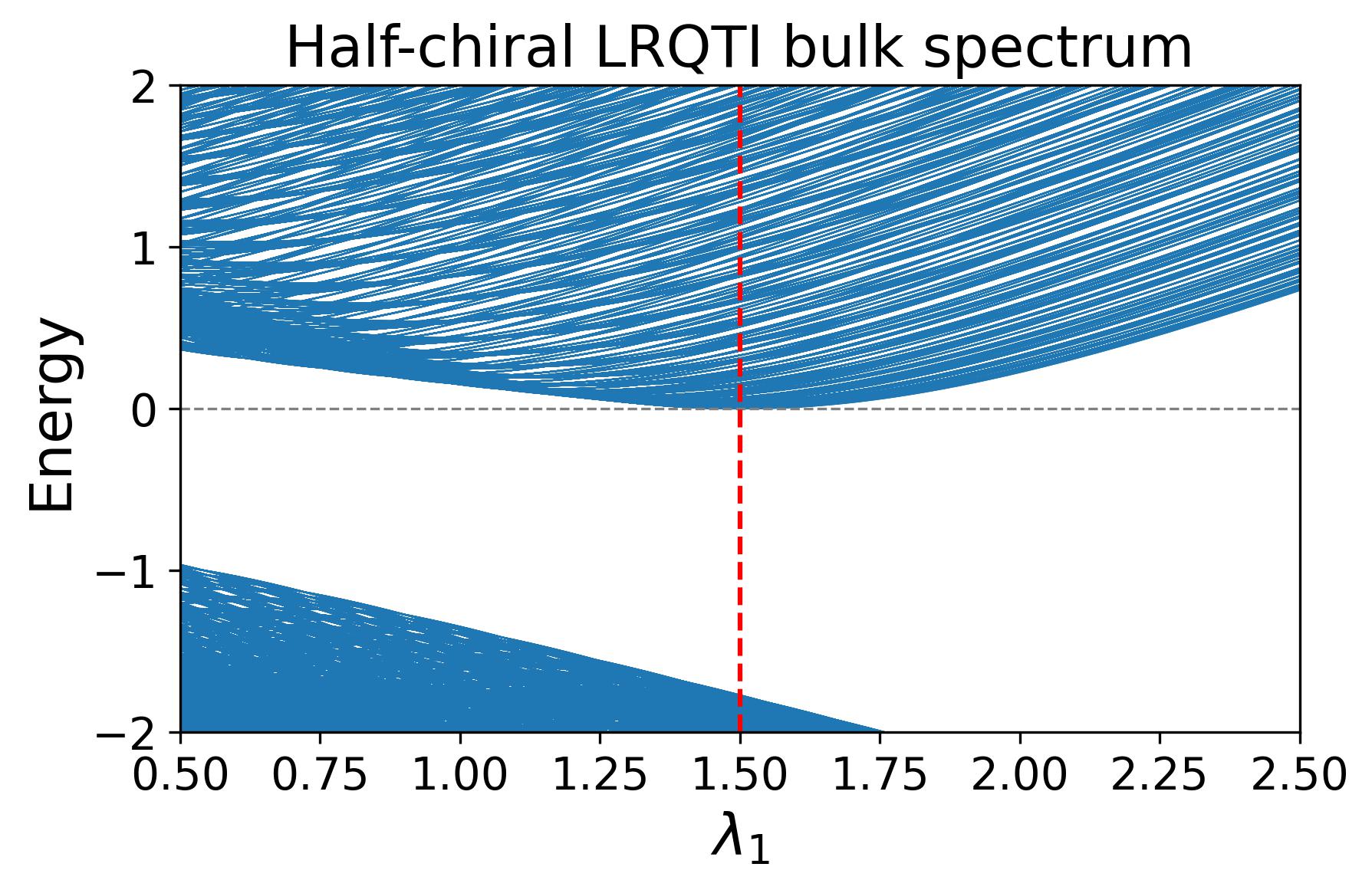}\label{exLRQTI_bulk_spectrum}}\hskip 0.5cm
\subfloat[]{\includegraphics[width=0.35\textwidth,valign=c]{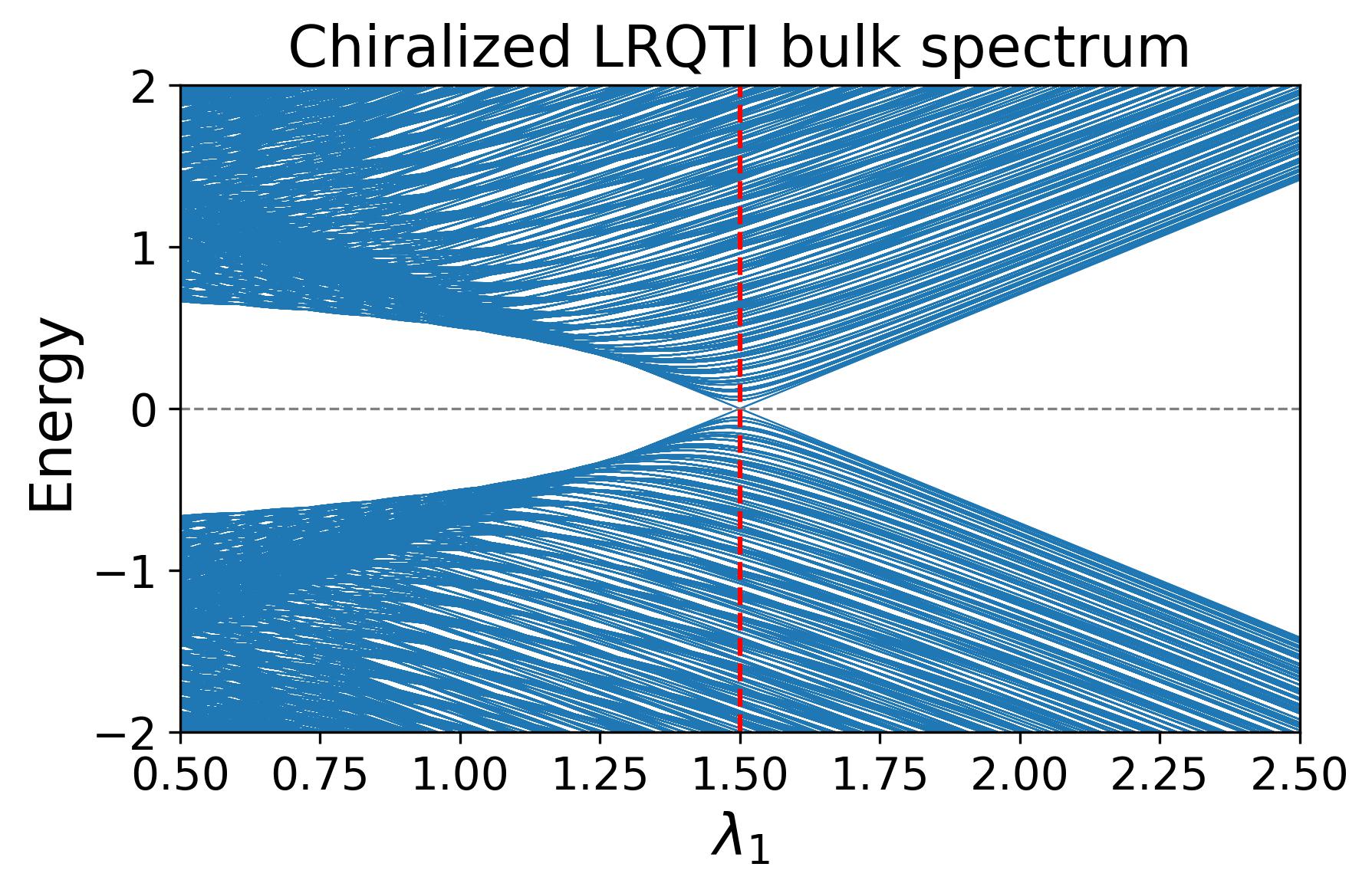}\label{LRQTI_chiralized_bulk_spectrum}}
\caption{Spectral evolution of the half-chiral LRQTI and its chiralized Hamiltonian $H_{\text{chiralized}}=H_{\text{LRQTI}}$ as $\lambda_1$ is varied, with $\gamma=1$, $\lambda_2=0.5$, and $g=0.5$. The red dashed lines mark the thermodynamic transition at $\lambda_1=1.5$ from the trivial phase to the HOT phase with $N^{xy}_{\text{half-chiral}}=1$. (a) OBC spectrum of the half-chiral LRQTI for a $20\times20$-unit-cell system, showing the emergence of two topological $E=0$ corner states around the transition. (b) Corresponding OBC spectrum of the chiralized Hamiltonian, showing the emergence of four topological $E=0$ corner states around the transition. (c) Momentum-space bulk spectrum of the half-chiral LRQTI, which touches $E=0$ at the transition while its bulk gap remains open. (d) Corresponding bulk spectrum of the chiralized Hamiltonian, whose bulk gap closes at the transition.}
\label{exLRQTI_spectral_evolution}
\end{figure*}

\begin{figure*}[t]
\centering
\begin{adjustbox}{max totalsize={\textwidth}{0.72\textheight},center}
\begin{minipage}{\textwidth}
\centering

\subfloat[]{
\includegraphics[width=0.3\textwidth]{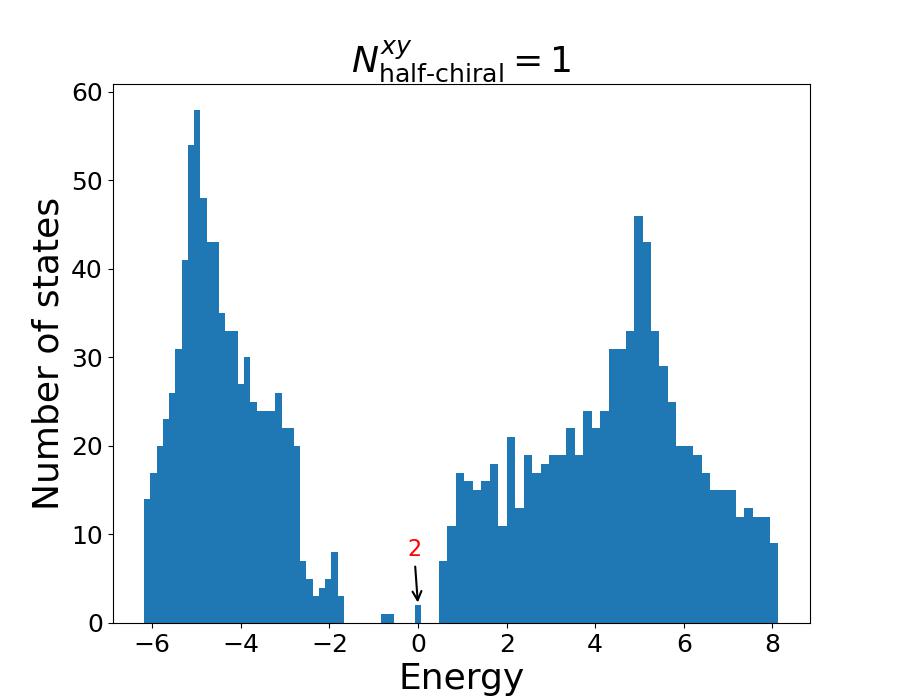}\hfill
\includegraphics[width=0.3\textwidth]{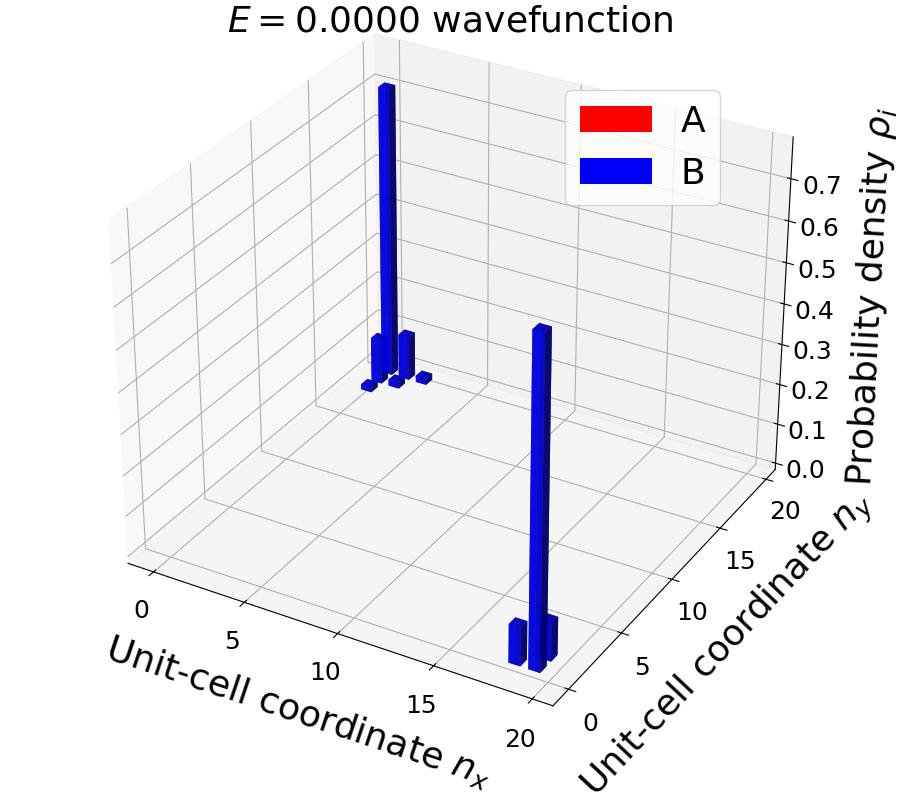}
\label{disordedexLRQTI_N=1}} \\
\subfloat[]{
\includegraphics[width=0.3\textwidth]{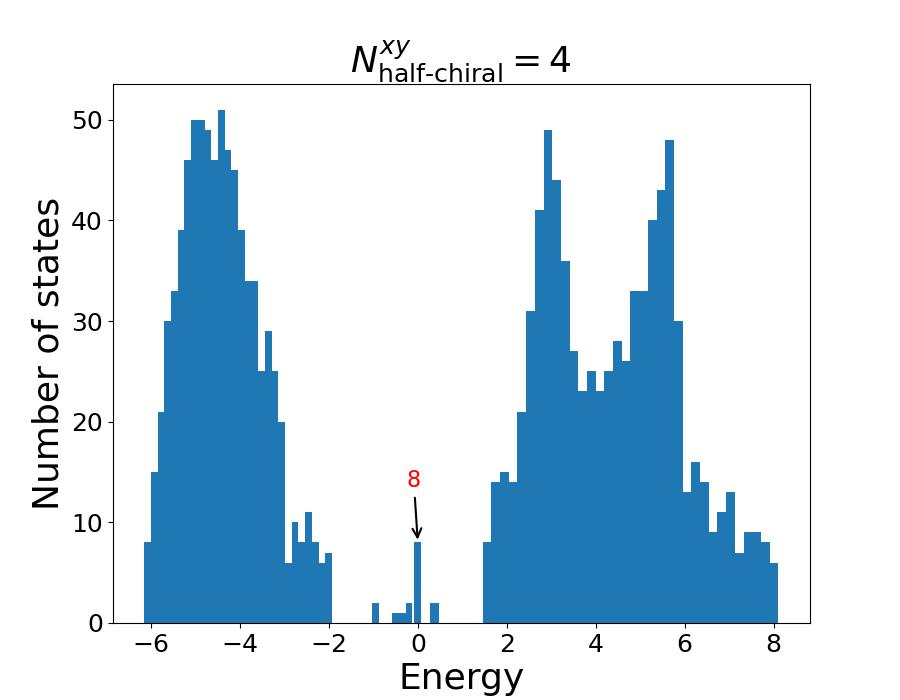}\hfill
\includegraphics[width=0.3\textwidth]{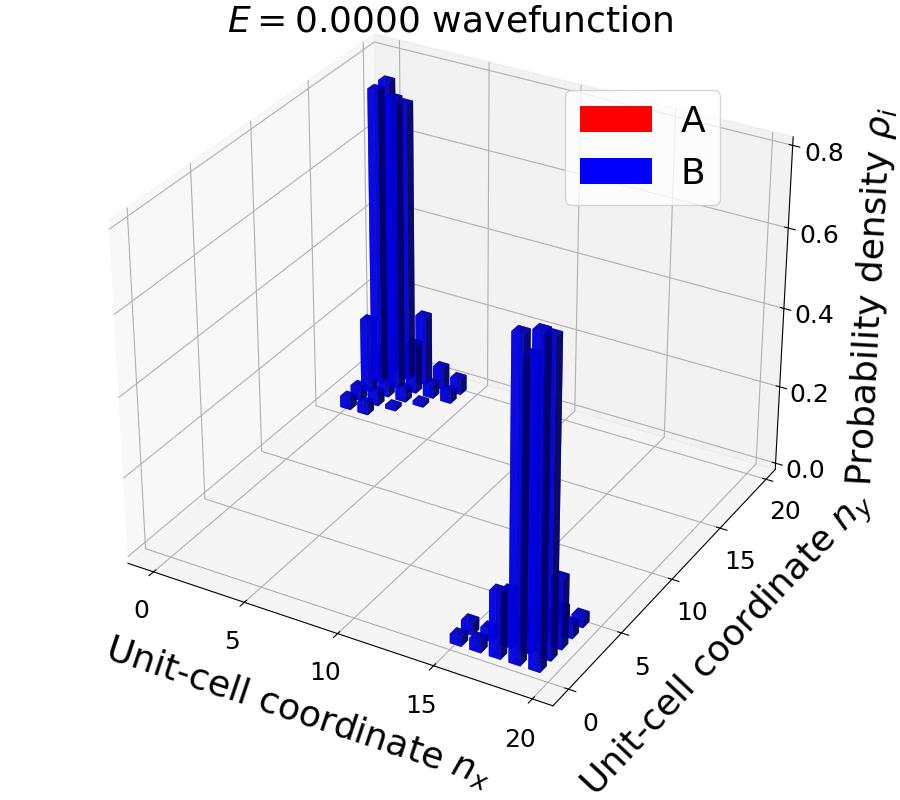}
\label{disordedexLRQTI_N=4}} \\
\end{minipage}
\end{adjustbox}
\par
\caption{Topological $E=0$ corner states of the disordered half-chiral LRQTI in the $N^{xy}_{\text{half-chiral}}=1$ and $N^{xy}_{\text{half-chiral}}=4$ half-chiral HOT phase. (a) Energy distribution of states (left) and total probability density of the $E=0$ states (right) for a $20\times20$-unit-cell square lattice in the $N^{xy}_{\text{half-chiral}}=1$ phase. The parameters are $\gamma=1$, $\lambda_1=3$, $\lambda_2=1$, and $g=0.5$. The disorder terms $\delta \gamma_n$, $\delta \lambda_{1,n}$, $\delta \lambda_{2,n}$, and $\delta g_{n}$ are independently drawn from a distribution in $[-0.2,0.2]$. There are two $E=0$ corner states. (b) Same as (a) but in the $N^{xy}_{\text{half-chiral}}=4$ phase with $\gamma=1$, $\lambda_1=1$, $\lambda_2=3$, and $g=0.5$. There are eight $E=0$ corner states.}
\end{figure*}

Using the perturbative results in Eq.~\eqref{domain wall half-chiral perturbation} together with the localization properties in Eq.~\eqref{localization position}, we can now understand why only the $E=0$ localized state of the domain-wall half-chiral SSH model is reflected as a $\xi\approx0$ mode of $\tilde{L}_{\text{half-chiral}}(x,0,H)$.
We first apply $\Tilde{L}_{\text{half-chiral}}(x_R,0,H)$ in Eq.~\eqref{half-chiral reduece spectral localizer} to the $E=0$ state $\psi_-^R$. We obtain
\bea
\begin{aligned}
&\Tilde{L}_{\text{half-chiral}}(x_R,0,H)\psi_-^R\\
&=\Big[iH_{\text{chiralized}}-\kappa(X-x_R\mathbbm{1})\Big]S\psi_-^R\\
&=-iH_{\text{chiralized}}\psi_-^R+\kappa(X\psi_-^R-x_R\psi_-^R)\\
&=\kappa(X\psi_-^R-x_R\psi_-^R)\\
&\approx\kappa(x_R\psi_-^R-x_R\psi_-^R)=0.
\end{aligned}
\eea
Therefore, the $E=0$ state of the domain-wall half-chiral SSH model is indeed reflected as a $\xi\approx0$ mode of $\tilde{L}_{\text{half-chiral}}(x,0,H)$. We next apply $\Tilde{L}_{\text{half-chiral}}(x_L,0,H)$ to the other localized in-gap state $\psi_{+}^{L}$. We obtain
\bea
\begin{aligned}
&\Tilde{L}_{\text{half-chiral}}(x_L,0,H)\psi_+^L\\
&=\Big[iH_{\text{chiralized}}-\kappa(X-x_L\mathbbm{1})\Big]S\psi_+^L\\
&\approx iH_{\text{chiralized}}\psi_+^L-\kappa(X\psi_+^L-x_L\psi_+^L)\\
&\approx i(E_L-\Delta E)\psi_+^L-\kappa(x_L\psi_+^L-x_L\psi_+^L)\\
&\approx i(E_L-\Delta E)\psi_+^L.
\end{aligned}
\eea
Therefore, although $\psi_+^L$ is localized near the position where another $\xi\approx0$ mode of $\tilde{L}_{\text{half-chiral}}(x,0,H)$ appears, it is not reflected as that $\xi\approx0$ mode. This is because $\tilde{L}_{\text{half-chiral}}(x,0,H)$ is a Hermitian operator whose eigenvalues must be real, whereas the above result would correspond to a purely imaginary eigenvalue.

\section{2D half-chiral HOT phases characterized by MCNs}\label{2D half-chiral HOT phases with MCNs}

In this section, we apply the framework developed in Sec.~\ref{bulk-boundary correspondecn for Hermitian half-chiral} to construct 2D half-chiral higher-order topological (half-chiral HOT) phases from the MCN-characterized chiral-symmetric HOT phases reviewed in Sec.~\ref{2D Chiral-symmetric Higher-order topological phases}. Following the general invariant construction in Eq.~\eqref{general half-chiral invariant}, we define the half-chiral MCN as
\bea\label{half-chiral MCN}
N^{xy}_{\text{half-chiral}}(H)=N^{xy}_{\text{chiral}}(H_{\text{chiralized}}).
\eea
This invariant characterizes the topological $E=0$ corner states of the resulting half-chiral HOT phases, whose number is half that of their corresponding chiralized Hamiltonians, according to Eq.~\eqref{BBC_half-chiral topological}.

These phases can also be viewed as a real-space extension of 1D half-chiral topological phases, mirroring the construction of MCN-characterized 2D chiral HOT phases from 1D chiral-symmetric SPT phases. Specifically, using sublattice-resolved operators, the winding number of $H_{\text{chiralized}}$, which characterizes 1D half-chiral topological phases, can be reformulated in real space. Extending this real-space invariant to 2D yields the MCN of $H_{\text{chiralized}}$ and hence the half-chiral MCN defined above. This connection shows that the real-space construction of 1D half-chiral topological phases serves not only to characterize systems lacking translation symmetry, but also to construct higher-order topological phases.

To see the bulk-boundary correspondence in this class, we consider the following half-chiral LRQTI, as shown in Fig.~\ref{exLRQTI_figure},
\bea\label{exLRQTI}
H_{\text{LRQTI}}^{\text{half-chiral}}=H_{\text{LRQTI}}+H_{\text{cb}}.
\eea
where $H_{\text{LRQTI}}$ is the LRQTI defined in Eq.~\eqref{LRQTI}, and the chiral-breaking term $H_{\text{cb}}$ is given by
\bea
\begin{aligned}
H_{\text{cb}}
&=g\sum_{\mathbf R}\Bigl(
c_{\mathbf R,1}^\dagger c_{\mathbf R+\hat{x},1}+c_{\mathbf R,1}^\dagger c_{\mathbf R+\hat{y},1}\\
&\hphantom{=g\sum_{\mathbf R}\Bigl(}{}
+c_{\mathbf R,3}^\dagger c_{\mathbf R+\hat{x},3}+c_{\mathbf R,3}^\dagger c_{\mathbf R+\hat{y},3}\Bigr)+\text{h.c.}.
\end{aligned}
\eea
The model satisfies the half-chiral symmetry $SHS^{-1}P_{-} = -HP_{-}$, where
\bea
\begin{aligned}
&S=\sum_{\mathbf R}\Bigl(
c_{\mathbf R,1}^\dagger c_{\mathbf R,1}-c_{\mathbf R,2}^\dagger c_{\mathbf R,2}\\
&\hphantom{S=\sum_{\mathbf R}\Bigl(}{}
+c_{\mathbf R,3}^\dagger c_{\mathbf R,3}-c_{\mathbf R,4}^\dagger c_{\mathbf R,4}\Bigr),\\
&P_-=\frac{1}{2}(1-S),
\end{aligned}
\eea
and the corresponding chiralized Hamiltonian is
\bea
H_{\text{chiralized}}[H_{\text{LRQTI}}^{\text{half-chiral}}]=H_{\text{LRQTI}}.
\eea
Therefore, the half-chiral MCN defined in Eq.~\eqref{half-chiral MCN} is obtained by evaluating the conventional MCN of $H_{\text{LRQTI}}$, yielding the phase diagram in Fig.~\ref{exLRQTI_phase}. The physical Hamiltonian inherits half of the topological $E=0$ corner states of $H_{\text{LRQTI}}$, according to Eq.~\eqref{BBC_half-chiral topological}. Combining this inheritance relation with the bulk-boundary correspondence for conventional chiral HOT phases reviewed in Sec.~\ref{2D Chiral-symmetric Higher-order topological phases}, we predict two topological $E=0$ corner states in the $N^{xy}_{\text{half-chiral}}=1$ phase and eight in the $N^{xy}_{\text{half-chiral}}=4$ phase. The numerical results in Figs.~\ref{exLRQTI_N=1} and~\ref{exLRQTI_N=4} confirm these predictions.

A half-chiral HOT phase transition can be associated with the boundary or bulk spectrum of the physical Hamiltonian touching $E=0$, without necessarily closing the corresponding gap. Here, we illustrate the latter case by fixing $\gamma=1$, $\lambda_2=0.5$, and $g=0.5$ and varying $\lambda_1$. Along this parameter path, the system undergoes a transition from the trivial phase to the HOT phase with $N^{xy}_{\text{half-chiral}}=1$ at $\lambda_1=1.5$. Figs.~\ref{exLRQTI_OBC_spectrum} and~\ref{LRQTI_chiralized_OBC_spectrum} show the OBC spectra of the physical and chiralized Hamiltonians for a $20\times20$-unit-cell system. In the physical Hamiltonian, the bulk spectrum approaches $E=0$ around $\lambda_1=1.5$ while the bulk gap remains open, followed by the emergence of two topological $E=0$ corner states. In the chiralized Hamiltonian, the bulk gap closes around the same transition, and four topological $E=0$ corner states emerge. To rule out finite-size effects, we further confirm this spectral behavior using the momentum-space bulk spectra shown in Figs.~\ref{exLRQTI_bulk_spectrum} and~\ref{LRQTI_chiralized_bulk_spectrum}.

These half-chiral HOT phases can also be characterized within the spectral-localizer framework. The corresponding half-chiral invariant, whose detailed construction will be presented in Sec.~\ref{2D half-chiral HOT phases without MCNs}, is obtained by evaluating the chiral spectral-localizer invariant for $H_{\text{chiralized}}$. Since $H_{\text{chiralized}}=H_{\text{LRQTI}}$ here, the results in Fig.~\ref{LRQTI_SL} also characterize the topological $E=0$ corner states shown in Fig.~\ref{exLRQTI_N=4}.

As in 1D half-chiral topological phases, the robustness of 2D half-chiral HOT phases against disorder can be illustrated by adding small, site-dependent perturbations to the half-chiral LRQTI. We introduce the parameter vector
\bea
\boldsymbol{\eta}=(\gamma,\lambda_1,\lambda_2,g)
\eea
and its site-dependent perturbation
\bea
\delta\boldsymbol{\eta}_n=
(\delta\gamma_n,\delta\lambda_{1,n},
\delta\lambda_{2,n},\delta g_n),
\eea
where the components of $\delta\boldsymbol{\eta}_n$ are small, site-dependent disorder terms added to the corresponding hopping parameters. We then write the disordered Hamiltonian as
\bea\label{disordered extended LRQTI}
\Tilde{H}_{\text{LRQTI}}^{\text{half-chiral}}
\equiv H_{\text{LRQTI}}^{\text{half-chiral}}
\left(\boldsymbol{\eta}+\delta\boldsymbol{\eta}_n\right).
\eea
As shown in Figs.~\ref{disordedexLRQTI_N=1} and~\ref{disordedexLRQTI_N=4}, $\Tilde{H}_{\text{LRQTI}}^{\text{half-chiral}}$ hosts two and eight $E=0$ corner states in the $N^{xy}_{\text{half-chiral}}=1$ and $N^{xy}_{\text{half-chiral}}=4$ half-chiral HOT phases, respectively, demonstrating the robustness of these corner states. Moreover, while the clean half-chiral LRQTI $H_{\text{LRQTI}}^{\text{half-chiral}}$ respects $C_2$ symmetry, defined by
\bea
\begin{aligned}
\hat C_2=\sum_{\mathbf R}\Big(
&c_{-\mathbf R,3}^\dagger c_{\mathbf R,1}
+c_{-\mathbf R,4}^\dagger c_{\mathbf R,2}\\
&+c_{-\mathbf R,1}^\dagger c_{\mathbf R,3}
+c_{-\mathbf R,2}^\dagger c_{\mathbf R,4}
\Big),
\end{aligned}
\eea
the site-dependent perturbations introduced here explicitly break this symmetry. The persistence of the $E=0$ corner states under such conditions confirms that their existence depends solely on the half-chiral symmetry~\eqref{type1SCS}, independently of crystalline symmetries.

Our results on half-chiral HOT phases should be contrasted with the recent work in Ref.~\cite{kang_higherorder_2026}. In that study, a QTI subjected to a quasi-periodic potential hosts three $E=0$ corner states in its topological phase, characterized by quantized state-resolved quadrupole moments. Although this model also satisfies half-chiral symmetry, the presence of all three $E=0$ corner states is not protected by half-chiral symmetry alone. Only two of these states are protected by half-chiral symmetry. The third is an additional $E=0$ state whose protection relies on the more restrictive subsymmetry condition; it can be shifted away from zero energy by perturbations that preserve half-chiral symmetry but break that condition.

\section{2D chiral HOT phases with spectral-localizer characterization but vanishing MCNs}\label{2D Chiral-symmetric Higher-order topological phases without MCNs}
\begin{figure*}[t]
\centering
\edef\Kekulerowheight{\the\dimexpr0.25\textwidth\relax}

\begin{minipage}[t]{0.33\textwidth}
\vspace{0pt}
\centering
\begin{minipage}[t][\Kekulerowheight][t]{\linewidth}
\vspace{0pt}
\centering
\subfloat[]{
\includegraphics[width=0.95\linewidth]{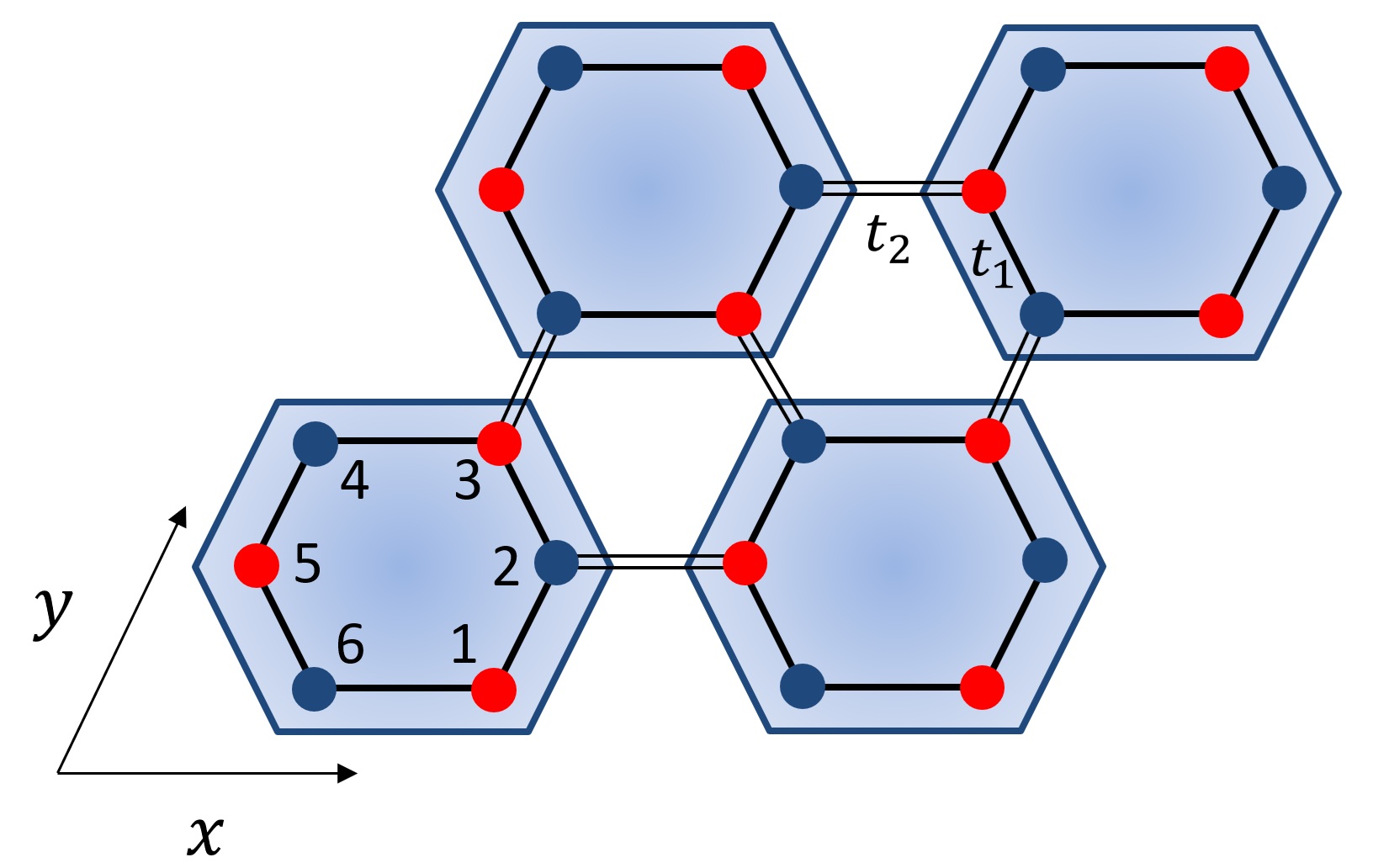}
\label{Fig_Kekule}
}
\end{minipage}
\par\vspace{6pt}
\subfloat[]{
\includegraphics[width=0.95\linewidth]{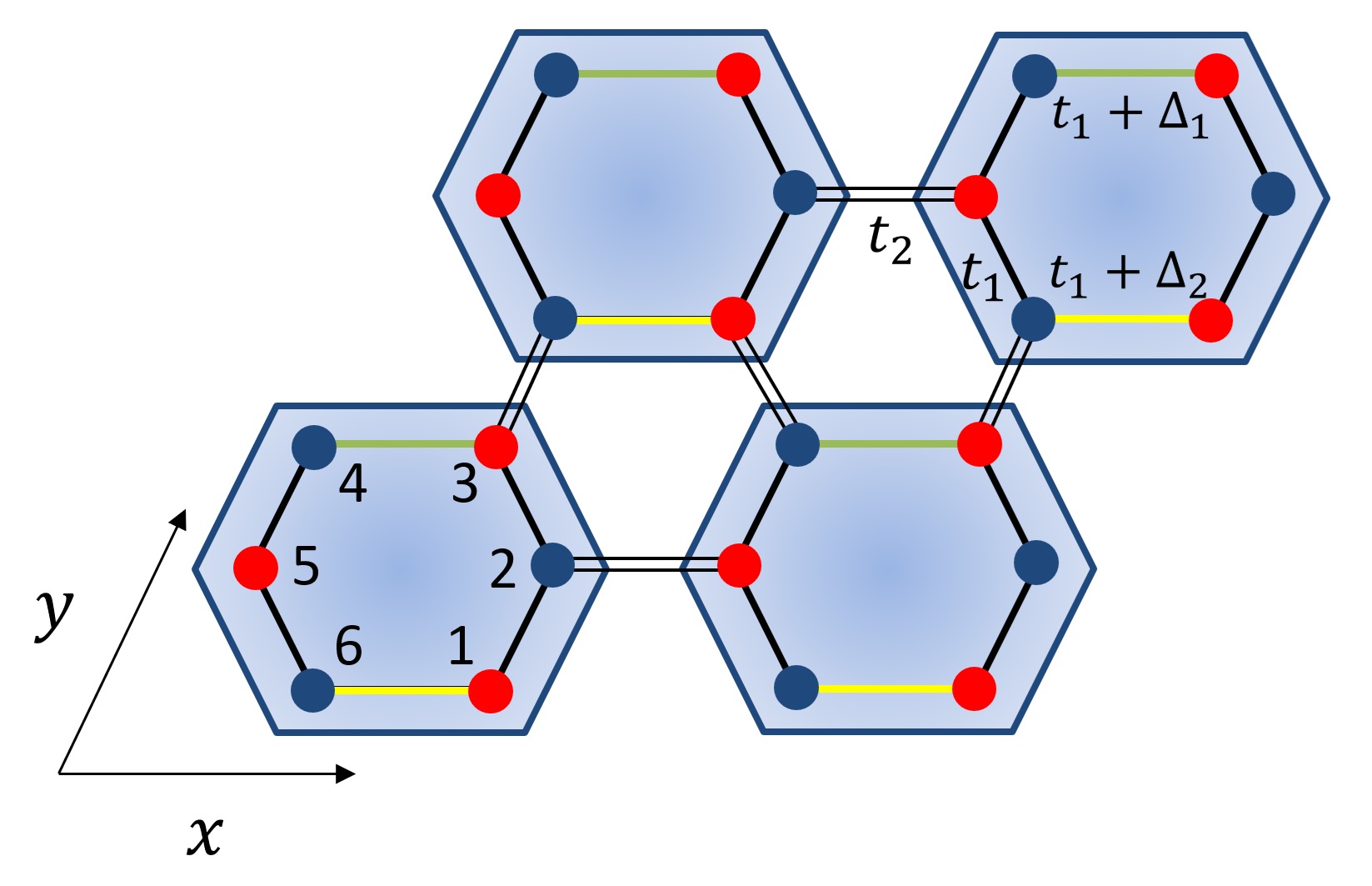}
\label{Fig_Kekule_delta}
}
\end{minipage}%
\hspace{0.01\textwidth}%
\begin{minipage}[t]{0.6\textwidth}
\vspace{0pt}
\centering
\begin{minipage}[t][\Kekulerowheight][t]{\linewidth}
\vspace{0pt}
\centering
\subfloat[]{
\begin{minipage}{\linewidth}
\centering
\includegraphics[width=0.46\linewidth,valign=c]{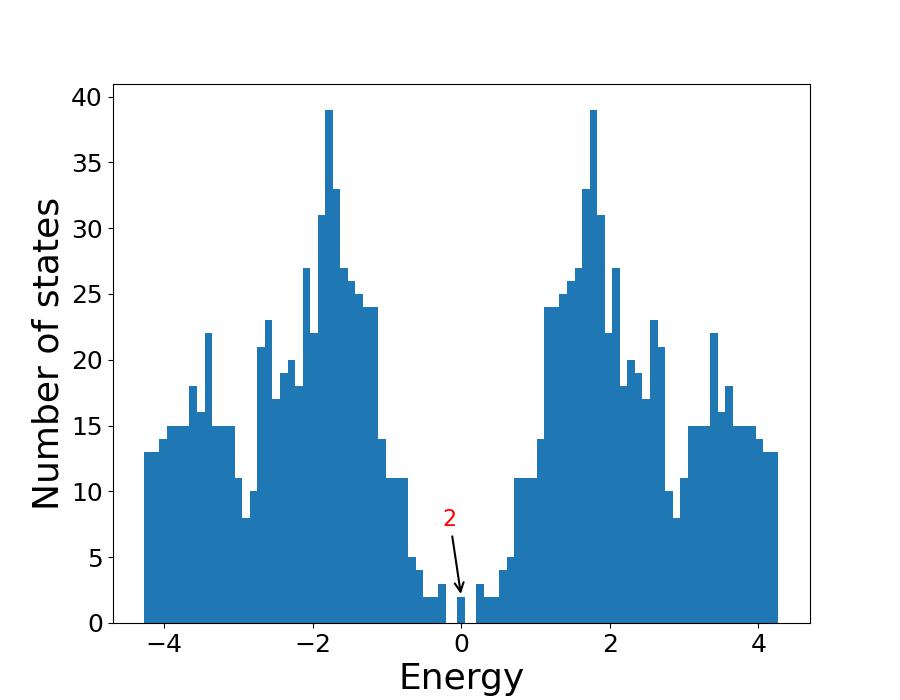}%
\hspace{0.1cm}%
\includegraphics[width=0.46\linewidth,valign=c]{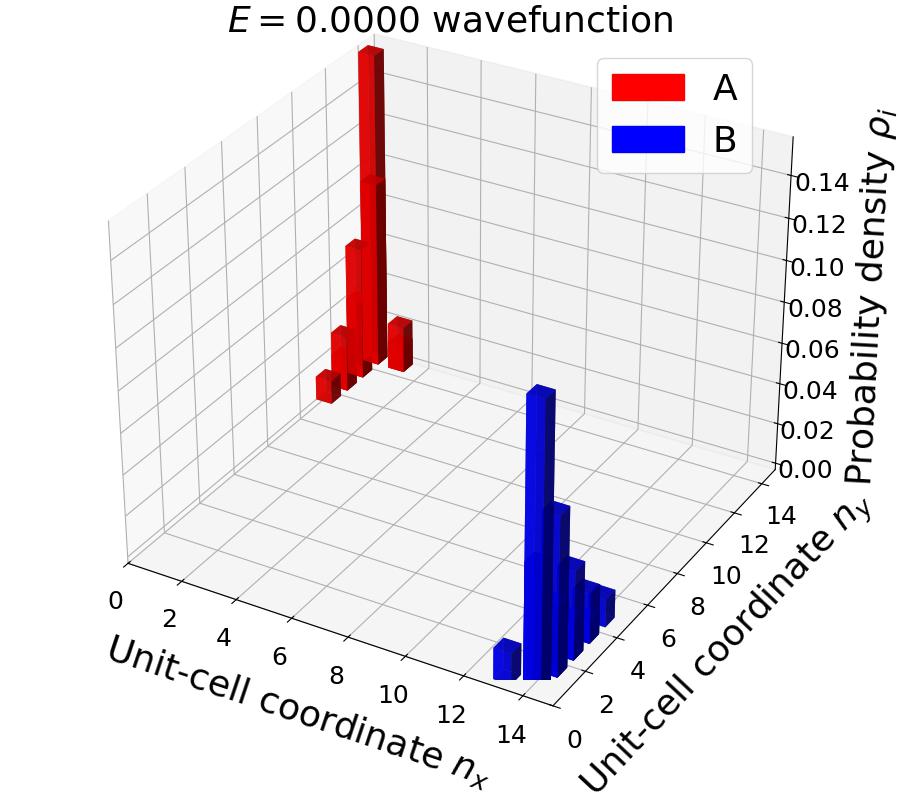}
\end{minipage}
\label{Kekule_delta_corner}
}
\end{minipage}
\par\vspace{6pt}
\subfloat[]{
\begin{minipage}{\linewidth}
\centering
\includegraphics[width=0.47\linewidth,valign=c]{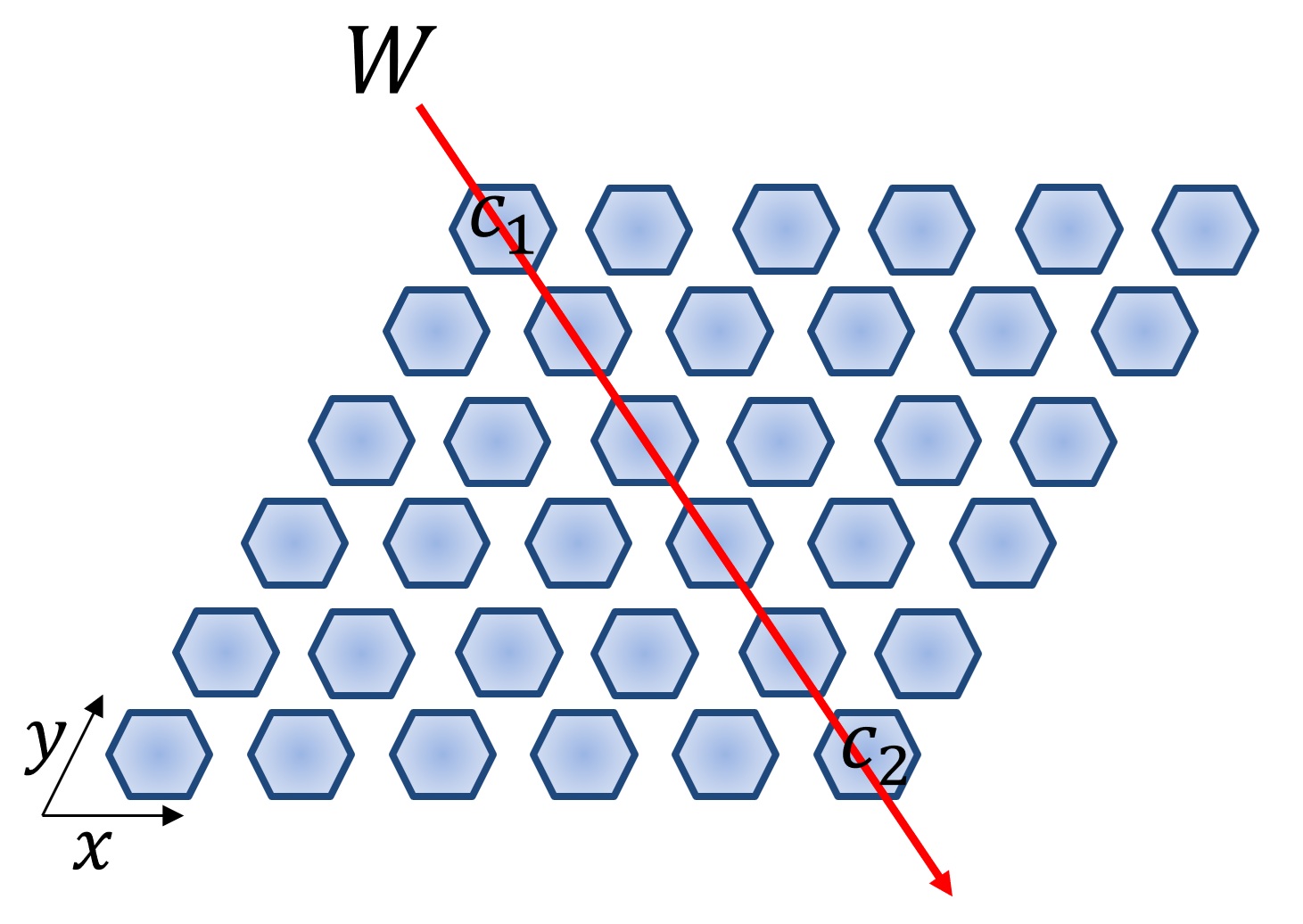}%
\hspace{0.1cm}%
\includegraphics[width=0.47\linewidth,valign=c]{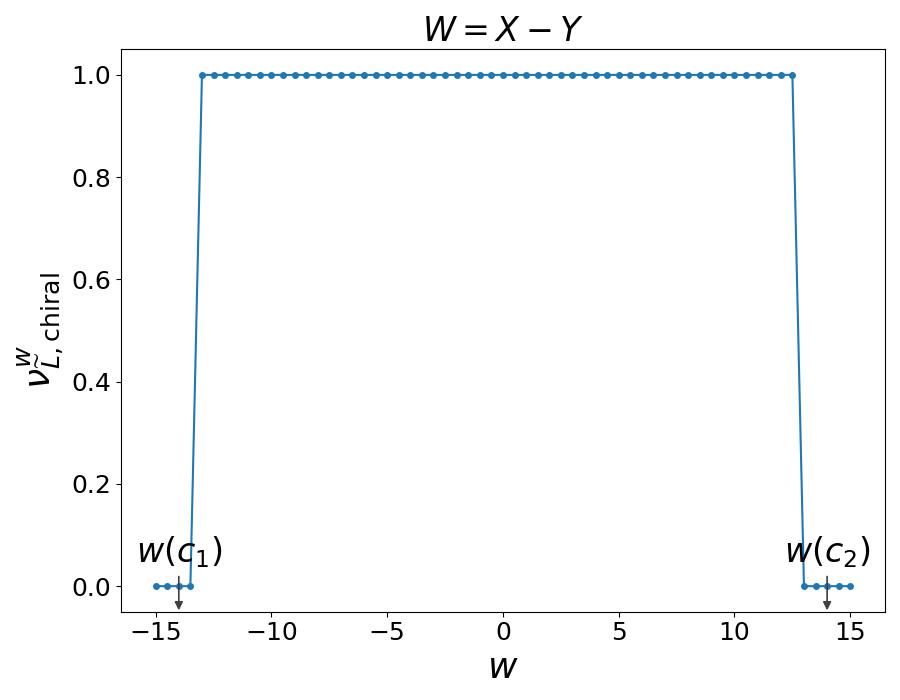}
\end{minipage}
\label{Kekule_delta_SL_schematic}
\label{Kekule_delta_SL}
}
\end{minipage}
\par
\caption{ (a) Schematic illustration of the conventional Kekulé-honeycomb model $H_\text{Kekulé}$ defined in Eq.~\eqref{Kekule} on a parallelogram. The model preserves both chiral symmetry and $C_2$ rotational symmetry. (b) Schematic illustration of the modified Kekulé model $H_\text{Kekulé,$\Delta$}$ defined in Eq.~\eqref{kekule_delta} on the same geometry. Choosing $\Delta_1\ne\Delta_2$ breaks $C_2$ rotational symmetry while preserving chiral symmetry. (c) Energy spectrum (left) and total probability density of the $E=0$ states (right) for a $15\times15$-unit-cell system with parameters $t_1=1$, $t_2=2$, $\Delta_1=0.3$, and $\Delta_2=0.5$. Even in the absence of $C_2$ symmetry, two robust $E=0$ corner states persist. (d) Schematic illustration of the projection onto $W=X-Y$ (left) and the corresponding spectral-localizer invariant $\nu^{w}_{\Tilde{L},\text{chiral}}$ defined in Eq.~\eqref{spectral_localizer invariant for HOT} (right). The two corners hosting the $E=0$ states, $c_1$ and $c_2$, are mapped to $w(c_1)=-14$ and $w(c_2)=14$, respectively. The invariant is evaluated using the same system size and model parameters as in (c), with $\kappa=0.05$. The unit increase in $\nu^{w}_{\Tilde{L},\text{chiral}}$ near $w(c_1)=-14$ and its unit decrease near $w(c_2)=14$ identify one topological $E=0$ corner state with chirality $S=+1$ at $c_1$ and $S=-1$ at $c_2$, respectively.}
\label{Kekule_all_figs}
\end{figure*}

In this section, we identify a class of 2D chiral-symmetric HOT phases that can be characterized by spectral-localizer invariants but not by MCNs. We demonstrate this using the Kekulé-honeycomb model on a parallelogram-shaped finite system. By introducing crystalline-symmetry-breaking but chiral-symmetry-preserving terms, we show that its $E=0$ corner states persist, indicating that these states can be protected solely by chiral symmetry. Despite this chiral protection, numerical calculations of the MCNs, together with an analytical evaluation in the topological dimer limit, yield vanishing MCNs. In contrast, the spectral-localizer invariant introduced in Sec.~\ref{2D Chiral-symmetric Higher-order topological phases and spectral localizer} successfully characterizes their chiral topological protection.

We consider a modified Kekulé-honeycomb model on a parallelogram-shaped finite system, as illustrated in Fig.~\ref{Fig_Kekule_delta}. Its Hamiltonian is
\bea\label{kekule_delta}
H_\text{Kekulé,$\Delta$} = H_\text{Kekulé} + H_{\Delta},
\eea
where the conventional Kekulé-honeycomb Hamiltonian is given by~\cite{noh_topological_2018}
\begin{widetext}
\bea\label{Kekule}
\begin{aligned}
H_\text{Kekulé}
&=\sum_{\mathbf R}\Bigl\{
t_1\bigl(c_{\mathbf R,1}^\dagger c_{\mathbf R,2}+c_{\mathbf R,2}^\dagger c_{\mathbf R,3}+c_{\mathbf R,3}^\dagger c_{\mathbf R,4}+c_{\mathbf R,4}^\dagger c_{\mathbf R,5}+c_{\mathbf R,5}^\dagger c_{\mathbf R,6}+c_{\mathbf R,6}^\dagger c_{\mathbf R,1}\bigr)\\
&\hphantom{=\sum_{\mathbf R}\Bigl\{}{}
+t_2\bigl(c_{\mathbf R,2}^\dagger c_{\mathbf R+\mathbf a_1,5}+c_{\mathbf R,3}^\dagger c_{\mathbf R+\mathbf a_2,6}+c_{\mathbf R,4}^\dagger c_{\mathbf R+\mathbf a_2-\mathbf a_1,1}\bigr)+\text{h.c.}\Bigr\}.
\end{aligned}
\eea
\end{widetext}
and the additional hopping terms are
\bea
H_{\Delta}=
\sum_{\mathbf R} \left(
\Delta_1 c_{\mathbf R,3}^\dagger c_{\mathbf R,4}
+\Delta_2 c_{\mathbf R,6}^\dagger c_{\mathbf R,1}
+ \text{h.c.}
\right).
\eea
Here, $\mathbf R$ labels the unit cells, and $c_{\mathbf R,\alpha}$ annihilates a fermion at site $\alpha=1,\ldots,6$ within unit cell $\mathbf R$. The primitive lattice vectors are $\mathbf a_1=(3,0)$ and $\mathbf a_2=(3/2,3\sqrt{3}/2)$. The hopping amplitudes $t_1$ and $t_2$ describe intracell and intercell hoppings, respectively, while $\Delta_1$ and $\Delta_2$ modify two intracell bonds.

The sites $1,3,5$ belong to sublattice $A$, and the sites $2,4,6$ belong to sublattice $B$. Since all hopping terms, including those in $H_{\Delta}$, connect opposite sublattices, the modified Hamiltonian preserves chiral symmetry with the chiral operator written as
\bea
\begin{aligned}
S&=\sum_{\mathbf R}\Bigl(
c_{\mathbf R,1}^\dagger c_{\mathbf R,1}-c_{\mathbf R,2}^\dagger c_{\mathbf R,2}+c_{\mathbf R,3}^\dagger c_{\mathbf R,3}\\
&\hphantom{=\sum_{\mathbf R}\Bigl(}{}
-c_{\mathbf R,4}^\dagger c_{\mathbf R,4}+c_{\mathbf R,5}^\dagger c_{\mathbf R,5}-c_{\mathbf R,6}^\dagger c_{\mathbf R,6}\Bigr).
\end{aligned}
\eea
Thus, sublattices $A$ and $B$ constitute the $S=+1$ and $S=-1$ chiral subspaces, respectively.

For $\Delta_1=\Delta_2=0$, the model reduces to the conventional Kekulé-honeycomb model shown in Fig.~\ref{Fig_Kekule}, which preserves $C_2$ rotational symmetry in addition to chiral symmetry. Taking the rotation center as the spatial origin, the corresponding operator is
\bea
\begin{aligned}
\hat C_2
&=\sum_{\mathbf R}\Bigl(
c^\dagger_{-\mathbf R,4}c_{\mathbf R,1}+c^\dagger_{-\mathbf R,5}c_{\mathbf R,2}+c^\dagger_{-\mathbf R,6}c_{\mathbf R,3}\\
&\hphantom{=\sum_{\mathbf R}\Bigl(}{}
+c^\dagger_{-\mathbf R,1}c_{\mathbf R,4}+c^\dagger_{-\mathbf R,2}c_{\mathbf R,5}+c^\dagger_{-\mathbf R,3}c_{\mathbf R,6}\Bigr).
\end{aligned}
\eea
In this limit, the system is in a trivial phase without corner states when $t_1>t_2$. Conversely, when $t_2>t_1$, it hosts two $E=0$ corner states localized at a pair of diagonally opposite corners. These corner states can be understood from two perspectives within the $C_2$-protected crystalline topology. In Ref.~\cite{noh_topological_2018}, the authors compute the $C_2$ rotational invariant $[M]$ for this model. They find that the system is trivial with $[M]=0$ when $t_1>t_2$, whereas it enters a topological phase with $[M]=2$ when $t_2>t_1$. This transition between different values of $[M]$ corresponds to the emergence of the $E=0$ corner states. Another perspective is provided by Ref.~\cite{benalcazar_quantization_2019}, which develops a general construction for fractional corner charges in systems with vanishing polarization, quantized in multiples of $e/n$ by $C_n$ rotational symmetry. Specifically, they note that when additional chiral symmetry is present, $e/2$ corner charges can manifest as $E=0$ corner states. Within the Kekulé-honeycomb model, the $C_2$ symmetry supports $e/2$ corner charges for $t_2>t_1$; together with chiral symmetry, this provides a complementary interpretation of the $E=0$ corner states.

However, the persistence of the $E=0$ corner states does not require $C_2$ rotational symmetry. Under the rotation, the two bonds modified by $\Delta_1$ and $\Delta_2$ are exchanged. Choosing $\Delta_1\ne\Delta_2$ therefore explicitly breaks $C_2$ symmetry while preserving chiral symmetry. For $t_1=1$, $t_2=2$, $\Delta_1=0.3$, and $\Delta_2=0.5$, our numerical results in Fig.~\ref{Kekule_delta_corner} show that two $E=0$ corner states remain localized at the diagonal corners. Their persistence demonstrates that $C_2$ symmetry is unnecessary for their zero-energy protection and motivates a characterization based on chiral symmetry alone.

The MCN, however, does not detect this phase. Numerically, we find $N^{xy}_{\text{chiral}}=0$ for both $H_\text{Kekulé}$ and $H_\text{Kekulé,$\Delta$}$ throughout the parameter regimes examined, including those supporting the two $E=0$ corner states. For the unmodified model, this numerical result is also supported by an analytical calculation in the topological dimer limit $t_1=0$, presented in Appendix~\ref{App_Kekule_MCN}. In this limit, the Hamiltonian contains only intercell hoppings, and the MCN can be evaluated exactly using the translation operators associated with the three intercell hopping channels. Their contributions cancel, yielding $N^{xy}_{\text{chiral}}=0$, even though the open-boundary system still has $E=0$ corner states.

Although the MCN fails to characterize these corner states, we find that their chiral topology can still be captured by the spectral-localizer framework reviewed in Sec.~\ref{2D Chiral-symmetric Higher-order topological phases and spectral localizer}. To demonstrate this, we evaluate the invariant $\nu^{w}_{\Tilde{L},\text{chiral}}$ defined in Eq.~\eqref{spectral_localizer invariant for HOT} for $H_\text{Kekulé,$\Delta$}$ and examine its changes across the projected positions of the two corners hosting the $E=0$ states.

We choose the projected position operator $W=X-Y$, with the corresponding coordinate $w=x-y$, to evaluate $\nu^{w}_{\Tilde{L},\text{chiral}}$. As illustrated in Fig.~\ref{Kekule_delta_SL_schematic}, this projection maps the two corners hosting the $E=0$ states, denoted by $c_1$ and $c_2$, onto distinct values of $w$, allowing their topological contributions to be resolved separately. We use a $15\times15$-unit-cell system with $t_1=1$, $t_2=2$, $\Delta_1=0.3$, and $\Delta_2=0.5$, the same parameters as those used for the corner-state results in Fig.~\ref{Kekule_delta_corner}, and set $\kappa=0.05$. The resulting invariant is shown as a function of $w$ in Fig.~\ref{Kekule_delta_SL}. The invariant exhibits a unit increase near $w(c_1)=-14$ and a unit decrease near $w(c_2)=14$. Following the spectral-localizer counting relation in Eq.~\eqref{1D localizer invariant jump}, these changes identify one topological $E=0$ state with chirality $S=+1$ at $c_1$ and one with chirality $S=-1$ at $c_2$, consistent with the numerical results in Fig.~\ref{Kekule_delta_corner}. Thus, the spectral-localizer invariant characterizes the chiral topological protection of these corner states despite the vanishing MCN and the absence of $C_2$ rotational symmetry.

The modified Kekulé-honeycomb model therefore illustrates a class of chiral-symmetric HOT phases characterized by spectral-localizer invariants despite vanishing MCNs. In Ref.~\cite{cerjan_local_2022}, the spectral-localizer invariant for chiral-symmetric HOT phases was conjectured to be a local version of the corresponding MCN. Here, we show that this local invariant can capture topological $E=0$ corner states even when the MCN vanishes. This demonstrates that the spectral-localizer framework can identify higher-order topological information beyond that encoded by the MCN.

\section{2D half-chiral HOT phases with spectral-localizer characterization but vanishing MCNs}\label{2D half-chiral HOT phases without MCNs}

\begin{figure*}[t]
\centering
\edef\HalfKekulerowheight{\the\dimexpr0.25\textwidth\relax}

\begin{minipage}[t]{0.35\textwidth}
\vspace{0pt}
\centering
\begin{minipage}[t][\HalfKekulerowheight][t]{\linewidth}
\vspace{0pt}
\centering
\subfloat[]{
\includegraphics[width=0.95\linewidth]{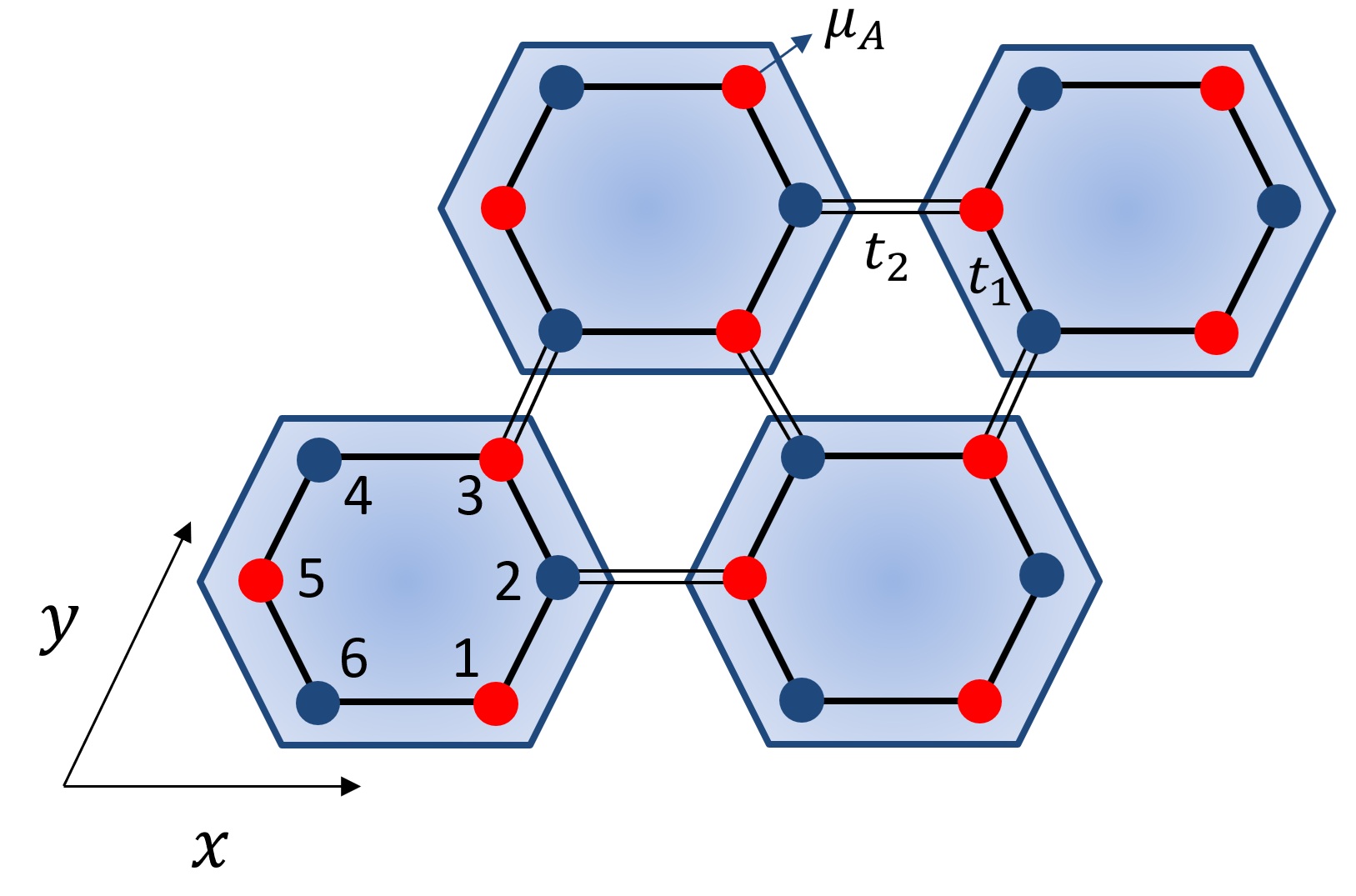}
\label{Fig_minimal_exKekule}
}
\end{minipage}
\par\vspace{6pt}
\subfloat[]{
\includegraphics[width=0.95\linewidth]{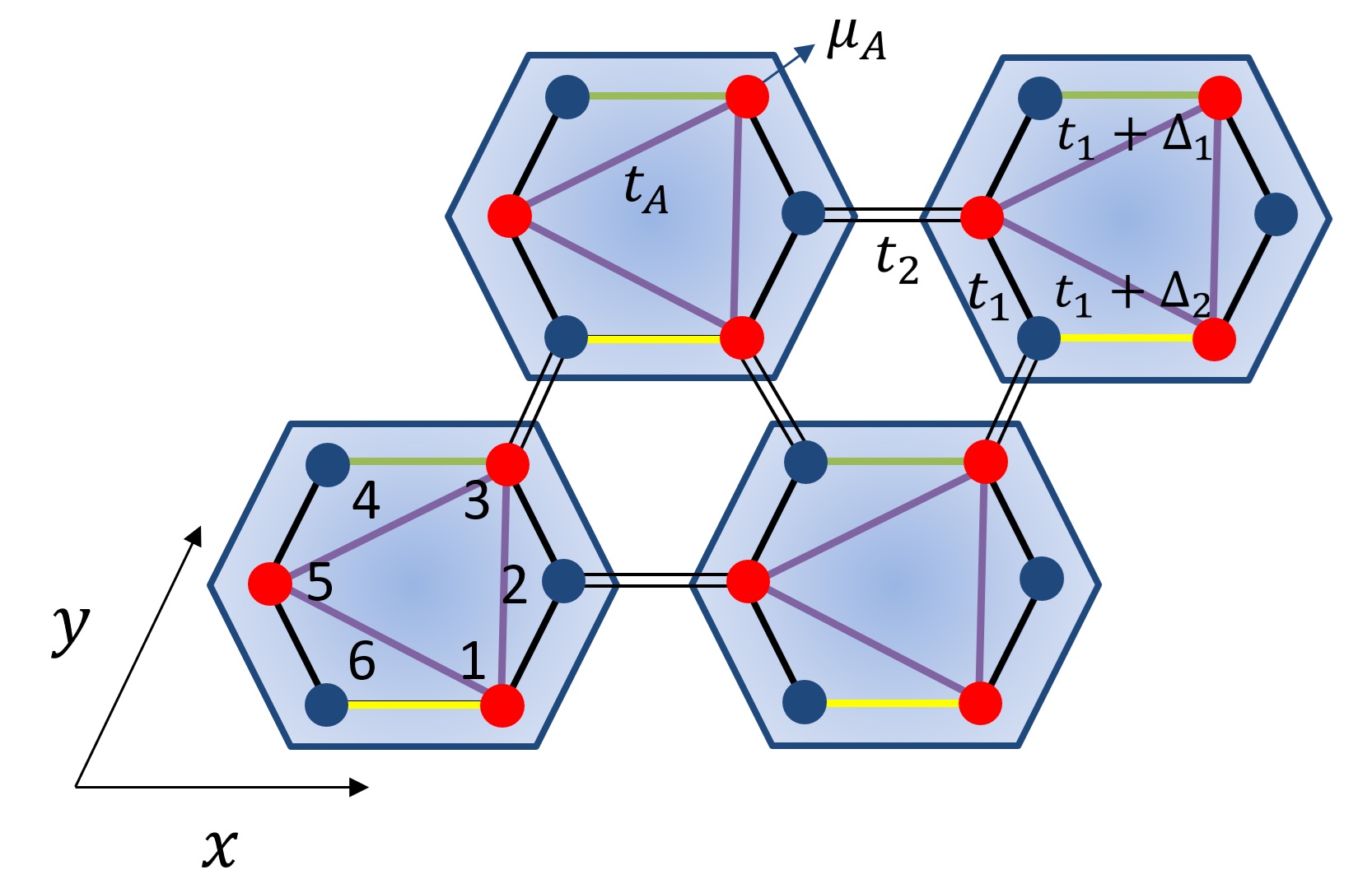}
\label{Fig_exKekule}
}
\end{minipage}%
\hspace{0.01\textwidth}%
\begin{minipage}[t]{0.60\textwidth}
\vspace{0pt}
\centering
\begin{minipage}[t][\HalfKekulerowheight][t]{\linewidth}
\vspace{0pt}
\centering
\subfloat[]{
\begin{minipage}{\linewidth}
\centering
\includegraphics[width=0.47\linewidth,valign=c]{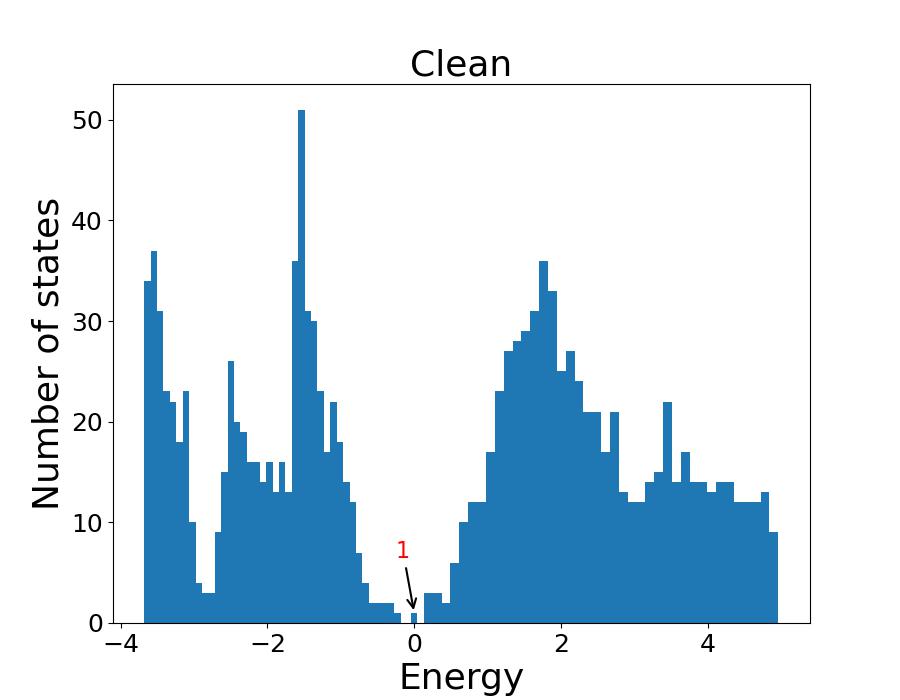}%
\hspace{0.1cm}%
\includegraphics[width=0.47\linewidth,valign=c]{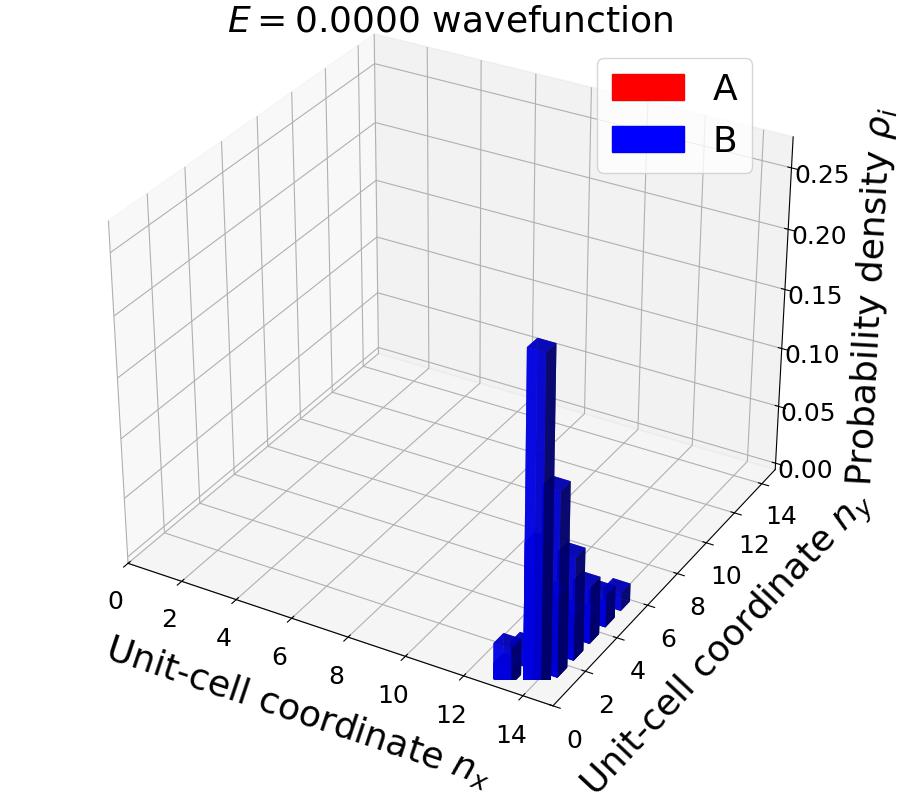}
\end{minipage}
\label{exKekule_corner}
}
\end{minipage}
\par\vspace{6pt}
\subfloat[]{
\begin{minipage}{\linewidth}
\centering
\includegraphics[width=0.47\linewidth,valign=c]{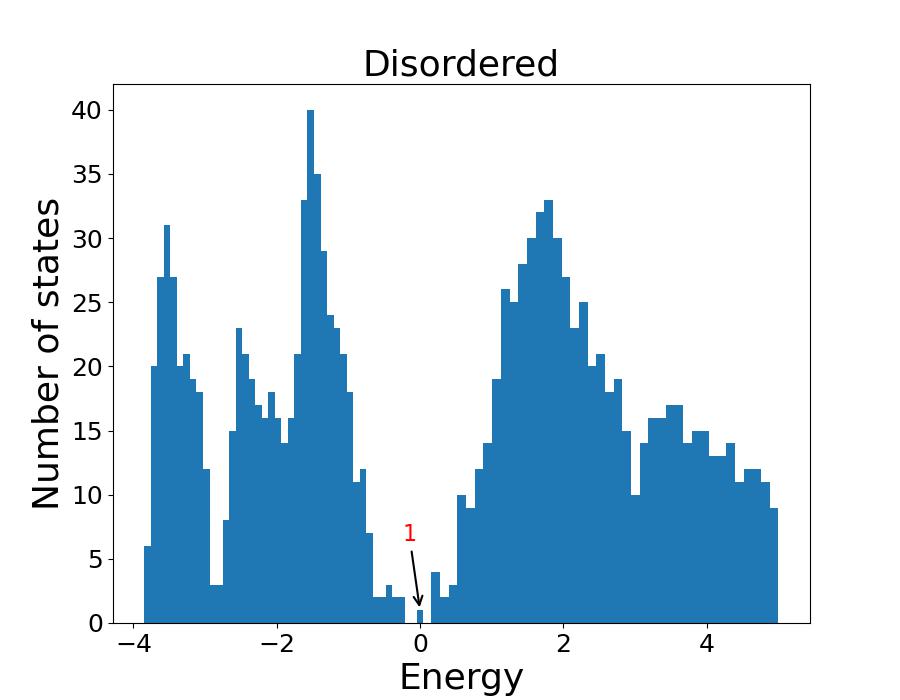}%
\hspace{0.1cm}%
\includegraphics[width=0.47\linewidth,valign=c]{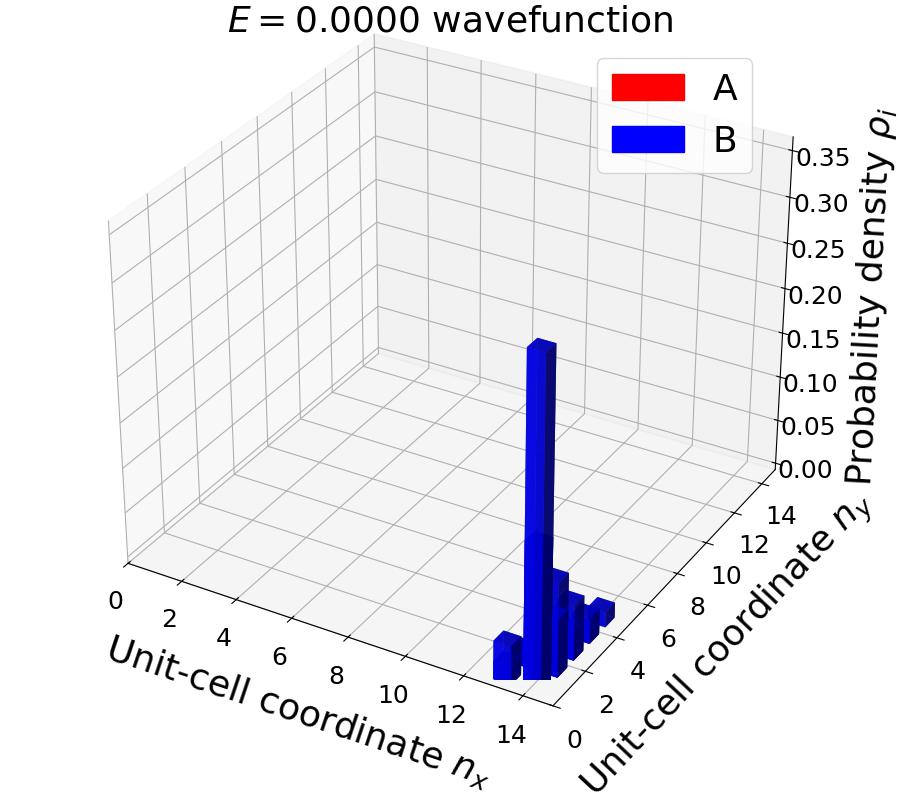}
\end{minipage}
\label{disordered_exKekule_corner}
}
\end{minipage}
\par
\caption{(a) Minimal realization of the half-chiral Kekulé model using only an on-site potential $\mu_A$ on sublattice $A$ ($t_A=0$). (b) Half-chiral Kekulé model defined in Eq.~\eqref{exKekule} on a parallelogram, with both the on-site potential $\mu_A$ and the intra-unit-cell hoppings $t_A$ on sublattice $A$. (c) Energy spectrum (left) and total probability density of the $E=0$ state (right) for a clean $15\times15$-unit-cell system with $t_1=1$, $t_2=2$, $\Delta_1=0.3$, $\Delta_2=0.5$, $t_A=0.5$, and $\mu_A=0.3$. The system hosts a single topological $E=0$ corner state. (d) Same as (c), but with the disorder defined in Eq.~\eqref{disordered exKekule}, with each perturbation independently sampled from $[-0.2,0.2]$. The single topological $E=0$ corner state persists.}
\end{figure*}

\begin{figure*}[t]
\centering
\subfloat[]{\includegraphics[width=0.35\textwidth,valign=c]{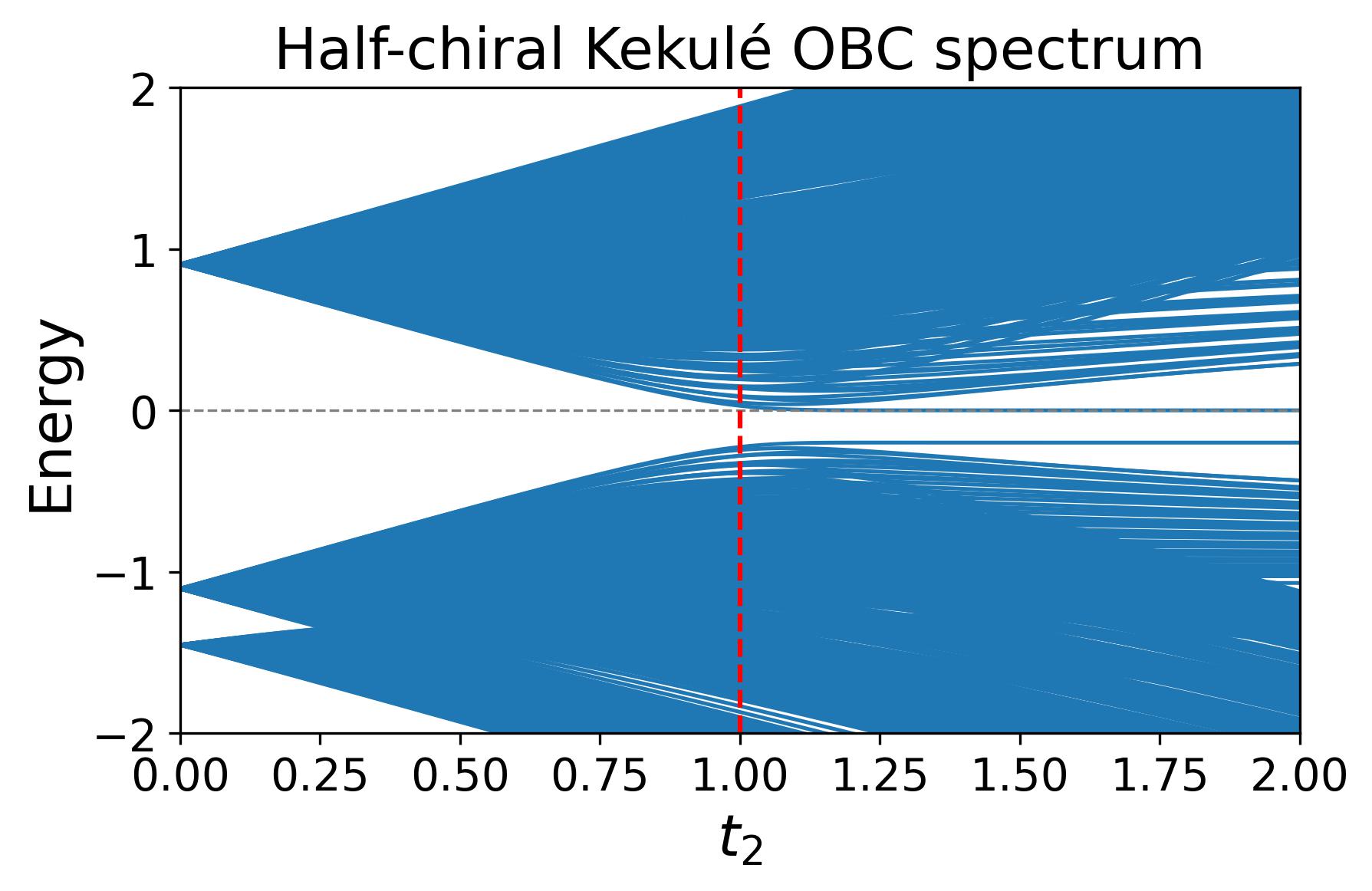}\label{exKekule_OBC_spectrum}}\hskip 0.5cm
\subfloat[]{\includegraphics[width=0.35\textwidth,valign=c]{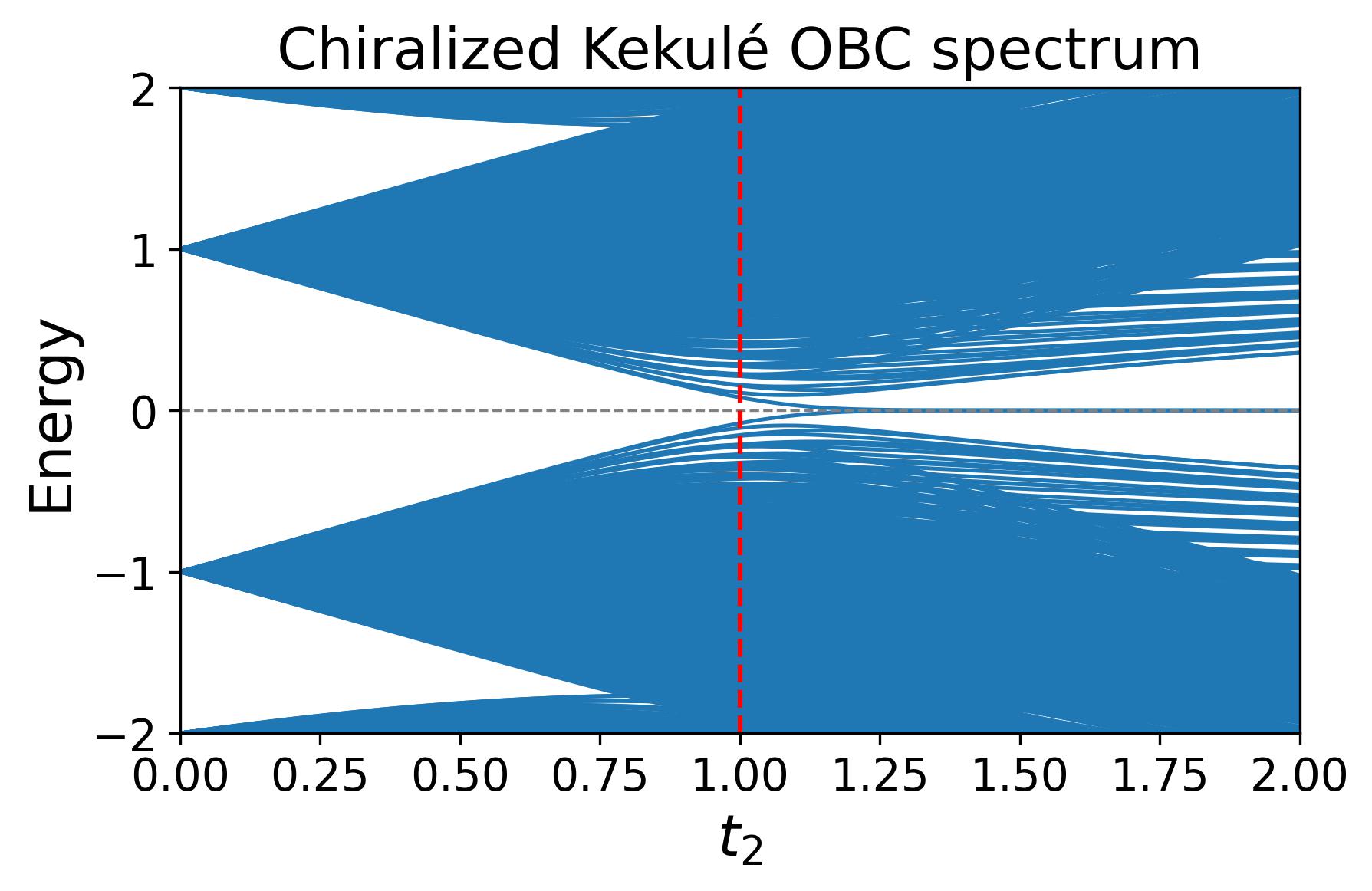}\label{Kekule_chiralized_OBC_spectrum}}\\
\subfloat[]{\includegraphics[width=0.35\textwidth,valign=c]{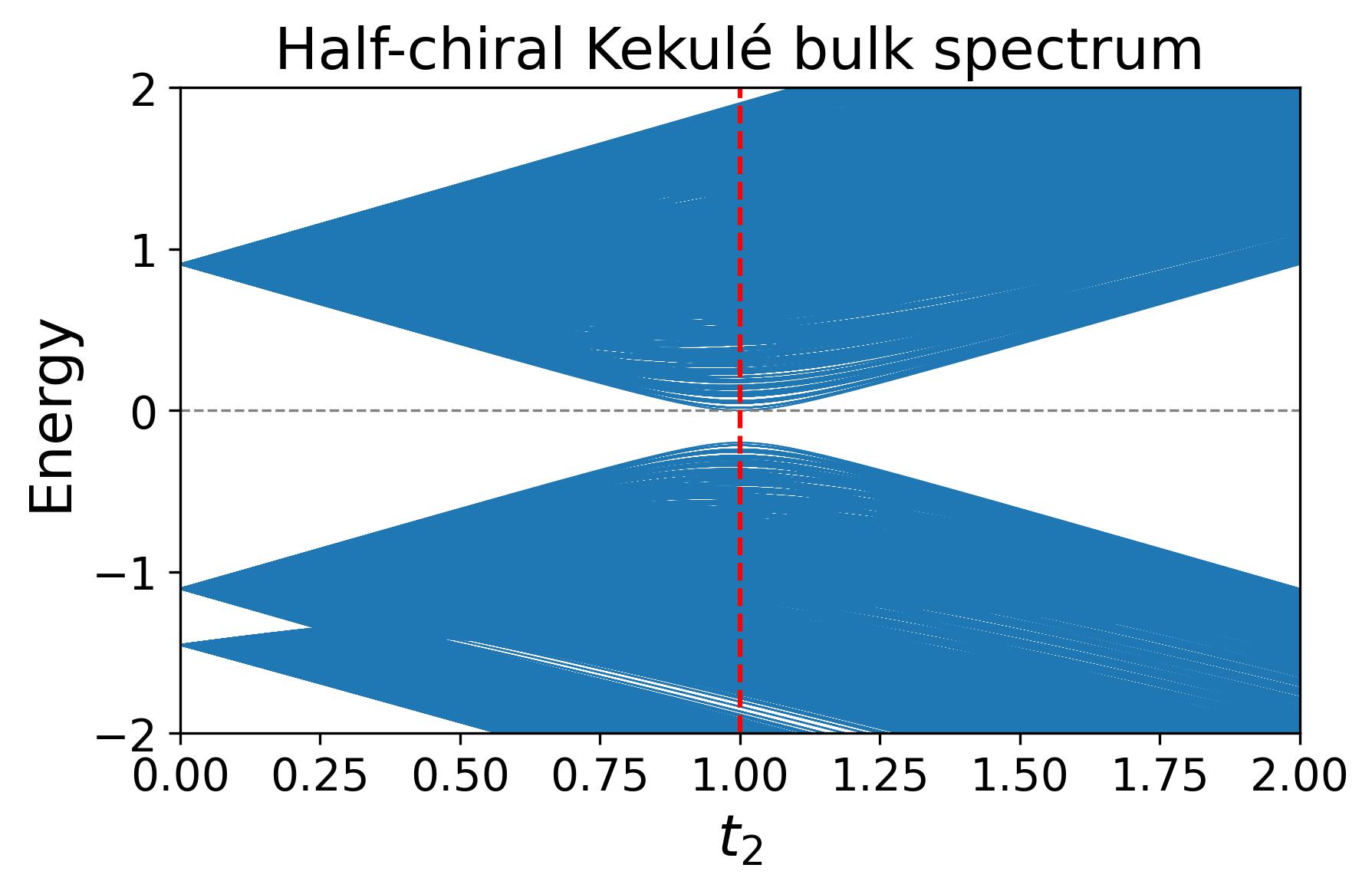}\label{exKekule_bulk_spectrum}}\hskip 0.5cm
\subfloat[]{\includegraphics[width=0.35\textwidth,valign=c]{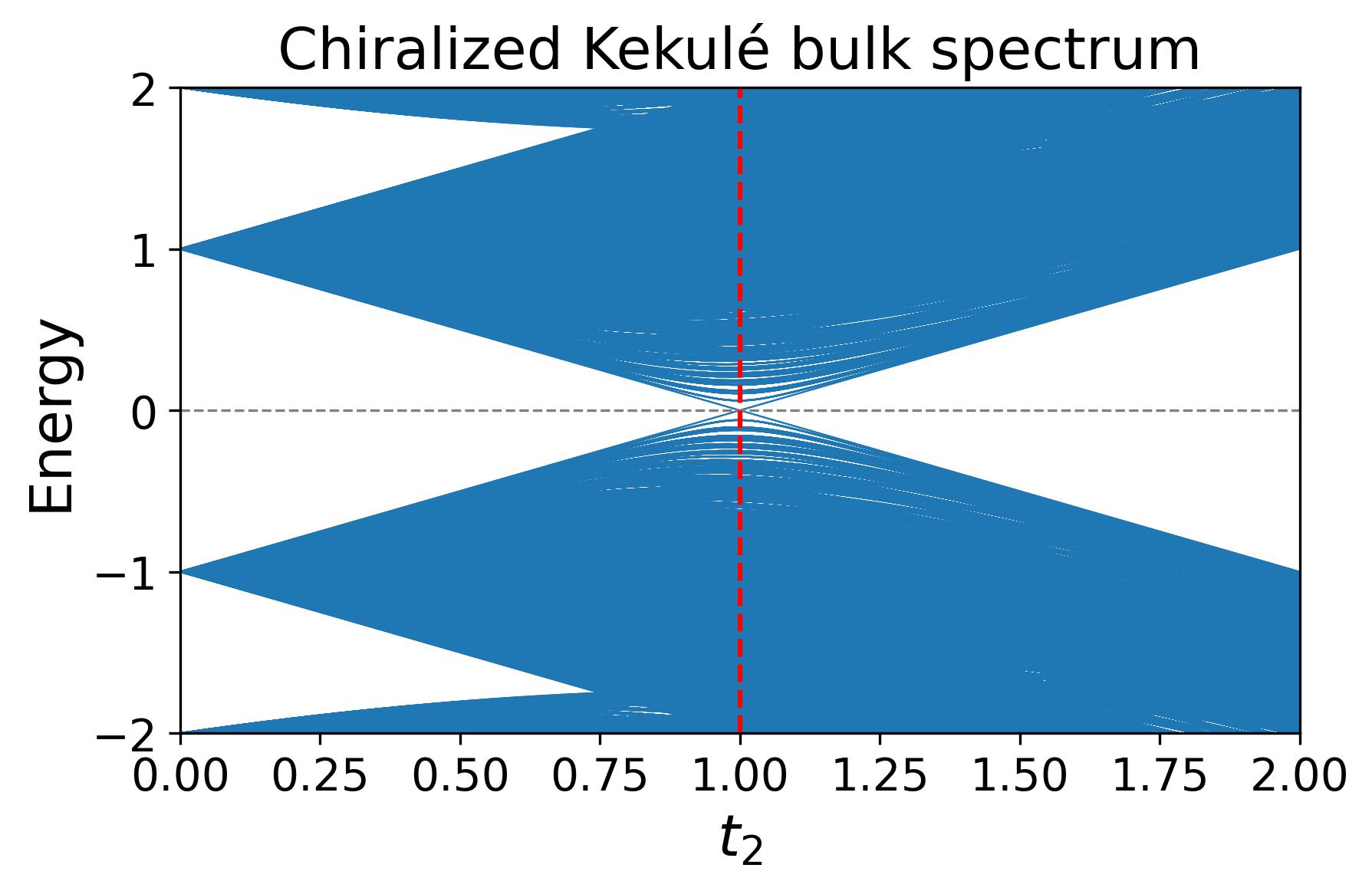}\label{Kekule_chiralized_bulk_spectrum}}
\caption{OBC and bulk spectra of the half-chiral Kekulé model and its chiralized Hamiltonian as functions of $t_2$, with $t_1=1$, $\Delta_1=\Delta_2=0$, $t_A=0.5$, and $\mu_A=0.3$. The red dashed lines indicate the thermodynamic transition at $t_2=1$ from the trivial phase to the half-chiral HOT phase. (a) OBC spectrum of the physical Hamiltonian for a $15\times15$-unit-cell system, showing the emergence of one topological $E=0$ corner state around the transition. (b) Corresponding OBC spectrum of the chiralized Hamiltonian, showing the emergence of two topological $E=0$ corner states around the transition. (c) Momentum-space bulk spectrum of the physical Hamiltonian, which touches $E=0$ at the transition while its bulk gap remains open. (d) Corresponding bulk spectrum of the chiralized Hamiltonian, whose bulk gap closes at the transition.}
\label{exKekule_spectral_evolution}
\end{figure*}

The framework developed in Sec.~\ref{bulk-boundary correspondecn for Hermitian half-chiral} applies generally to topological states protected by chiral symmetry and does not depend on which invariant characterizes the underlying chiral topology. It therefore allows us to construct half-chiral counterparts of the 2D chiral-symmetric HOT phases discussed in Sec.~\ref{2D Chiral-symmetric Higher-order topological phases without MCNs}, whose topological $E=0$ corner states are characterized by spectral-localizer invariants despite vanishing MCNs. In particular, this construction allows us to realize a half-chiral HOT phase hosting only a single topological $E=0$ corner state, as demonstrated below.

To characterize these half-chiral HOT phases and their topological $E=0$ corner states, we construct the corresponding half-chiral invariant using Eq.~\eqref{general half-chiral invariant}. Specifically, we evaluate the 2D chiral spectral-localizer invariant defined in Eq.~\eqref{spectral_localizer invariant for HOT} in terms of $H_{\text{chiralized}}$, giving
\bea
\begin{aligned}
\nu^{w}_{\Tilde{L},\text{half-chiral}}(H)
&=\nu^{w}_{\Tilde{L},\text{chiral}}(H_{\text{chiralized}})\\
&=\frac{1}{2}\text{sig}\Big[\Tilde{L}_{\text{half-chiral}}(w,0,H)\Big],
\end{aligned}
\eea
where
\bea
\begin{aligned}
&\Tilde{L}_{\text{half-chiral}}(w,0,H)\\
&=\Tilde{L}_{\text{chiral}}(w,0,H_{\text{chiralized}})\\
&=\Big[iH_{\text{chiralized}}-\kappa(W-w\mathbbm{1})\Big]S\\
&=\frac{i}{2}(HS-SH)-\kappa(W-w\mathbbm{1})S.
\end{aligned}
\eea
When the projection $W$ resolves the corners of interest separately, the changes in this invariant can be interpreted using the criterion in Eq.~\eqref{half-chiral localizer selection rule}, with $(X,x)$ replaced by $(W,w)$. For the half-chiral symmetry $SHS^{-1}P_+=-HP_+$ ($SHS^{-1}P_-=-HP_-$), only increases (decreases) as $w$ increases correspond to physical topological $E=0$ corner states. The magnitude of each selected change gives the number of these states at the corresponding corner. According to Eq.~\eqref{BBC_half-chiral topological}, the total number of topological $E=0$ corner states of $H$ is half that of $H_{\text{chiralized}}$.

As a concrete example, we construct a half-chiral Kekulé model from the modified chiral-symmetric Kekulé model studied above. The system is defined on a parallelogram, as illustrated in Fig.~\ref{Fig_exKekule}, with Hamiltonian
\bea\label{exKekule}
H_{\text{Kekulé,$\Delta$}}^{\text{half-chiral}}=H_{\text{Kekulé,$\Delta$}}+H_{\text{cb}},
\eea
where $H_{\text{Kekulé,$\Delta$}}$ is the modified Kekulé model defined in Eq.~\eqref{kekule_delta}, which describes the Kekulé-honeycomb model with an additional $C_2$-breaking term, and the chiral-breaking term $H_{\text{cb}}$ is given by
\newpage
\begin{widetext}
\bea
\begin{aligned}
H_{\text{cb}}
&=\sum_{\mathbf R}\Big\{
t_A\Big(c_{\mathbf R,1}^\dagger c_{\mathbf R,3}
+c_{\mathbf R,3}^\dagger c_{\mathbf R,5}
+c_{\mathbf R,5}^\dagger c_{\mathbf R,1}\Big)+\text{h.c.}\\
&\hphantom{=\sum_{\mathbf R}\Big\{}{}
+\mu_A\Big(c_{\mathbf R,1}^\dagger c_{\mathbf R,1}
+c_{\mathbf R,3}^\dagger c_{\mathbf R,3}
+c_{\mathbf R,5}^\dagger c_{\mathbf R,5}\Big)\Big\}.
\end{aligned}
\eea
\end{widetext}
where $t_A$ denotes the triangle hoppings among sublattice A sites within a unit cell, and $\mu_A$ represents the on-site potential specific to these sites. Either term alone breaks full chiral symmetry while preserving half-chiral symmetry. We include both terms in the model to illustrate the generality of our construction. A minimal realization of the half-chiral HOT phase retains only the on-site potential $\mu_A$ on sublattice $A$, as shown in Fig.~\ref{Fig_minimal_exKekule}. This model respects the half-chiral symmetry $S H S^{-1} P_{-} = -H P_{-}$ with
\bea
\begin{aligned}
&S=\sum_{\mathbf R}\Bigl(
c_{\mathbf R,1}^\dagger c_{\mathbf R,1}-c_{\mathbf R,2}^\dagger c_{\mathbf R,2}+c_{\mathbf R,3}^\dagger c_{\mathbf R,3}\\
&\hphantom{S=\sum_{\mathbf R}\Bigl(}{}
-c_{\mathbf R,4}^\dagger c_{\mathbf R,4}+c_{\mathbf R,5}^\dagger c_{\mathbf R,5}-c_{\mathbf R,6}^\dagger c_{\mathbf R,6}\Bigr),\\
&P_-=\frac{1}{2}(1-S).
\end{aligned}
\eea
The corresponding chiralized Hamiltonian is
\bea
H_{\text{chiralized}}[H_{\text{Kekulé,$\Delta$}}^{\text{half-chiral}}]=H_{\text{Kekulé,$\Delta$}}.
\eea
Using the half-chiral spectral-localizer invariant introduced earlier in this section, we obtain
\bea
\begin{aligned}
&\nu^{w}_{\Tilde{L},\text{half-chiral}}
(H_{\text{Kekulé,$\Delta$}}^{\text{half-chiral}})\\
&=\nu^{w}_{\Tilde{L},\text{chiral}}
(H_{\text{Kekulé,$\Delta$}}).
\end{aligned}
\eea
The invariant on the right-hand side has already been evaluated in Fig.~\ref{Kekule_delta_SL}. As $w$ increases, it rises from $0$ to $1$ near $w(c_1)=-14$ and decreases from $1$ to $0$ near $w(c_2)=14$. Since the present model satisfies the half-chiral symmetry $SHS^{-1}P_-=-HP_-$, only the decrease near $c_2$ corresponds to physical topological $E=0$ corner states, following Eq.~\eqref{half-chiral localizer selection rule}. This unit decrease predicts a single topological $E=0$ corner state at $c_2$, consistent with the numerical results in Fig.~\ref{exKekule_corner}.

To examine the spectral behavior of half-chiral HOT phases characterized by spectral-localizer invariants but not by MCNs, we consider the half-chiral Kekulé model with $t_1=1$, $\Delta_1=\Delta_2=0$, $t_A=0.5$, and $\mu_A=0.3$ and vary $t_2$. For these parameters, the system has two phases: a trivial phase for $t_2<t_1$ and a half-chiral HOT phase hosting a single topological $E=0$ corner state for $t_2>t_1$. The OBC spectra for a $15\times15$-unit-cell system are shown in Figs.~\ref{exKekule_OBC_spectrum} and~\ref{Kekule_chiralized_OBC_spectrum}. For the physical Hamiltonian, the bulk spectrum approaches $E=0$ around the transition while the bulk gap remains open, and a single topological $E=0$ corner state emerges. For the chiralized Hamiltonian, the bulk gap closes around the transition, accompanied by the emergence of two topological $E=0$ corner states. We further confirm this spectral behavior independently of finite-size effects through the momentum-space bulk spectra in Figs.~\ref{exKekule_bulk_spectrum} and~\ref{Kekule_chiralized_bulk_spectrum}.

To test the robustness of these 2D half-chiral HOT phases, we add small, site-dependent perturbations to the half-chiral Kekulé model. To describe these perturbations, we collect the coupling parameters into the vector
\bea
\boldsymbol{\lambda}=
(t_1,t_2,\Delta_1,\Delta_2,t_A,\mu_A),
\eea
and introduce the corresponding site-dependent disorder
\bea
\delta\boldsymbol{\lambda}_n=
(\delta t_{1,n},\delta t_{2,n},
\delta\Delta_{1,n},\delta\Delta_{2,n},
\delta t_{A,n},\delta\mu_{A,n}),
\eea
where the components of $\delta\boldsymbol{\lambda}_n$ are small, independent perturbations to the corresponding couplings at site $n$. The disordered Hamiltonian is then written as
\bea\label{disordered exKekule}
\Tilde{H}_\text{Kekulé,$\Delta$}^{\text{half-chiral}}\equiv H_\text{Kekulé,$\Delta$}^{\text{half-chiral}}\left(\boldsymbol{\lambda}+\delta\boldsymbol{\lambda}_n\right).
\eea
As shown in Fig.~\ref{disordered_exKekule_corner}, a single topological $E=0$ corner state persists under these perturbations when the system is in the half-chiral HOT phase.

It's worth noting that single $E=0$ corner states have also been reported in the breathing kagome lattice~\cite{xue_acoustic_2019}, where their topological interpretation was based on crystalline-symmetry-quantized Wannier centers, while generalized chiral symmetry was proposed to protect the $E=0$ corner modes in related studies~\cite{ni_observation_2019}. However, subsequent work showed that perturbations preserving the threefold rotational, mirror, and generalized chiral symmetries can still shift the corner modes away from zero energy~\cite{vanmiert_topological_2020}, demonstrating that these symmetries alone do not guarantee $E=0$ corner states. Their robustness has instead been discussed under additional hopping constraints. Ref.~\cite{herrera_corner_2022} showed that additional hoppings from the sublattice hosting the corner state can preserve its zero energy when they connect to the other sublattices along specific lattice directions. Ref.~\cite{wang_subsymmetryprotected_2023} studied a different restriction: in addition to preserving the corresponding subsymmetry, certain long-range hoppings from that sublattice must have equal amplitudes. Although these studies explain how such restrictions can keep corner states at zero energy, neither develops a topological theory that yields an invariant guaranteeing $E=0$ corner states under these restrictions.
These restrictions cannot be understood as arising by any symmetry, cannot be expressed in a generic manner, and do not offer any topological invariants.

In contrast, in the half-chiral Kekulé model of Eq.~\eqref{exKekule}, the topological $E=0$ corner state is characterized by the half-chiral spectral-localizer invariant $\nu^{w}_{\Tilde{L},\text{half-chiral}}$. Its protection requires neither crystalline symmetries nor additional equalities between hopping amplitudes. It persists as long as the half-chiral condition $SHS^{-1}P_-=-HP_-$ is preserved and the invariant retains a decrease as $w$ passes through the projected corner position.

\section{Outlook}\label{Conclusion and discussion}

In this work we introduced the concept of half-chiral symmetry and investigated the resulting topological phenomena. Unlike conventional symmetries, this symmetry cannot be described as a unitary operator within the framework of traditional group theory. Instead, its action is constructed using a chiral operator together with projection operators. This symmetry supports new topological and HOT phases, which we call half-chiral topological and half-chiral HOT phases, respectively. In Sec.~\ref{Hermitian Subspace-Chiral Symmetry and Topological Phases}, we developed a general framework for constructing half-chiral topological invariants and established a bulk-boundary correspondence linking these invariants to the topological $E=0$ boundary states of half-chiral systems. Our framework further shows that every conventional chiral topological phase admits a half-chiral counterpart, providing a systematic route to constructing these phases from conventional chiral-symmetric systems.

Using this framework, we constructed 1D half-chiral topological phases from conventional 1D chiral-symmetric SPT phases in Sec.~\ref{1D half-chiral topological phases}. We defined a momentum-space half-chiral winding number and introduced a new real-space half-chiral spectral-localizer invariant to characterize their topological $E=0$ edge states. The real-space formulation allows this characterization without requiring translational symmetry, and the spatial dependence of the half-chiral spectral-localizer invariant further identifies the localization positions of the edge states.

Extending the framework to 2D, we constructed half-chiral HOT phases from conventional chiral-symmetric HOT phases characterized by MCNs and defined the corresponding half-chiral MCN to characterize their topological $E=0$ corner states in Sec.~\ref{2D half-chiral HOT phases with MCNs}. These phases can also be obtained by generalizing the real-space construction of 1D half-chiral topological phases to 2D.

Moreover, we identified 2D chiral-symmetric HOT phases whose topological $E=0$ corner states are characterized by spectral-localizer invariants despite vanishing MCNs, as demonstrated in Sec.~\ref{2D Chiral-symmetric Higher-order topological phases without MCNs}. These phases show that the spectral-localizer framework can capture higher-order topology beyond MCN characterization. Using our general framework, we constructed their half-chiral counterparts and the corresponding half-chiral spectral-localizer invariant in Sec.~\ref{2D half-chiral HOT phases without MCNs}. In particular, the resulting half-chiral Kekulé model hosts a single topological $E=0$ corner state.

The general ideas proposed here could be observed across various experimental platforms, as we discussed in the introduction.
Specializing to the specific 2D models we propose, since standard quadrupole topological insulators have been implemented in topoelectric circuits~\cite{imhof_topolectricalcircuit_2018} and microwave resonator lattices~\cite{peterson_quantized_2018}, and the Kekulé model has been realized in photonic lattices~\cite{noh_topological_2018}, extensions to the half-chiral HOT examples proposed here should be quite doable.
Interestingly, the half-chiral Kekulé-honeycomb HOT phase proposed in Sec.~\ref{2D half-chiral HOT phases without MCNs} could also be realized in solid-state systems or 2D materials, specifically in honeycomb-based systems with a Kekulé pattern and an $A$--$B$ sublattice imbalance, such as a chemical potential difference between the two sublattices. 
Heterostructures of graphene and hexagonal boron nitride (hBN) or similar honeycomb materials, with kekule bond-density-wave patterns, would be a natural candidate to realize this novel state with its unusual prediction: a single $E=0$ state localized at a single corner.

Let us outline potential theoretical extensions of this work. 
One direction is to investigate half-chiral topology in gapless systems, where topological $E=0$ boundary states may coexist with gapless bulk states. Half-chiral symmetry may protect these $E=0$ boundary states without requiring the nonzero-energy gapless bulk states to occur in $E$ and $-E$ pairs.
Another direction is to explore similar constructions for other unitary symmetries. Projectors onto selected eigenspaces of the symmetry operator could be used to preserve the symmetry relation within those eigenspaces. One could then construct an associated ``symmetrized Hamiltonian'' and investigate whether the system satisfying this projected symmetry relation inherits some of its topological properties. This approach does not extend directly to antiunitary symmetries such as time-reversal or particle-hole symmetries, because their eigenspaces do not generally form complex vector subspaces onto which such projectors can be defined.
Finally, we note that half-chiral symmetry lacks a corresponding inversion operator, so they may represent a type of non-invertible symmetry (see, e.g., Refs.~\cite{shao_whats_2023,schafer-nameki_ictp_2024}), depending on whether an underlying field-theoretic structure can be found.

\section{Acknowledgments}
We thank Alexander Cerjan for helpful discussions.
This work was supported by the U.S. Department of Energy, Office of Science, Basic Energy Sciences, under Early Career Award Number DE-SC0025478.

\appendix

\section*{APPENDIX}

\section{Real-space invariants for 1D chiral-symmetric topological phases and their finite-size effects}\label{Real-space invariants for 1D chiral symmetric topological phases}
In this section, we review two different real-space formulations for defining the winding number in 1D. We then compute the corresponding topological invariants, as well as the spectral-localizer invariant in Eq.~\eqref{local winding number}, for finite-size systems. Our results show that the spectral localizer exhibits the smallest finite-size effects among the three approaches.

The first method is based on projectors onto the occupied states, where the winding number is defined directly in real space as an analogue of the momentum-space invariant~\cite{mondragon-shem_topological_2014}. This formulation characterizes the bulk topology through the flattened Hamiltonian and its commutator with the position operator. The real-space winding number with this construction is given by
\bea
\nu_{\mathcal T} = -\,\mathcal T \!\left\{ Q_{-+}[X,Q_{+-}] \right\},
\eea
where $X$ is the position operator, $X\ket{x} = x\ket{x}$, and
\bea
\mathcal T\{A\} = \lim_{L\to\infty} \frac{1}{2L+1} \sum_{x=-L}^{L} \bra{x}A\ket{x}
\eea
denotes the trace per unit length of the operator $A$. Here $Q_{+-}$ and $Q_{-+}$ are the off-diagonal components of the flattened Hamiltonian $Q$ with respect to the chiral decomposition. More specifically, the flattened Hamiltonian $Q$ is constructed from the spectral projections onto the positive and negative energy sectors,
\bea
\begin{gathered}
P_+ = \sum_{E_n>0} \ket{\psi_n}\bra{\psi_n} , \\ P_- = \sum_{E_n<0} \ket{\psi_n}\bra{\psi_n} ,
\end{gathered}
\eea
as $Q = P_+ - P_- = \text{sgn}(H)$, and the off-diagonal components are defined explicitly by $Q_{+-}=S_+ Q S_-$ and $Q_{-+}=S_- Q S_+$, where $S_\pm = \frac{1}{2}(1\pm S)$ are the chiral projectors.

The second method is based on the singular value decomposition (SVD) approach, which provides a real-space representation of the winding number~\cite{lin_realspace_2021}. The winding number is given by
\bea
\nu_{\text{SVD}} = \frac{1}{2\pi i} \text{Tr} \Big[ \ln \big( \mathcal{X}_+ \mathcal{X}_{-}^{-1} \big) \Big],
\eea
where $\text{Tr}$ denotes the trace over the Hilbert space and $\ln$ represents the principal branch of the matrix logarithm. The projected position operator $\mathcal{X}_{\pm}$ is defined as
\bea
\mathcal{X}_{\pm} = U_{\pm}^{-1}\Gamma_{\pm} \mathcal{X} \Gamma_{\pm} U_{\pm},
\eea
where $\Gamma_{+} = \sum_{l=1}^{L} \ket{l,+} \bra{l,+}$ and $\Gamma_{-} = \sum_{l=1}^{L} \ket{l,-} \bra{l,-}$ are the projectors onto the $S=+1$ and $S=-1$ sublattices. The exponentiated position operator $\mathcal{X}$ is expressed as
\bea
\mathcal{X} = \sum_{l=1}^{L} e^{i \frac{2\pi}{L} l} \left( \ket{l,+} \bra{l,+} + \ket{l,-} \bra{l,-} \right),
\eea
with $L$ denoting the total number of unit cells and $l$ the cell index. The unitary matrix $U_{\pm}$ is derived from the chiral decomposition of the system. In the basis where the Hamiltonian takes an off-diagonal form,
\bea
H = \begin{pmatrix} 0 & M \\ M^\dagger & 0 \end{pmatrix},
\eea
we can perform the decomposition
\bea
M = U_{+} \Sigma U_{-}^\dagger,
\eea
where
\bea
\begin{gathered}
U_{+} = \left( \ket{\phi^{+}_1}, \ket{\phi^{+}_2}, \dots \right), \\ U_{-} = \left( \ket{\phi^{-}_1}, \ket{\phi^{-}_2}, \dots \right).
\end{gathered}
\eea
The states $\ket{\phi^{+}_n}$ and $\ket{\phi^{-}_n}$ are obtained from the eigenvectors of $MM^\dagger$ and $M^\dagger M$, respectively. The eigenstates of the Hamiltonian $H$ can be constructed using these two states. Specifically, the eigenstate with energy $\epsilon_n$ is given by $\ket{\psi_{n}} = \frac{1}{\sqrt{2}}(\ket{\phi^{+}_n}, \ket{\phi^{-}_n})^T$. The corresponding state with energy $-\epsilon_n$ is expressed as $S\ket{\psi_{n}} = \frac{1}{\sqrt{2}}(\ket{\phi^{+}_n}, -\ket{\phi^{-}_n})^T$. Notably, the multipole chiral number $N_{xy}$ in Eq.~\eqref{MCN} can be viewed as a higher-dimensional generalization of the SVD-based winding number $\nu_{\text{SVD}}$.

\begin{figure*}[t]
\centering
\begin{adjustbox}{max totalsize={\textwidth}{0.72\textheight},center}
\begin{minipage}{\textwidth}
\centering

\subfloat[]{\includegraphics[width=0.4\textwidth,valign=c]{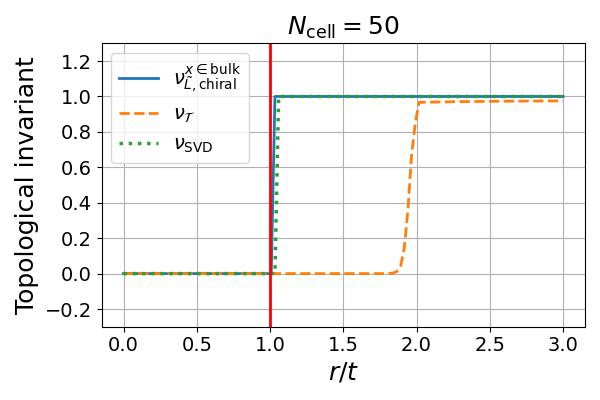}\label{chiral_invariants_50}}\hskip 0.5cm
\subfloat[]{\includegraphics[width=0.4\textwidth,valign=c]{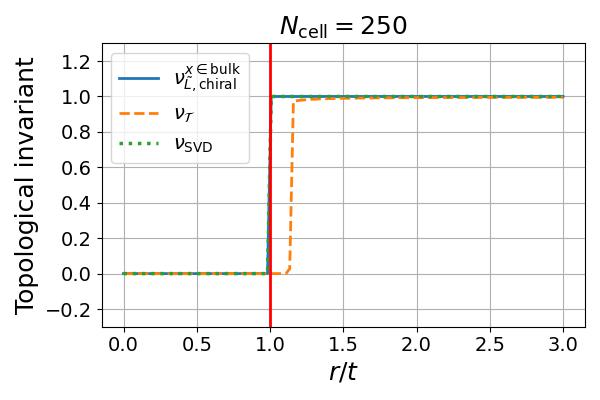}\label{chiral_invariants_250}}\\
\end{minipage}
\end{adjustbox}
\par
\caption{Numerical calculation of the three chiral-symmetric invariants for the SSH model given by Eq.~\eqref{SSH} under different system sizes. Different line styles denote different invariants, and the vertical red line indicates the exact transition position $(r/t=1)$. (a) Topological invariants as functions of the parameter ratio $r/t$ for $N_{\text{cell}}=50$, where both $\nu^{x}_{\tilde{L}}$ with $\kappa=0.05$ and $\nu_{\text{SVD}}$ show close agreement with the exact phase transition point at $t=r$. (b) The same invariants calculated for a larger system size of $N_{\text{cell}}=250$, highlighting that the transition location for $\nu_{\mathcal{T}}$ remains discernibly deviated from the exact transition point.}
\end{figure*}

To evaluate finite-size effects, we calculate the three chiral-symmetric invariants, the projector-based winding number $\nu_{\mathcal{T}}$, the SVD-based winding number $\nu_{\text{SVD}}$, and the spectral-localizer invariant $\nu^{x\in\text{bulk}}_{\tilde{L},\text{chiral}}$, for the SSH model with various $N_{\text{cell}}$ and track their transition locations against the exact phase transition at $r/t=1$. The numerical results are summarized in Figs.~\ref{chiral_invariants_50} and \ref{chiral_invariants_250}. At $N_{\text{cell}}=50$, both the spectral localizer (with a small $\kappa$) and the SVD approach yield transition points close to the exact value, with the spectral localizer showing the highest precision. The projector-based invariant, however, suffers from severe finite-size effects and fails to achieve quantization even when the parameter reaches $r/t=3$. Furthermore, even when the system size is increased to $N_{\text{cell}}=250$, the transition location computed via the projector-based method remains discernibly deviated from $r/t=1$. These results demonstrate that the spectral localizer has the smallest finite-size effects, closely followed by the SVD approach, while the projector-based method is the most sensitive to finite-size effects.

\section{Vanishing multipole chiral number of the Kekulé model in the topological limit}
\label{App_Kekule_MCN}

In this Appendix, we analytically show that the multipole chiral number of the Kekulé model vanishes in the topological limit $t_1=0$. We consider an $L\times L$ system with periodic boundary conditions along the two primitive lattice directions. The unit cells are labeled by $(x,y)$,
\bea
\mathbf R=x\mathbf a_1+y\mathbf a_2,\qquad x,y=0,\ldots,L-1,
\eea
where $x$ and $y$ are the unit-cell coordinates entering the operator $Q_{xy}^\pm$ in Eq.~\eqref{Q_xy def}.

In the $t_1=0$ limit, only the three intercell hopping channels proportional to $t_2$ remain. Since each of these hoppings connects unit cells separated by a fixed lattice displacement, it is convenient to represent them as operators acting on the unit-cell coordinates. We therefore introduce the translation operator
\bea
T_{(m,n)}\ket{x,y}=\ket{x+m,y+n},
\label{translation_operator}
\eea
where the unit-cell indices are understood modulo $L$.
Under periodic boundary conditions, we have
\bea
\begin{gathered}
T_{(m,n)}^\dagger=T_{(-m,-n)},\\ T_{(m,n)}^\dagger T_{(m,n)}=I.
\label{translation_unitary}
\end{gathered}
\eea

To express the intercell hoppings in terms of these translation operators, we first specify the complete single-particle basis. A basis state is denoted by $\ket{x,y;\alpha}$, where $(x,y)$ labels the unit cell and $\alpha=1,\ldots,6$ labels the internal orbital. According to the chiral operator defined in the main text, the internal orbitals are separated into the two chiral sectors
\bea
\begin{gathered}
S=+1:\quad \alpha\in\{1,3,5\},\\ S=-1:\quad \alpha\in\{2,4,6\}.
\end{gathered}
\eea
We therefore choose the complete basis ordering
\begin{widetext}
\bea
\big(
\{\ket{x,y;1}\}_{x,y},
\{\ket{x,y;3}\}_{x,y},
\{\ket{x,y;5}\}_{x,y};
\{\ket{x,y;2}\}_{x,y},
\{\ket{x,y;4}\}_{x,y},
\{\ket{x,y;6}\}_{x,y}
\big),
\eea
\end{widetext}
where $\{\ket{x,y;\alpha}\}_{x,y} \equiv \{\ket{x,y;\alpha}\mid x,y=0,\ldots,L-1\}$ denotes the $L^2$ unit-cell states associated with a fixed internal orbital $\alpha$. In this basis, the Hamiltonian takes the form
\bea
H_{\mathrm{Kekul\acute e}}=
\begin{pmatrix}
0&M\\
M^\dagger&0
\end{pmatrix}.
\eea
To construct $M$, we select the hopping terms in the
orientation $c^\dagger_{S=+1}c_{S=-1}$. The three $t_2$
hopping channels are then represented by
\begin{equation*}
\begin{aligned}
c_{\mathbf R,3}^\dagger c_{\mathbf R+\mathbf a_2,6}
&:\quad 6_{\mathbf R}\rightarrow3_{\mathbf R-\mathbf a_2}
\quad\Rightarrow\quad T_{(0,-1)},\\
c_{\mathbf R+\mathbf a_1,5}^\dagger c_{\mathbf R,2}
&:\quad 2_{\mathbf R}\rightarrow5_{\mathbf R+\mathbf a_1}
\quad\Rightarrow\quad T_{(1,0)},\\
c_{\mathbf R+\mathbf a_2-\mathbf a_1,1}^\dagger c_{\mathbf R,4}
&:\quad 4_{\mathbf R}\rightarrow1_{\mathbf R+\mathbf a_2-\mathbf a_1}
\quad\Rightarrow\quad T_{(-1,1)}.
\end{aligned}
\end{equation*}
Here, the second and third terms are the Hermitian-conjugate
partners of the corresponding terms written explicitly in
Eq.~\eqref{Kekule}. It follows that
\bea\label{M_Kekule_topological}
M=t_2
\begin{pmatrix}
0&T_{(-1,1)}&0\\
0&0&T_{(0,-1)}\\
T_{(1,0)}&0&0
\end{pmatrix}.
\eea
Here, $M$ is a $3\times3$ block matrix in the internal-orbital
space, with each translation operator $T_{(m,n)}$ represented
by an $L^2\times L^2$ matrix in the unit-cell space.

The above representation of $M$ in terms of translation operators allows the MCN to be evaluated straightforwardly. Since each $T_{(m,n)}$ is unitary and $M/t_2$ contains exactly one nonzero translation block in each block row and block column, $M/t_2$ is itself unitary,
\bea
\left(\frac{M}{t_2}\right)^\dagger\left(\frac{M}{t_2}\right)
=
\left(\frac{M}{t_2}\right)\left(\frac{M}{t_2}\right)^\dagger
=
I.
\eea
Therefore, for $t_2>0$, an exact singular-value decomposition
of $M$ is
\bea
M=U_+\Sigma U_-^\dagger
\eea
with
\bea
U_+=\frac{M}{t_2},\qquad U_-=I,\qquad\Sigma=t_2I.
\eea

According to Eq.~\eqref{MCN}, evaluating the MCN requires the transformed operators $\bar Q_{xy}^{\pm}=U_\pm^\dagger Q_{xy}^{\pm}U_\pm$. Their explicit forms can be obtained using the singular-value decomposition above. We first note that $Q_{xy}^{+}$ and $Q_{xy}^{-}$ depend only on the unit-cell coordinates and act identically on the internal orbitals within each chiral sector. They therefore have the same matrix representation,
\bea
Q_{xy}^{+}=Q_{xy}^{-}=I_3\otimes Q_{xy},
\eea
where $I_3$ is the identity matrix acting on the three internal orbitals within each chiral sector, and
\bea
Q_{xy}=\sum_{x,y}\ket{x,y}\exp\left(-i\frac{2\pi xy}{L^2}\right)\bra{x,y}.
\eea
Using $U_+=M/t_2$ and $U_-=I$, we then obtain
\bea
\begin{aligned}
\bar Q_{xy}^{+}
&=
\left(\frac{M}{t_2}\right)^\dagger
\left(I_3\otimes Q_{xy}\right)
\left(\frac{M}{t_2}\right),\\
\bar Q_{xy}^{-}
&=I_3\otimes Q_{xy}.
\end{aligned}
\eea
Substituting these transformed operators into Eq.~\eqref{MCN} gives
\bea
\begin{aligned}
N_{\mathrm{chiral}}^{xy}
&=\frac{1}{2\pi i}\Tr\log\Bigl[
\left(\frac{M}{t_2}\right)^\dagger(I_3\otimes Q_{xy})\\
&\hphantom{=\frac{1}{2\pi i}\Tr\log\Bigl[}{}
\times\left(\frac{M}{t_2}\right)(I_3\otimes Q_{xy}^\dagger)\Bigr].
\end{aligned}
\eea
The explicit block form of $M/t_2$ in Eq.~\eqref{M_Kekule_topological} now allows us to separate the three intercell hopping channels. In particular, the matrix inside the logarithm becomes block diagonal,
\begin{widetext}
\bea
\begin{aligned}
&\left(\frac{M}{t_2}\right)^\dagger
\left(I_3\otimes Q_{xy}\right)
\left(\frac{M}{t_2}\right)
\left(I_3\otimes Q_{xy}^\dagger\right)\\
&\qquad=
\begin{pmatrix}
T_{(1,0)}^\dagger Q_{xy}T_{(1,0)}Q_{xy}^\dagger&0&0\\
0&T_{(-1,1)}^\dagger Q_{xy}T_{(-1,1)}Q_{xy}^\dagger&0\\
0&0&T_{(0,-1)}^\dagger Q_{xy}T_{(0,-1)}Q_{xy}^\dagger
\end{pmatrix}.
\end{aligned}
\eea
\end{widetext}
Each diagonal block is associated with one translation. For a general translation $(m,n)$, we define
\bea
W_{(m,n)}=T_{(m,n)}^\dagger Q_{xy}T_{(m,n)}Q_{xy}^\dagger,
\eea
with the corresponding contribution to the MCN given by
\bea\label{B_delta_def}
\mathcal B_{(m,n)}=\frac{1}{2\pi i}\Tr\log W_{(m,n)}.
\eea
Since the trace of the logarithm separates into the contributions from the three diagonal blocks, the MCN reduces to
\bea\label{MCN_channel_sum}
N_{\mathrm{chiral}}^{xy}
=\mathcal B_{(0,-1)}+\mathcal B_{(1,0)}+\mathcal B_{(-1,1)}.
\eea

To evaluate the three terms in Eq.~\eqref{MCN_channel_sum}, we derive two relations among the translation contributions $\mathcal B_{(m,n)}$. First, we consider reversing the translation direction. Let $\boldsymbol{\delta}=(m,n)$. Using $T_{-\boldsymbol{\delta}}=T_{\boldsymbol{\delta}}^\dagger$, we have
\bea
\begin{aligned}
W_{-\boldsymbol{\delta}}
&=T_{-\boldsymbol{\delta}}^\dagger Q_{xy}T_{-\boldsymbol{\delta}}Q_{xy}^\dagger\\
&=T_{\boldsymbol{\delta}}Q_{xy}T_{\boldsymbol{\delta}}^\dagger Q_{xy}^\dagger.
\end{aligned}
\eea
On the other hand,
\bea
W_{\boldsymbol{\delta}}^\dagger
=
Q_{xy}T_{\boldsymbol{\delta}}^\dagger Q_{xy}^\dagger T_{\boldsymbol{\delta}}.
\eea
Therefore,
\bea
T_{\boldsymbol{\delta}}W_{\boldsymbol{\delta}}^\dagger T_{\boldsymbol{\delta}}^\dagger
=
W_{-\boldsymbol{\delta}},
\eea
so $W_{-\boldsymbol{\delta}}$ and
$W_{\boldsymbol{\delta}}^\dagger$ are unitarily equivalent.
Hence, if the eigenvalues of $W_{\boldsymbol{\delta}}$ are
$e^{i\theta_j}$, those of $W_{-\boldsymbol{\delta}}$ are
$e^{-i\theta_j}$, which gives
\bea\label{B_inverse}
\mathcal B_{(-m,-n)}=-\mathcal B_{(m,n)}.
\eea

The second relation follows from exchanging the two unit-cell coordinates. Since the system contains the same number $L$ of unit cells along both primitive lattice directions, we define the unitary coordinate-exchange operator
\bea
\mathcal P_{xy}\ket{x,y}=\ket{y,x}.
\eea
Because the phase entering $Q_{xy}$ is symmetric under $x\leftrightarrow y$,
\bea
\mathcal P_{xy}Q_{xy}\mathcal P_{xy}^\dagger=Q_{xy}.
\eea
At the same time,
\bea
\mathcal P_{xy}T_{(m,n)}\mathcal P_{xy}^\dagger=T_{(n,m)}.
\eea
Therefore,
\bea
W_{(n,m)}=\mathcal P_{xy}W_{(m,n)}\mathcal P_{xy}^\dagger.
\eea
Thus, $W_{(m,n)}$ and $W_{(n,m)}$ are unitarily equivalent and have identical eigenvalues, which gives
\bea\label{B_swap}
\mathcal B_{(m,n)}=\mathcal B_{(n,m)}.
\eea

We now apply Eqs.~\eqref{B_inverse} and \eqref{B_swap} to the three contributions in Eq.~\eqref{MCN_channel_sum}. For the first two terms,
\bea
\mathcal B_{(0,-1)}=\mathcal B_{(-1,0)}=-\mathcal B_{(1,0)},
\eea
where the first equality follows from Eq.~\eqref{B_swap} and the second from Eq.~\eqref{B_inverse}. Hence,
\bea
\mathcal B_{(0,-1)}+\mathcal B_{(1,0)}=0.
\eea
For the remaining diagonal contribution,
\bea
\mathcal B_{(-1,1)}=\mathcal B_{(1,-1)}=-\mathcal B_{(-1,1)},
\eea
where we again used Eqs.~\eqref{B_swap} and \eqref{B_inverse}. Therefore,
\bea
\mathcal B_{(-1,1)}=0.
\eea
Substituting these results into Eq.~\eqref{MCN_channel_sum}, we obtain
\bea
N_{\mathrm{chiral}}^{xy}=\mathcal B_{(0,-1)}+\mathcal B_{(1,0)}+\mathcal B_{(-1,1)}=0.
\eea
Thus, the multipole chiral number of the Kekulé model vanishes in the topological limit,
\bea
N_{\mathrm{chiral}}^{xy}(t_1=0)=0.
\eea
Nevertheless, for the open-boundary parallelogram geometry considered in the main text, the same $t_1=0$ limit supports topological $E=0$ corner states protected by chiral symmetry. This provides an explicit example in which chiral-symmetric corner states coexist with a vanishing MCN.

\bibliographystyle{apsrev4-2}
\bibliography{references}

\end{document}